\documentclass[lettersize,journal]{IEEEtran}
\normalsize
\ifCLASSINFOpdf
\else
\fi
\usepackage{graphicx}
\usepackage{amsmath}
\usepackage{amssymb}
\usepackage{amsthm}
\usepackage{array}
\usepackage{xcolor}
\usepackage{multirow}
\usepackage{float}
\usepackage{algorithm,algorithmic}
\usepackage{hyperref}
\usepackage{balance}
\usepackage{flushend}
\usepackage{microtype}
\usepackage{subfig}
\usepackage{cite}

\DisableLigatures[f]{encoding = *, family = * }

\hypersetup{pdftex,colorlinks=true,citecolor=blue,linkcolor=blue,urlcolor=blue,allcolors=blue}

\begin{document}

\title{When Future Communications Shift Toward Narrow Beams: A Forward-Looking Survey on Pointing Errors and Alignment Limits}

\author{Meysam~Ghanbari,~Mohammad~Taghi~Dabiri,
    ~Osamah~S.~Badarneh,
	~Mazen~Hasna,~{\it Senior Member,~IEEE},\\
    ~Yazan~H.~Al\text{-}Badarneh,~{\it Member,~IEEE},
	~Mustafa~K.~Alshawaqfeh,
	~and~Khalid~Qaraqe,~{\it Senior Member,~IEEE}

\thanks{M. Ghanbari and M.T. Dabiri are with the College of Science and Engineering, Hamad Bin Khalifa University, Doha, Qatar (Email: megh89467@hbku.edu.qa; mdabiri@hbku.edu.qa).}
\thanks{Osamah~S.~Badarneh and Mustafa~K.~Alshawaqfeh are with the Electrical Engineering Department, German-Jordanian University, Amman 11180, Jordan (Email: osamah.badarneh@gju.edu.jo; mustafa.alshawaqfeh@gju.edu.jo).}
	\thanks{Mazen Hasna is with the Department of Electrical Engineering, Qatar University, Doha, Qatar (Email: hasna@qu.edu.qa).}
\thanks{Yazan~H.~Al-Badarneh is with the Electrical Engineering Department, Jordan University, Amman 11942, Jordan (Email: yalbadarneh@ju.edu.jo).}
	\thanks{Khalid A. Qaraqe is with the College of Science and Engineering, Hamad Bin Khalifa University, Doha, Qatar, and also with the Department of Electrical Engineering, Texas A\&M University at Qatar, Doha, Qatar (Email: kqaraqe@hbku.edu.qa).}
	\thanks{This publication was made possible by NPRP14C-0909-210008 from the Qatar Research, Development and Innovation (QRDI) Fund (a member of Qatar Foundation).}

}


\maketitle

\begin{abstract}
Directional links in free-space optical (FSO), millimeter-wave (mmWave), and terahertz (THz) systems are a cornerstone of emerging 6G networks, yet their reliability is fundamentally limited by pointing errors and misalignment. Existing studies address this impairment using technology-specific definitions, models, and mitigation approaches, which hinders cross-domain comparison and transferable design insight. This survey provides a unified treatment of pointing errors across optical and high-frequency wireless communications. We establish consistent terminology and a cross-technology taxonomy of pointing errors, review angular misalignment and statistical distribution models, and analyze their impact on system performance. Mitigation techniques are systematically surveyed with emphasis on optical systems and their connection to underlying pointing-error models. The survey further provides a detailed examination of pointing-error effects in orbital angular momentum (OAM) links and quantum optical communications, and surveys the corresponding mitigation approaches tailored to mode-dependent impairments and quantum measurement constraints. The survey also outlines open challenges and future research directions. By consolidating fragmented literature into a coherent framework, this work supports consistent analysis and robust design of next-generation directional communication systems.
\end{abstract}


\section{Introduction}

\subsection{Background and Motivation}
\IEEEPARstart{F}{uture} wireless generations, most notably 6G, are expected to extend connectivity beyond today’s “best-effort broadband” into regimes where links must be simultaneously high-capacity, low-latency, and highly reliable under mobility and dynamic topologies. This vision is tightly coupled to a broader expansion of operational environments, including dense urban deployments with aggressive spatial reuse, factory and campus networks requiring deterministic performance, and wide-area coverage through aerial and non-terrestrial platforms, including unmanned aerial vehicles (UAVs), high-altitude platforms (HAPs), and low-Earth-orbit/geostationary-orbit (LEO/GEO) systems \cite{Fayad2025COMST,Ozturk2025IEEE_Network,Guirado2025TGCN,Dabiri2025HardLimiter,Dabiri2025JSAC_MRRUAV}. Across these scenarios, a recurring architectural shift is visible: 6G links are becoming increasingly beam-centric, relying on narrow beams and high-gain apertures to close the link budget and manage interference. The beam-centric paradigm is most evident at millimeter-wave (mmWave) and terahertz (THz), where severe path loss, molecular absorption (at THz), and hardware constraints make high-directional beamforming a necessity rather than a feature \cite{Liu2025NPJ,Zaman2021JMASS,Dabiri2025TAES_THzUAV}. It is also central to optical wireless and free-space optical (FSO), where the carrier physics enables very large bandwidths but the link fundamentally depends on precise spatial coupling between transmitter and receiver \cite{Wang2016LPT}, \cite{Kaushal2017COMST}. In both radio frequency (RF) and optical domains, the move toward narrower beams is not incremental: as frequency increases and aperture/array gains scale, beamwidth contracts, and the tolerance to misalignment can drop from degrees to millidegrees (RF) and to microradians (optical) depending on geometry, range, and optics \cite{Liu2025NPJ,Zaman2021JMASS,Wang2016LPT,Kaushal2017COMST}. This contraction transforms alignment from an implementation detail into a first-order system constraint.

What makes this shift particularly consequential in 6G is that high-directionality is arriving together with mobility and platform dynamics. Many target 6G scenarios involve moving endpoints or moving reflectors (UAV relays, drone base stations, vehicle-to-infrastructure, maritime links, and NTN backhaul), where the pointing state evolves continuously and can be dominated by vibration, jitter, attitude variations, and control-loop limitations \cite{Ozturk2025IEEE_Network}, \cite{Priyadarshani2024TAES}, \cite{Jahed2023OWCFSO}. Even in ground-based deployments, beam tracking must operate under user motion, blockage, and frequent topology changes, which forces repeated beam search, training, and refinement. The result is a design landscape where alignment accuracy and tracking agility directly trade off against training overhead, latency, energy consumption, and protocol complexity \cite{Pradhan2019TVT}, \cite{Singh2024LAWP}. In FSO and optical wireless, pointing-related impairments manifest through beam/spot displacement at the receiver, angle-of-arrival fluctuations, and coupling efficiency variations driven by platform motion and atmospheric effects (e.g., turbulence-induced beam wander). High-capacity optical links are therefore often inseparable from acquisition–tracking–pointing (ATP) mechanisms and receiver design choices such as aperture size and field-of-view (FoV). These design choices introduce nontrivial tradeoffs: increasing FoV and aperture can improve robustness to misalignment, but may also raise background noise and interference susceptibility, and can change the security and detectability characteristics of the link depending on the scenario \cite{Wang2016LPT,Kaushal2017COMST,Priyadarshani2024TAES}. As FSO extends from static terrestrial backhaul into UAV and satellite links, the pointing problem becomes even more central because the relative motion and attitude dynamics are no longer negligible.

In mmWave and THz systems, the same underlying phenomenon appears with different engineering primitives: directional array patterns, beam codebooks, hybrid beamforming constraints, beam squint in wideband systems, and beam management procedures operating under mobility and blockage \cite{Liu2025NPJ}, \cite{Zaman2021JMASS}, \cite{Pradhan2019TVT}. Misalignment here is not only a channel impairment; it is also a system overhead driver. Frequent beam re-training and tracking can consume time-frequency resources, inflate latency, and increase energy usage, while overly conservative strategies (e.g., widening beams) sacrifice peak gain and reduce spatial reuse. Consequently, accurate characterization of misalignment dynamics and their impact on effective gain is necessary to produce credible performance predictions and to inform beam management and scheduling policies \cite{Pradhan2019TVT}, \cite{Singh2024LAWP}. A key reason pointing errors deserve dedicated treatment is that they behave differently from classical fading models. In many directional links, misalignment acts as a multiplicative, geometry-driven attenuation whose statistics are governed by platform motion, tracking error, and beam geometry, often yielding distinctive distributions and time correlation structures. This makes system evaluation sensitive to the modeling choices used for misalignment (e.g., whether the error is treated as purely angular, as a combined angular–lateral displacement, as a quasi-static block process, or as a dynamic process driven by control/estimation loops) \cite{Kaushal2017COMST,Priyadarshani2024TAES,Pradhan2019TVT,Singh2024LAWP}. Because these choices directly shape outage, rate, and reliability predictions, they also influence how mitigation techniques should be compared and selected.

Beyond conventional RF and optical links, pointing and misalignment effects become even more consequential in emerging 6G technologies that encode information in the spatial structure of the field. In orbital angular momentum (OAM)-based communications, for example, misalignment can induce mode coupling and crosstalk, so that small pointing perturbations translate into a pronounced loss of modal purity and degraded detection performance \cite{Liu2025TAP,Dabiri2025LCOMM_OAMSecurity,Dabiri2025OAMISL}. In quantum optical communications, including quantum key distribution (QKD), pointing stability directly influences photon collection efficiency and received count rates, and therefore constrains achievable key rates and operating margins, particularly in free-space and platform-based deployments \cite{ArteagaDiaz2022ACCESS,Dabiri2025UAVQuantumFSO,Dabiri2025EntanglementLimits}. These regimes reinforce a broader 6G theme: tighter spatial selectivity can unlock new capabilities, but it simultaneously elevates alignment sensitivity and makes accurate modeling and robust mitigation indispensable.

Taken together, these trends establish pointing errors as a cross-technology impairment that spans high-frequency RF and optical systems, extends from terrestrial to non-terrestrial scenarios, and couples physical-layer performance with acquisition and tracking dynamics and protocol overhead \cite{Fayad2025COMST,Ozturk2025IEEE_Network,Guirado2025TGCN,Dabiri2025HardLimiter,Dabiri2025JSAC_MRRUAV,Liu2025NPJ,Zaman2021JMASS,Dabiri2025TAES_THzUAV,Jahed2023OWCFSO,Wang2016LPT,Kaushal2017COMST,Priyadarshani2024TAES,Pradhan2019TVT,Singh2024LAWP,Liu2025TAP,Dabiri2025LCOMM_OAMSecurity,Dabiri2025OAMISL,ArteagaDiaz2022ACCESS,Dabiri2025UAVQuantumFSO,Dabiri2025EntanglementLimits}. The practical question is therefore no longer whether directional links can deliver high capacity, but whether they can do so reliably under realistic mobility, platform dynamics, and tracking constraints. Addressing this reliability requires a clear and consistent understanding of what “pointing error” means across technologies. Optical and FSO studies often describe misalignment through spot displacement, angle-of-arrival fluctuations, and coupling efficiency, while mmWave and THz studies commonly express it through beam misalignment, beam training and tracking, and codebook-driven beam management. These differences are not merely linguistic; they imply different state variables, different statistical or dynamic models, and different performance endpoints such as outage, achievable rate, error probability, acquisition latency, and retraining overhead. Without a unified interpretation of definitions and modeling assumptions, results become difficult to compare and design insights do not transfer cleanly between directional technology stacks. This need for unification becomes even more pressing once mitigation is considered, because mitigation is inherently multi-layer and tightly coupled to the assumed pointing-error model and its timescale. Practical solutions range from stabilization and ATP, to beam management and tracking procedures, to diversity and signal-processing techniques, and increasingly learning-enabled estimation and control. Yet many existing works illuminate only one side of the problem, either analyzing misalignment under narrow modeling assumptions with limited mitigation discussion, or proposing mitigation methods evaluated in restricted settings that obscure broader applicability. The same fragmentation is amplified in emerging 6G technologies, where misalignment can manifest as structural distortion rather than simple power loss, for instance through OAM mode coupling, and can directly constrain operating margins in quantum optical links. This motivates an integrated survey treatment that connects definitions, models, performance analysis, and mitigation tradeoffs across mmWave, THz, and FSO, and extends the discussion to emerging technologies; accordingly, the next subsection standardizes key pointing-error concepts and frames the cross-technology viewpoint that guides the remainder of the paper.

\subsection{Pointing Errors in 6G Directional Links}
In this survey, pointing error refers to the mismatch between the intended and the realized spatial alignment of a directional link, resulting in a loss of effective coupling or array gain. Depending on the underlying technology and hardware, this misalignment may be expressed as an angular deviation between transmitter and receiver boresights, a lateral displacement of the beam footprint (or optical spot) relative to the receiver aperture, or a combined effect driven by platform motion and tracking imperfections. We focus on misalignment phenomena that materially affect link performance and system operation in beam-centric 6G, including their interaction with acquisition and tracking procedures; propagation effects are discussed insofar as they contribute to, or compound, misalignment-induced degradation. To enable a consistent discussion across FSO/optical wireless and mmWave/THz systems, we use a common classification of pointing-related effects that recur throughout the literature: angular misalignment (boresight error), beam/spot displacement at the receiver, orientation-induced mismatch (including polarization-related effects when relevant), and misalignment induced by beam training and tracking imperfections under mobility. In addition, emerging spatial-structure technologies introduce misalignment manifestations beyond pure power loss, such as mode coupling and crosstalk in OAM links, while wideband high-frequency RF systems may exhibit effects that functionally resemble mis pointing in beam space. Throughout the paper, we emphasize the shared system narrative across these domains: narrow-beam operation requires acquisition and tracking; residual misalignment determines the effective gain or coupling loss; and mitigation strategies trade robustness against overhead, complexity, and energy. The detailed taxonomy and the modeling foundations used to analyze pointing errors and their performance impact are developed in Section II.

\subsection{Related Surveys and Identified Gaps}

A growing body of survey literature has examined directionality and alignment-related challenges across different 6G-relevant communication technologies. These efforts are largely technology-specific, with separate survey streams for FSO communications, mmWave systems, and THz links. As a result, pointing errors and misalignment are typically discussed within isolated domains using technology-dependent terminology and perspectives. In this subsection, we review representative surveys in the FSO domain to position the present work within the broader survey landscape. In the FSO literature, early survey efforts mainly address alignment from a terminal and control perspective. The survey in \cite{Kaymak2018COMST} focuses on ATP mechanisms for mobile FSO systems, emphasizing hardware architectures, sensing approaches, and control strategies required to maintain alignment under mobility and platform dynamics. Other works adopt a more organizational viewpoint. The classification framework in \cite{Hamza2019COMST_Class} categorizes FSO links and systems according to deployment environments, configurations, and applications, providing a structured overview of the FSO landscape.

System-level perspectives are presented in \cite{AlGailani2021ACCESS}, which surveys FSO communication systems, links, and networks across multiple layers, including channel effects, system components, and network considerations. In this context, pointing and misalignment are discussed alongside other physical impairments such as atmospheric attenuation and turbulence. At higher protocol layers, the survey in \cite{LePham2022COMST} investigates retransmission-based error-control protocols for FSO communications, where alignment-related degradations are reflected indirectly through error statistics and reliability behavior. Network-centric views are emphasized in \cite{Tarhouni2025OJCOMS}, which surveys FSO mesh networks and discusses architectural design, connectivity, and robustness in multi-hop optical deployments. Environmental and deployment factors are highlighted in the systematic survey of outdoor FSO systems in \cite{Altakhaineh2025ACCESS}, which reviews how weather conditions, link distance, and atmospheric effects influence overall system performance. Finally, application-specific perspectives are considered in \cite{Karmous2025COMST}, which surveys optical communications for deep-space scenarios, where extremely narrow beams make precise pointing and tracking a critical practical concern and where advanced optical techniques such as OAM and quantum communications are discussed. Overall, existing FSO surveys provide valuable insights into alignment mechanisms, system design, and deployment challenges, but they remain fragmented across layers, applications, and technologies, limiting cross-scenario comparability and unified understanding of pointing-related effects.

Building on the FSO-focused surveys, a parallel body of literature has examined alignment and directionality challenges in mmWave communications, where narrow beams are essential to counteract severe path loss. In this domain, pointing-related effects are typically discussed in terms of beam misalignment, beam training overhead, and tracking robustness, rather than through explicit geometric pointing-error models. Early surveys emphasize the physical-layer foundations of mmWave systems. The work in \cite{Hemadeh2018COMST} provides a comprehensive overview of mmWave channel characteristics, antenna constructions, and link-budget considerations, highlighting the dependence on highly directional transmission and the associated sensitivity to beam misalignment. Broader physical-layer perspectives are offered in \cite{He2021JPROC}, which surveys mmWave transmission technologies, including beamforming architectures and antenna arrays, and discusses beam alignment primarily as an operational requirement within system design.

\begin{figure*}[t]
\centering
\includegraphics[width=7in]{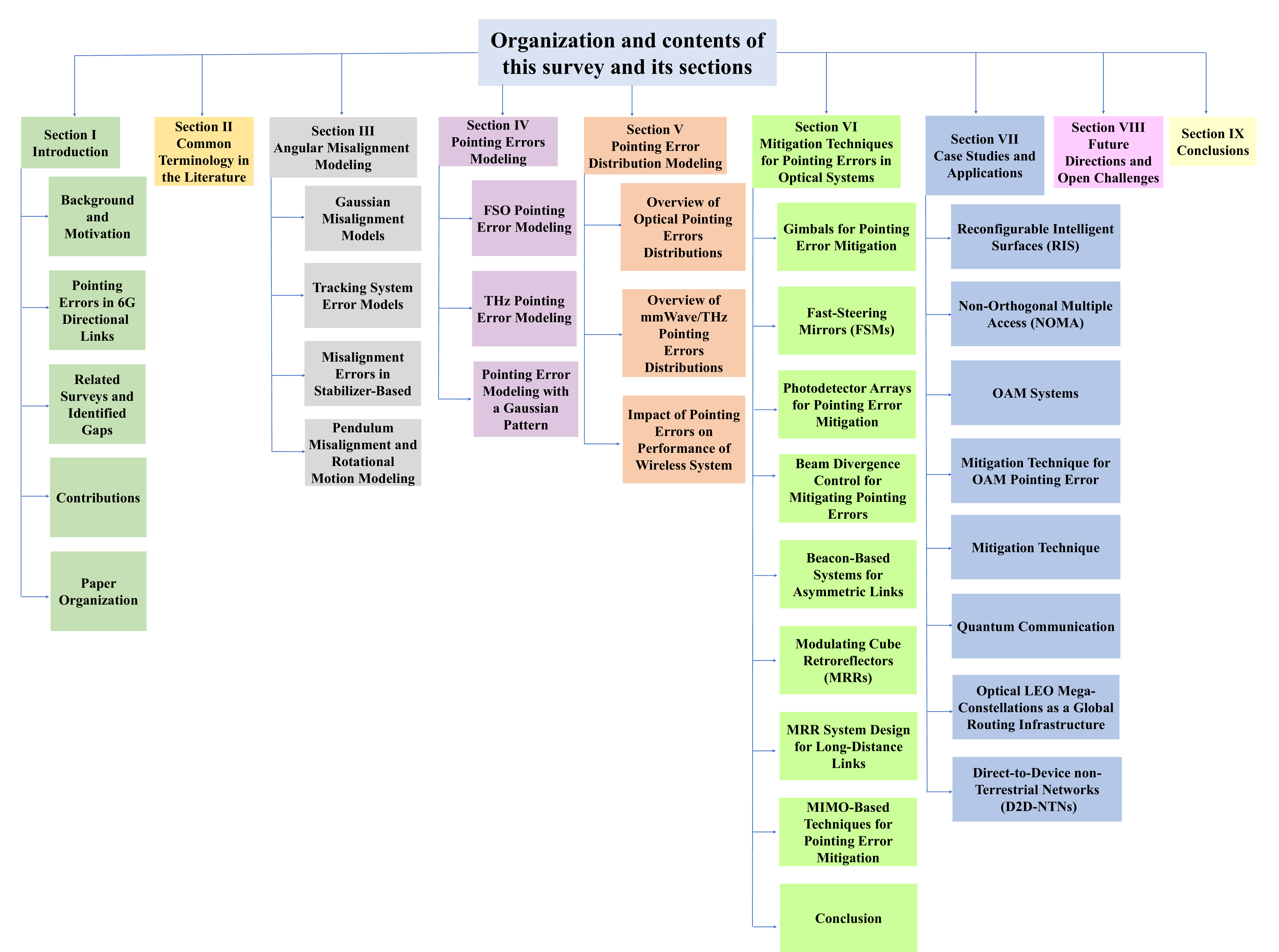}
\caption{Organization of the paper}
\label{fig:roadmap}
\end{figure*}

Mobility-centric challenges are addressed in \cite{Li2022COMST}, which focuses on mobility support for mmWave communications and identifies beam misalignment and frequent beam reconfiguration as key obstacles under user movement. More recent surveys shift attention to beam management as a central system function. The survey in \cite{Xue2024COMST} reviews beam management procedures for mmWave and THz systems toward 6G, covering initial access, beam training, tracking, and recovery, with emphasis on protocol design and overhead reduction. Application-driven perspectives further refine the treatment of alignment. The survey in \cite{Tan2024COMST} concentrates on beam alignment in mmWave vehicle-to-everything (V2X) communications, categorizing alignment techniques and performance metrics tailored to highly dynamic vehicular environments. Finally, the survey in \cite{Adnan2025ACCESS} examines positioning for mmWave distributed MIMO systems, focusing on localization techniques and architectures, while implicitly highlighting the tight coupling between positioning accuracy, beam alignment reliability, and link performance. Overall, mmWave surveys provide detailed coverage of beam-centric system design, mobility support, and protocol-level alignment procedures, but they largely frame misalignment operationally, limiting unified abstraction of pointing-related impairments across applications and technologies.

In the THz literature, survey efforts largely extend mmWave concepts to higher frequencies while addressing additional propagation, hardware, and architectural constraints. Consequently, alignment and directionality are discussed throughout THz surveys, but typically as part of broader treatments of beamforming, channel sparsity, and system feasibility rather than as standalone impairments. Several surveys focus on THz channel behavior and propagation fundamentals. The holistic survey in \cite{Han2022COMST} reviews THz channel measurements, modeling, and analysis, covering molecular absorption, near-field effects, sparse multipath characteristics, and highly directional propagation. A broader communication-oriented perspective is provided in \cite{Sharma2025OJCOMS}, which surveys THz communication technologies, including channel models, modulation schemes, signal processing methods, and application scenarios, with beamforming and beam steering highlighted as essential enablers.

Technology-driven viewpoints are emphasized in \cite{Thomas2025OJCOMS}, which reviews advances in THz circuits, antennas, and experimental demonstrations for 6G systems, underscoring the reliance on ultra-directive antennas and large-scale arrays. Environment-aware and reconfigurable solutions are discussed in \cite{Ahmed2024ACCESS}, which surveys reconfigurable intelligent surface (RIS) technologies for THz communications and their role in coverage enhancement and blockage mitigation in highly directional environments.Cross-band perspectives that include sub-THz links are presented in \cite{Chukhno2024COMST}, which surveys multicasting techniques for mmWave and sub-THz systems, emphasizing directional transmission and beam coordination. Finally, \cite{Zhang2026TNSE} considers an integrated view of THz sensing, communication, and networking, highlighting the role of high angular resolution and precise beam control for joint sensing–communication operation. Overall, while THz surveys consistently recognize the importance of narrow beams and alignment-sensitive operation, very few works explicitly discuss pointing and misalignment errors in any systematic or unified manner.
\begin{table*}[t]
\centering
\normalsize
\caption{Comparison of Representative Surveys Related to Pointing and Alignment Effects in FSO, mmWave, and THz Communications, Highlighting Coverage of Definitions (Defs.), Pointing-Error Modeling (Model), Performance Analysis (Perf.), Mitigation Strategies (Mitig.), Emerging Technologies (Emerg.), and Platform Scope (Platform).}
\label{tab:survey_comparison}

\begin{tabular}{|p{2.1cm}|c|c|c|c|c|c|c|}
\hline
\textbf{Domain} & \textbf{Ref.} & \textbf{Defs.} & \textbf{Model} & \textbf{Perf.} & \textbf{Mitig.} & \textbf{Emerg.} & \textbf{Platform} \\
\hline

\multirow{7}{*}{FSO}
 & \cite{Kaymak2018COMST} & $\checkmark$ & $\bullet$ & $\bullet$ & $\checkmark$ & X & mixed \\ \cline{2-8}
 & \cite{Hamza2019COMST_Class} & $\bullet$ & X & X & $\bullet$ & X & mixed \\ \cline{2-8}
 & \cite{AlGailani2021ACCESS} & $\bullet$ & $\bullet$ & $\checkmark$ & $\bullet$ & X & mixed \\ \cline{2-8}
 & \cite{LePham2022COMST} & X & X & $\checkmark$ & $\bullet$ & X & Terrestrial \\ \cline{2-8}
 & \cite{Tarhouni2025OJCOMS} & $\bullet$ & $\bullet$ & X & $\bullet$ & $\bullet$ & mixed \\ \cline{2-8}
 & \cite{Altakhaineh2025ACCESS} & $\checkmark$ & X & X & $\bullet$ & $\bullet$ & mixed \\ \cline{2-8}
 & \cite{Karmous2025COMST} & $\bullet$ & X & X & $\bullet$ & $\bullet$ & Non-terrestrial \\
\hline

\multirow{6}{*}{mmWave}
 & \cite{Hemadeh2018COMST} & X & X & $\checkmark$ & X & X & Terrestrial \\ \cline{2-8}
 & \cite{He2021JPROC} & $\bullet$ & $\bullet$ & $\bullet$ & $\checkmark$ & X & Terrestrial \\ \cline{2-8}
 & \cite{Li2022COMST} & $\bullet$ & X & $\bullet$ & $\checkmark$ & $\bullet$ & Terrestrial \\ \cline{2-8}
 & \cite{Xue2024COMST} & $\bullet$ & X & X & $\checkmark$ & X & mixed \\ \cline{2-8}
 & \cite{Tan2024COMST} & $\checkmark$ & $\bullet$ & X & X & X & Terrestrial \\ \cline{2-8}
 & \cite{Adnan2025ACCESS} & $\bullet$ & X & X & $\checkmark$ & X & Terrestrial \\
\hline

\multirow{6}{*}{THz}
 & \cite{Han2022COMST} & $\checkmark$ & X & X & X & X & Terrestrial \\ \cline{2-8}
 & \cite{Sharma2025OJCOMS} & $\bullet$ & X & X & X & $\bullet$ & mixed \\ \cline{2-8}
 & \cite{Thomas2025OJCOMS} & $\bullet$ & X & X & X & X & Terrestrial \\ \cline{2-8}
 & \cite{Ahmed2024ACCESS} & $\bullet$ & X & X & X & X & Terrestrial \\ \cline{2-8}
 & \cite{Chukhno2024COMST} & X & X & X & X & $\bullet$ & Terrestrial \\ \cline{2-8}
 & \cite{Zhang2026TNSE} & X & X & X & X & $\bullet$ & mixed \\
\hline

Hybrid & Our work & $\checkmark$ & $\checkmark$ & $\checkmark$ & $\checkmark$ & $\checkmark$ & mixed \\
\hline
\end{tabular}

\vspace{1mm}
\caption*{\textit{Note:} $\checkmark$ indicates comprehensive coverage of the considered aspect; 
$\bullet$ denotes partial or limited discussion; 
X indicates that the aspect is not addressed.}
\end{table*}

Despite the breadth and maturity of the surveyed literature, a unifying treatment of pointing errors across directional 6G technologies remains absent. Existing surveys predominantly address alignment-related issues within technology-specific silos: FSO surveys emphasize acquisition, tracking, and implementation constraints or embed misalignment among environmental impairments; mmWave surveys frame misalignment through beam training, tracking, and beam-management procedures; and THz surveys absorb alignment sensitivity into discussions of beamforming, channel sparsity, or hardware feasibility. Consequently, pointing errors are characterized using inconsistent definitions and taxonomies (Defs.), modeled with incompatible statistical abstractions (Model), and evaluated using heterogeneous performance metrics (Perf.), which limits cross-paper comparability and obstructs the transfer of design insights across frequency regimes.

Furthermore, while individual surveys discuss performance degradation or mitigation within their respective domains, the interdependence between pointing-error modeling, performance analysis, and mitigation design is rarely addressed in a systematic and unified manner. In particular, existing works do not consistently map pointing-error models and distributions to their implications for outage, achievable rate, error probability, or control overhead, nor do they organize mitigation strategies (Mitig.) across hardware, physical-layer, beam-management, and learning-assisted control mechanisms within a single framework. This fragmentation becomes even more pronounced for emerging technologies (Emerg.), and with respect to the underlying platform context (Platform), as existing surveys rarely distinguish between terrestrial, non-terrestrial, and mixed deployments, where misalignment characteristics and mitigation requirements can differ substantially. These gaps are summarized in Table~\ref{tab:survey_comparison}, which compares representative surveys across FSO, mmWave, and THz domains and highlights the absence of a cross-technology perspective. In contrast to prior survey efforts, the present work treats pointing error as a fundamental impairment shared across optical, mmWave, and THz systems, and provides a unified framework that harmonizes definitions and distribution models, systematically connects modeling assumptions to performance analysis and system design, organizes mitigation techniques within a coherent multi-layer taxonomy, and explicitly incorporates emerging 6G technologies and their corresponding mitigation strategies across terrestrial, non-terrestrial, and mixed platform deployments. This integrated viewpoint enables consistent comparison, informed design, and transferable insights for next-generation directional communication systems.

\subsection{Contributions of This Survey}

This survey makes several distinct and complementary contributions to the literature on directional communications, with a deliberate focus on pointing errors as a first-order impairment spanning optical, mmWave, and THz systems.

First, the survey establishes a cross-technology unification of pointing-error definitions and taxonomies. In contrast to prior surveys that rely on technology-specific terminology and isolated abstractions, this work harmonizes the definition and classification of pointing errors across FSO, mmWave, and THz domains. The unified framework clarifies the underlying state variables used to represent misalignment—such as angular deviation, beam displacement, and beam–receiver mismatch—and introduces consistent terminology that enables meaningful comparison across heterogeneous studies and frequency regimes.

Second, the survey provides a systematic consolidation of pointing-error modeling approaches, with particular emphasis on statistical, distribution-based, and dynamic models. Existing models are reviewed and organized according to their assumptions, temporal characteristics, and applicability to different platforms and propagation environments. By aligning modeling practices across optical and high-frequency RF systems, the survey reveals structural similarities and key differences, enabling the transfer of modeling insights across technologies and supporting more informed model selection.

Third, this work explicitly connects pointing-error models to performance analysis and system design considerations. Rather than treating modeling and evaluation in isolation, the survey synthesizes how pointing errors propagate into key performance metrics, including outage probability, achievable rate, error probability, and training or control overhead. This integrated perspective highlights the role of misalignment as a cross-layer impairment and clarifies its impact on end-to-end system behavior across diverse directional communication technologies.

Fourth, the survey presents a comprehensive and structured taxonomy of mitigation techniques, spanning hardware-level solutions (e.g., stabilization and acquisition–tracking–pointing mechanisms), physical-layer and beamforming strategies, beam management and tracking procedures, signal-processing techniques, and learning-assisted estimation and control methods. By explicitly linking mitigation strategies to underlying pointing-error models and performance objectives, this work moves beyond fragmented discussions in prior surveys and offers a holistic, design-oriented framework.

Fifth, the survey extends pointing-error analysis to emerging 6G paradigms, including spatially structured transmissions and quantum optical communications. To the best of our knowledge, this work is among the earliest structured survey efforts that systematically examine the role of pointing errors in OAM–based communications and free-space quantum communication systems, and that review corresponding mitigation strategies within a unified, cross-technology framework.

Sixth, this survey is one of the early works to explicitly discuss optical LEO mega-constellations as a global routing infrastructure and their role in enabling direct-to-device non-terrestrial networks (D2D-NTNs). The survey highlights how narrow-beam optical inter-satellite and space-to-device links fundamentally elevate pointing accuracy from an implementation detail to a network-level constraint, influencing routing reliability, connectivity continuity, and end-user accessibility in future global-scale architectures.

Finally, the survey provides a comparative overview of existing survey literature, summarized in Table~\ref{tab:survey_comparison}, which contrasts representative works across FSO, mmWave, and THz domains from a holistic perspective. The comparison emphasizes differences in scope, analytical depth, technological coverage, and treatment of misalignment-related phenomena, thereby highlighting limitations in prior surveys and clearly positioning the present work as a unifying, cross-technology reference.

Collectively, these contributions establish a coherent foundation for understanding, modeling, and mitigating pointing errors across diverse directional communication technologies, and provide a unified framework intended to support consistent analysis, informed system design, and future research in 6G and beyond.

\subsection{Organization of the Paper}
The remainder of this paper is organized as follows. Section II establishes a unified terminology and taxonomy for pointing errors across optical, mmWave, and THz systems. Section III reviews angular misalignment models and dynamic effects. Section IV presents detailed pointing-error modeling approaches for FSO and mmWave/THz systems. Section V consolidates the statistical distribution models of pointing errors for terrestrial and non-terrestrial scenarios and examines the resulting impact of pointing errors on system performance. Section VI surveys mitigation techniques. Section VII discusses representative case studies and applications in emerging technologies and analyzes the impact of pointing errors on these systems. Section VIII outlines open challenges and future research directions, and Section IX concludes the paper. A visual overview of the paper structure is provided in Fig.~\ref{fig:roadmap}.

\section{Common Terminology in the Literature}


Various types of alignment errors and distortions arise due to environmental factors, equipment limitations, and system design. This section provides an overview of the key terminologies used in the literature to describe these errors, with a focus on pointing errors, polarization-related misalignments, and other notable issues affecting system performance.

\subsubsection{Pointing Error}
Pointing error refers to the angular misalignment between the intended direction of the transmitted signal and the actual direction. This can be a fixed deviation or a dynamic error caused by environmental factors or platform movement. Pointing errors result in decreased signal strength and reduced communication efficiency.

\subsubsection{Beam Wobbling}
Beam wobbling is a specific case of pointing error that is often mentioned in the literature of millimeter-wave and terahertz communication systems \cite{azari2022thz, yang2022impact}. It describes rapid and continuous fluctuations in the beam’s direction, often caused by mechanical vibrations or movement of the transmission platform (e.g., UAVs or satellites). This error disrupts signal targeting and reduces the system's reliability by causing frequent deviations from the intended path.

\subsubsection{Beam Wandering}
Beam wandering refers to the slow, random displacement or deviation in the propagation axis of the beam, often caused by changes in air pressure or atmospheric turbulence \cite{berman2007beam,briantcev2023beam}. This phenomenon is particularly significant in long-distance links, such as ground-to-satellite communication, where variations in altitude lead to substantial changes in air pressure. These pressure changes cause variations in the refractive index of the atmosphere, resulting in random deviations in the beam’s propagation axis. 

\subsubsection{Spot Wander}
In FSO communications, to reduce background light interference from sources like the sun, the receiver's field of view is often narrowed using a lens \cite{trung2021performance,sridhar2022performance}. This lens focuses the incoming optical beam onto a focal point where a photodetector or optical fiber is placed. Spot wander occurs when random fluctuations in the angle of the incoming light cause the beam’s position on the focal point to shift randomly \cite{du2010measurements}. These random fluctuations, known as spot wandering, lead to variations in the location of the beam on the photodetector or fiber, resulting in decreased signal accuracy and potential data loss.

\subsubsection{Beam Squint}
Beam squint refers to the frequency-dependent deviation in the beam direction in wideband directional antennas, such as those used in mmWave or THz systems \cite{gao2023integrated,wan2021hybrid, zhang2020hybrid}. As the frequency changes, the angle of the beam shifts, leading to misalignment between the transmitted and received signals across different frequencies.

\subsubsection{Polarization Misalignment}
\begin{figure}
	\begin{center} 
		\includegraphics[width=3.35 in]{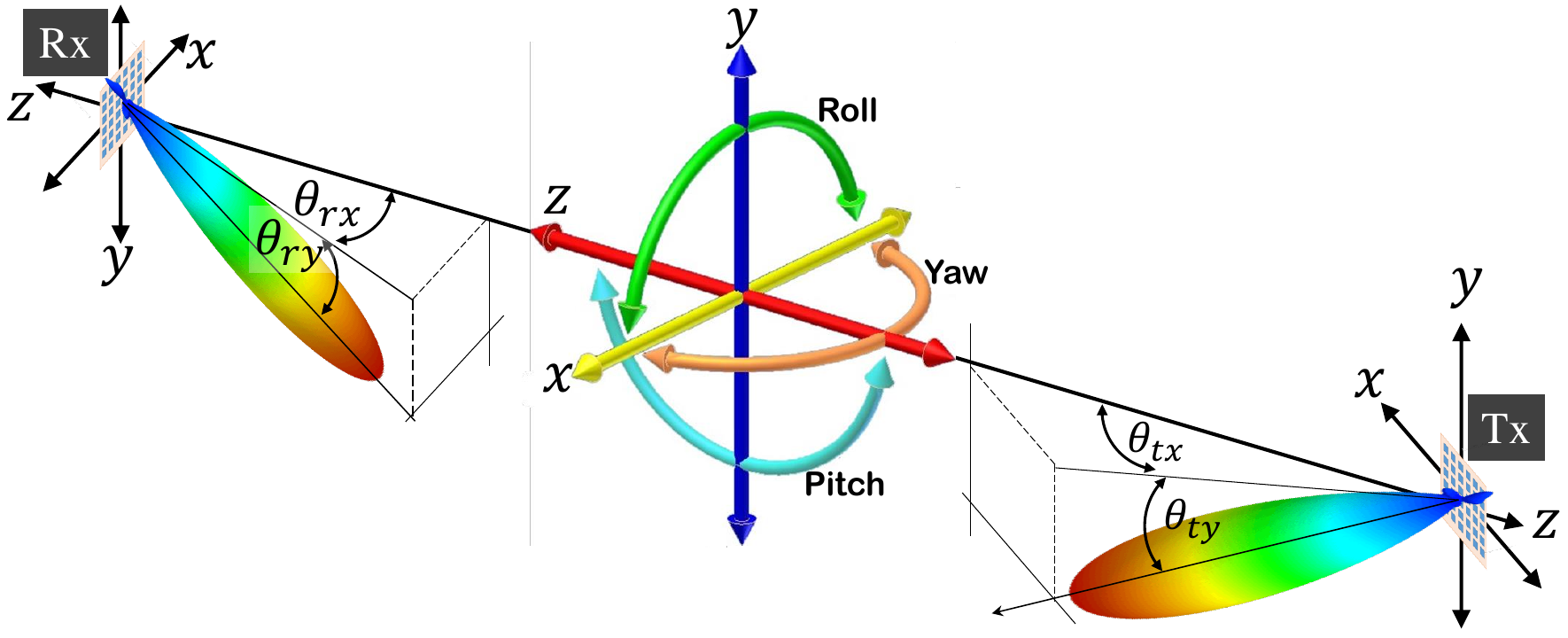}
		\caption{If the z-axis represents the propagation direction, fluctuations along the yaw and pitch angles (in the x-z and y-z planes) result in pointing errors or beam wobbling. Meanwhile, fluctuations along the roll axis cause polarization-related errors, affecting the polarization alignment between the transmitter and receiver \cite{dabiri2022general}.} 
		\label{po1}
	\end{center}
\end{figure}

Polarization misalignment occurs when the polarization state of the transmitted signal does not align with the receiver’s polarization axis \cite{zhang2009effects}. This leads to a reduction in the received signal power and can significantly degrade system performance, particularly in optical and mmWave communications. As illustrated in Fig. \ref{po1}, fluctuations along the roll axis result in polarization misalignment errors, while fluctuations along the yaw and pitch axes lead to pointing errors.

\subsubsection{Cross-Polarization Interference}
Cross-polarization interference occurs when signals with different polarization states interfere with each other, typically due to improper alignment of polarization states between the transmitter and receiver \cite{core2006cross, lankl1988cross}. This causes crosstalk between channels and reduces the efficiency of communication systems that rely on polarization diversity.

\subsubsection{Mode Crosstalk in OAM Systems}
In OAM systems, mode crosstalk happens when the transmitted OAM modes are not properly distinguished at the receiver due to misalignment or other distortions. This results in interference between different OAM modes, degrading the capacity and reliability of the system \cite{liu2023broadband, liu2018multiplexed, wang2023synthesizing}.

\subsubsection{Chromatic Aberration (Dispersion-Induced Pointing Error)}
Chromatic aberration occurs when different wavelengths within a signal experience varying angles of refraction as they pass through lenses or optical systems, leading to dispersion \cite{jaeken2011peripheral}. This can cause different parts of the signal spectrum to deviate, resulting in what is called dispersion-induced pointing error. It primarily affects systems with large bandwidths or wide spectral ranges.

\begin{table*}[h!]
\centering
\caption{Common Terminology in the Literature Related to Pointing Errors and Misalignment}

\begin{tabular}{|p{4cm}||p{12cm}|}
\hline
\textbf{Error Type} & \textbf{Description} \\ \hline \hline
Pointing Error & Angular misalignment between the intended direction and actual direction of the transmitted signal, caused by environmental factors or platform movement. \\ \hline

Beam Wobbling & Rapid and continuous fluctuations in the beam’s direction, often caused by mechanical vibrations or movement, especially in mmWave and terahertz systems. \\ \hline

Beam Wandering & Slow, random displacement in the beam’s propagation axis, typically caused by atmospheric turbulence or changes in air pressure. This is especially critical in long-distance links like ground-to-satellite communication. \\ \hline

Spot Wander & In FSO systems, random fluctuations in the angle of incoming light cause shifts in the beam’s focal point on the receiver, leading to decreased signal accuracy and potential data loss. \\ \hline

Beam Squint & Frequency-dependent deviation in the beam’s direction in wideband directional antennas, causing misalignment as frequency changes. \\ \hline

Polarization Misalignment & Misalignment between the polarization state of the transmitted signal and the receiver’s polarization axis, leading to reduced signal power and degraded system performance. \\ \hline

Cross-Polarization Interference & Occurs when signals with different polarization states interfere, typically due to improper polarization alignment, leading to crosstalk between channels. \\ \hline

Mode Crosstalk in OAM Systems & Misalignment or distortion in OAM systems, leading to interference between transmitted OAM modes and reduced system capacity and reliability. \\ \hline

Chromatic Aberration (Dispersion-Induced Pointing Error) & Occurs when different wavelengths experience varying angles of refraction, causing dispersion and deviations in the signal’s spectrum. Common in systems with large bandwidths. \\ \hline
\end{tabular}
\label{tab:1}
\end{table*}



\begin{table}[h!]
\centering
\caption{Notation Definitions}
\begin{tabular}{|c|p{5.5cm}|}
\hline
\textbf{Notation} & \textbf{Definition} \\ \hline
$\theta_{tx}$ & Transmitter angular fluctuation along the x-axis \\ \hline
$\theta_{ty}$ & Transmitter angular fluctuation along the y-axis \\ \hline
$\theta_{rx}$ & Receiver angular fluctuation along the x-axis \\ \hline
$\theta_{ry}$ & Receiver angular fluctuation along the y-axis \\ \hline
$\theta_{ax}$ & Angle of Arrival (AoA) along the x-axis \\ \hline
$\theta_{ay}$ & Angle of Arrival (AoA) along the y-axis \\ \hline
$\theta_{dx}$ & Angle of Departure (AoD) along the x-axis \\ \hline
$\theta_{dy}$ & Angle of Departure (AoD) along the y-axis \\ \hline
$r_a$ & Radius of the optical lens or aperture \\ \hline
$r_d$ & Radius of the optical fiber or photodetector cross-section \\ \hline
$Z$ & Length of the communication link \\ \hline
\end{tabular}
\label{tab:2}
\end{table}
To avoid ambiguity and ensure consistency across the surveyed literature, Table~\ref{tab:1} summarizes common terminology used to describe pointing errors and misalignment phenomena, while Table~\ref{tab:2} lists the notation and symbol definitions adopted throughout this paper.

\section{Angular Misalignment Modeling}  
In addition to the antenna pattern, angular misalignment is another critical parameter that significantly contributes to pointing error and is essential for accurate modeling of beam misalignment \cite{huang2017free, lopresti2007mitigating}. Angular misalignment refers to the deviation of the transmitter or receiver from its intended path as shown in Fig. \ref{po1}. Initially, misalignment models were primarily developed for fixed terrestrial links, where both the transmitter and receiver remained stationary \cite{farid2007outage, bhatnagar2016performance, yang2014free,sandalidis2008ber}. However, with advancements in technology and the implementation of sophisticated tracking methods, the use of FSO links and directional millimeter-wave or terahertz antennas for mobile devices has expanded significantly \cite{agheli2022high,khallaf2019uav,taheri2017provisioning, dabiri2019blind, khallaf2021comprehensive}. Given the high sensitivity of directional links, particularly laser-based FSO systems, the introduction of mobility adds new dimensions of error. Mobile platforms introduce additional errors associated with random fluctuations, changes in velocity, acceleration, and tracking inaccuracies. Each of these factors has led to the development of novel models for analyzing misalignment and pointing errors \cite{dabiri2019tractable}.

\subsection{Gaussian Misalignment Models}

One of the key challenges in FSO, terahertz, and millimeter-wave directional communications is the lack of obstacle penetration, necessitating a clear LoS between the transmitter and receiver \cite{dabiri2024coverage}. With the rapid advancement of technology, UAVs have become a crucial solution in establishing line-of-sight (LoS) links due to their high maneuverability, enabling them to relay signals between nodes \cite{lyu2019network, sabzehali20213d}. However, the inherent instability of UAV platforms often introduces both angular and positional misalignment errors \cite{dabiri2018channel}. These errors can arise from various factors, including random atmospheric pressure variations, control system inaccuracies (which themselves depend on multiple random parameters), tracking system errors, wind speed fluctuations, and mechanical vibrations of the UAV platform \cite{verbeke2016vibration, plasencia2012modelling}. Additionally, external environmental factors such as temperature changes and UAV body dynamics can further contribute to these errors \cite{banerjee2023vibration}.

Given the multiple sources of misalignment, the overall error is influenced by numerous independent random variables. According to the Central Limit Theorem, the sum of a large number of independent random variables tends to follow a Gaussian distribution  \cite{dabiri2018channel}. As a result, it is often shown that the distribution of misalignment errors typically follows a Gaussian profile, unless there is a dominant source of error with a specific distribution. Therefore, in much of the early work on UAV-based link modeling and performance analysis, Gaussian models have been widely adopted to represent misalignment errors.

\subsection{Tracking System Error Models}
In high-speed directional communication scenarios, such as communication with a fast-moving vehicle, the dominant source of misalignment often arises from the limitations and inaccuracies of the tracking system. Tracking errors occur due to the system's inability to precisely follow the rapid changes in the position and orientation of the mobile platform. These errors can significantly affect the overall performance of the communication link, and their impact depends heavily on the specific tracking technique employed. Different tracking techniques, such as mechanical tracking, electronic beam steering, or hybrid approaches, introduce varying types of errors. For example, mechanical tracking systems, which involve physical movement of the antenna or optical device, often suffer from latency and mechanical vibration, while electronic beam steering systems may introduce phase noise or resolution errors due to digital limitations. The distribution of tracking errors can vary based on the method and environmental conditions. Common models include \cite{Kaymak2018COMST}:
\begin{itemize}
    \item \textit{Gaussian Distribution}: In cases where small, random errors accumulate, the overall tracking error tends to follow a Gaussian distribution, particularly when the errors are the result of multiple independent sources.
    
    \item \textit{Laplace Distribution}: For tracking systems with sharp changes or impulsive behaviors (such as sudden shifts in vehicle speed or direction), the tracking error may be better modeled using a Laplace distribution, which captures the heavy tails and sharper transitions.
    
    \item \textit{Uniform Distribution}: In some simplified models, where the tracking error is assumed to be equally likely within a certain range, a uniform distribution is applied, especially when precision in tracking is constrained within specific boundaries.
    
    \item \textit{Markov Processes}: For dynamic systems where the error in one time step is dependent on the previous state (such as systems with feedback delay or prediction errors), tracking error can be modeled as a Markov process.
\end{itemize}

An additional key parameter influencing tracking system error is the update rate of the tracking system. Between updates, the misalignment error can evolve, and its model may change based on the dynamics of the mobile platform. This needs to be considered in performance analysis, as the time between tracking updates directly impacts the accuracy and stability of the link.

\subsection{Misalignment Errors in Stabilizer-Based Systems}

One effective solution for mitigating misalignment errors in directional communication systems is the use of stabilizers. Stabilizers are mechanical or electronic systems designed to maintain the alignment of the communication beam by compensating for angular fluctuations or platform movements. However, the performance of a stabilizer is determined by two critical factors: the response speed and angular accuracy. These parameters directly influence the effectiveness of the stabilizer in reducing pointing errors but also significantly impact the system's overall weight, power consumption, and cost. In this section, we will explore both factors and their trade-offs.

\subsubsection{Response Speed of Stabilizers}
The response speed of a stabilizer is a crucial parameter, especially in dynamic environments with rapid angular fluctuations, such as UAV-based communication systems. The speed of the stabilizer’s response must be significantly faster than the coherence time of the angular fluctuations to effectively counteract misalignment errors. For example, in a UAV system, if the fluctuations occur with a coherence time on the order of milliseconds, the stabilizer must be capable of responding on a timescale faster than that, ideally in the microsecond range. If the stabilizer's response is slower than the coherence time, it will introduce a delay in correcting the misalignment error, causing it to lag behind the fluctuations. This delay results in the loss of a portion of the transmitted data due to misalignment. The significance of this issue becomes even more critical in the next generation of wireless systems. As we move from 2G to 5G and beyond, pulse widths become shorter, meaning that even brief interruptions in the alignment of the communication channel can lead to the loss of a substantial portion of the transmitted data \cite{Li2024OJCOMS_LaserStabilizers}.

A key challenge in increasing the response speed is the exponential rise in weight, power consumption, and cost. Faster stabilizers require stronger actuators and higher precision sensors, which in turn demand more power and add substantial weight to the system. This increase in weight is directly related to the torque required to counteract rapid angular movements. To generate higher torque, the stabilizer's motors and mechanical components must be larger and more robust, leading to an overall increase in system mass. For instance, in UAV applications, if the angular fluctuations have a coherence time in the millisecond range, the stabilizer must be lightweight and fast. However, to achieve microsecond-level response times, the weight and power requirements of mechanical stabilizers can become impractical for UAV platforms. This additional weight can also reduce the UAV's overall stability. As a result, mechanical stabilizers are typically more suitable for environments with slower angular fluctuations (longer coherence times), where high-speed response is less critical. In analyzing stabilizer-based systems, the coherence time of angular fluctuations must be carefully considered, as it is often overlooked in analytical models found in the literature.

\subsubsection{Angular Accuracy of Stabilizers}
Similar to response speed, the angular accuracy of a stabilizer plays a significant role in reducing misalignment errors, particularly in high-precision communication links such as FSO systems. Increasing the angular accuracy of a stabilizer, however, comes with an exponential increase in cost, weight, and complexity. The most precise mechanical stabilizers available on the market today can achieve accuracies on the order of milliradians, which is sufficient for FSO communication over distances of a few kilometers. However, when the communication distance increases, such as in ground-to-satellite links, the challenges of maintaining such high accuracy become more pronounced. For example, in a 1000 km ground-to-satellite link, even a small misalignment of 1 milliradian in the transmitter’s direction would cause the beam to miss the satellite by 1000 meters. As the link distance increases, the required accuracy becomes more stringent, and the stabilizer alone may no longer be sufficient to maintain alignment. In such cases, additional optical tracking systems with higher precision must complement the stabilizer.

\begin{figure}
	\begin{center} 
		\includegraphics[width=2.4 in]{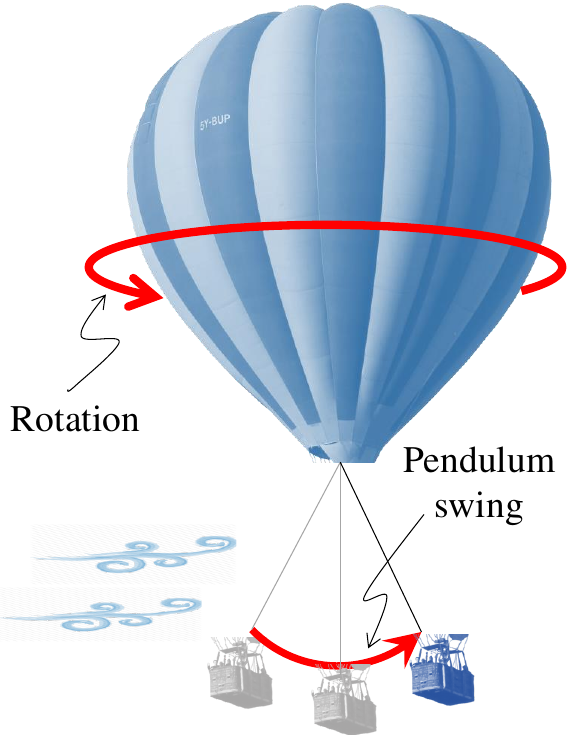}
		\caption{Illustration of pendulum misalignment and rotational motion in balloon-based communication systems. The pendulum swing and rotational movement of the balloon lead to periodic angular deviations in the communication beam, causing both pendulum-like motion and rotational misalignment.}

		\label{po2}
	\end{center}
\end{figure}

\begin{figure*}
	\begin{center} 
		\includegraphics[width=6 in]{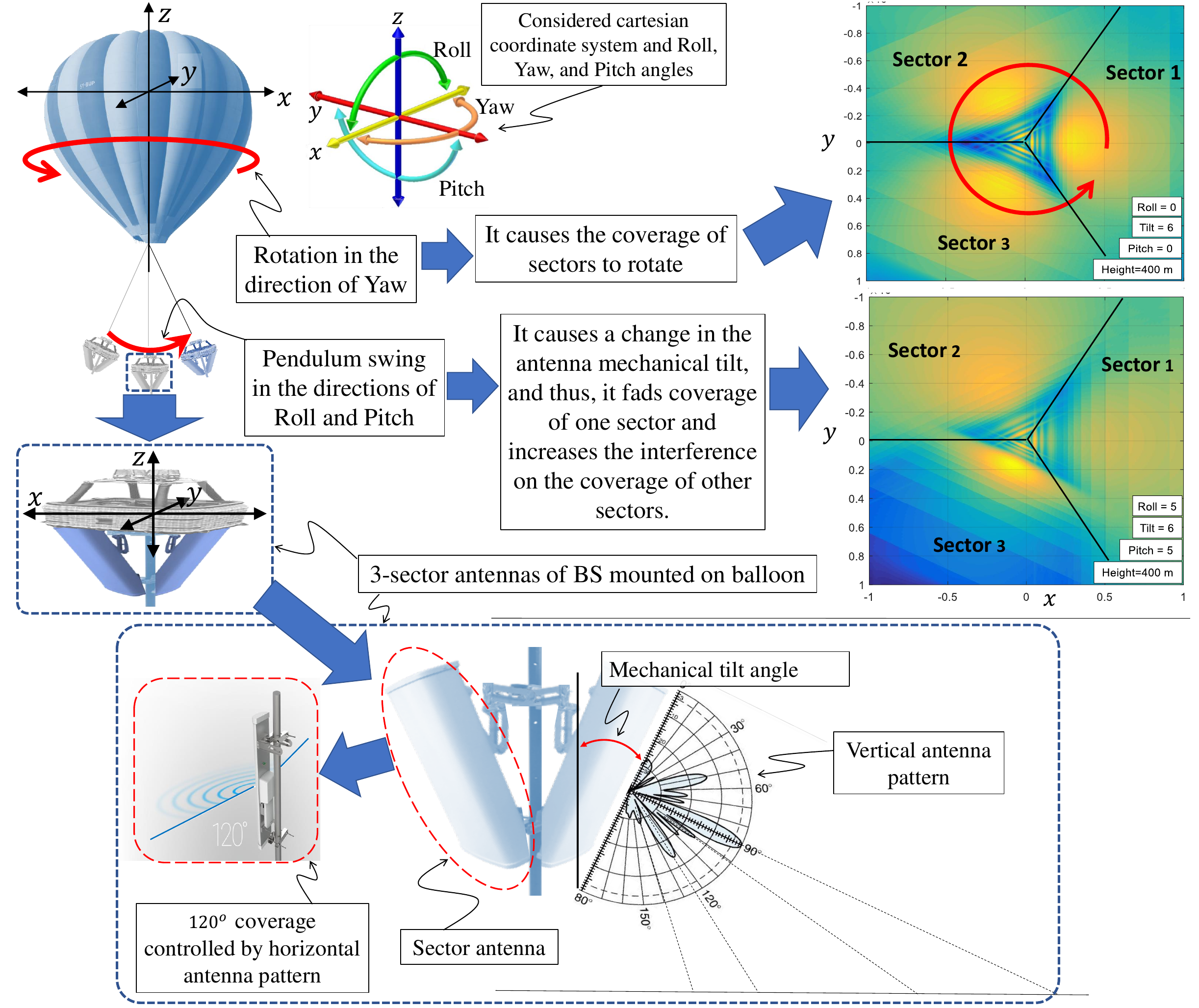}
		\caption{Depicting the 3D rotation and fluctuations of a balloon equipped with a three-sector antenna and its effect on the created three-sector coverage \cite{Dabiri2024BalloonUAV}.}

		\label{ballonx}
	\end{center}
\end{figure*}

\begin{figure*}
	\begin{center} 
		\includegraphics[width=6 in]{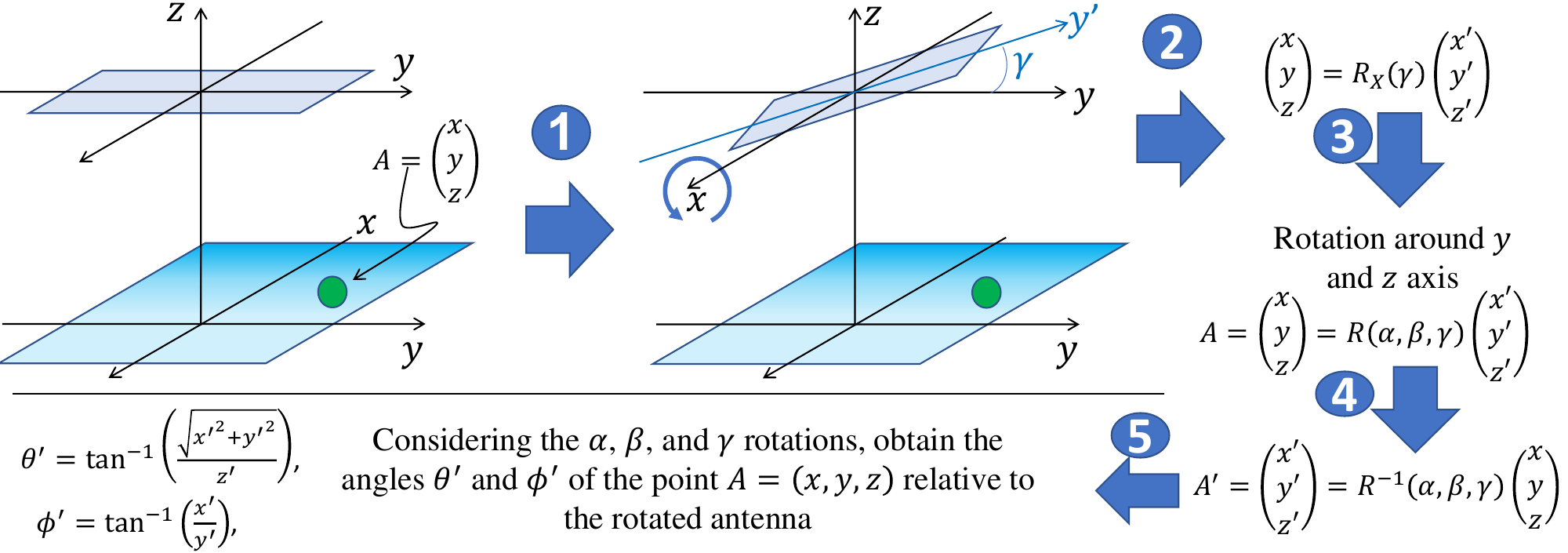}
		\caption{Illustration of the 3D rotational effects on antenna misalignment. The figure shows how sequential rotations around the \(z\), \(y\), and \(x\) axes result in transformed coordinates \((x', y', z')\), impacting the direction and alignment of the communication beam in balloon-based systems. }
		\label{ballon2}
	\end{center}
\end{figure*}

\subsection{Pendulum Misalignment and Rotational Motion Modeling}

In the new category of NTNs, there has been increasing interest in utilizing tethered balloons to provide coverage similar to the sector antennas used in traditional telecom towers. Unlike UAVs, balloons can carry heavier payloads and have higher power capabilities, making them suitable for longer-duration operations. The balloon is anchored to the ground with several cables. Along with these cables, a fiber-optic link is usually used to connect the balloon to the ground network. Additionally, DC power can be sent to the balloon through these cables \cite{Dabiri2024BalloonUAV}. However, two main categories of pointing errors arise in these balloon-based systems. First, as depicted in Fig. \ref{po2}, the balloon’s rotational motion about its axis introduces significant misalignment. Second, to mitigate the impact of wind-induced oscillations on the communication payload, the payload is usually suspended from the balloon by a rope or chain. While this suspension reduces the direct effect of the balloon’s movement on the payload, it introduces a pendulum-like swinging motion. This motion, which follows a pendulum distribution, as shown in Fig. \ref{ballonx}, results in additional angular deviations that must be carefully accounted for in system design.

\subsubsection{Impact on Antenna Coverage and Handovers}
As shown in Fig. \ref{ballon2}, there are two primary effects that impact antenna coverage and handovers in balloon-based systems:
\begin{itemize}
    \item \textbf{Rotation around the \(z\)-axis:} The rotational motion of the balloon around its \(z\)-axis causes the 3-sector antenna coverage to rotate. This rotation shifts the boundaries of the sectors, leading to frequent handovers for users located at the sector edges. As a result, users near the borders may experience increased handover frequency and potential disruptions in service.
    
    \item \textbf{Pendulum Motion and Mechanical Tilt:} The pendulum-like swinging motion of the suspended communication payload introduces a mechanical tilt in the antenna, leading to random variations in the coverage pattern. As the tilt angle oscillates, coverage in certain areas may fade while coverage in other areas is strengthened. In some cases, this can even lead to interference with neighboring cells, as the beam misalignment causes overlapping coverage areas.
\end{itemize}

These two effects significantly impact the stability and reliability of the communication link, especially for users near sector boundaries or at the edges of the coverage area.

\subsubsection{Three-Dimensional pendulum Transformation}
For modeling the pendulum effects on antenna misalignment in 3D space, we consider a point \((x, y, z)\) that undergoes sequential rotations about the \(z\), \(y\), and \(x\) axes. These rotations result in a transformed coordinate \((x', y', z')\), which can be computed using the following rotation matrices:
\begin{itemize}
    \item Rotation by angle \(\gamma\) about the \(z\)-axis \cite{Dabiri2024BalloonUAV}:
    \[
    R_z(\gamma) = \begin{bmatrix}
    \cos \gamma & -\sin \gamma & 0 \\
    \sin \gamma & \cos \gamma & 0 \\
    0 & 0 & 1
    \end{bmatrix}
    \]
    
    \item Rotation by angle \(\beta\) about the \(y\)-axis \cite{Dabiri2024BalloonUAV}:
    \[
    R_y(\beta) = \begin{bmatrix}
    \cos \beta & 0 & \sin \beta \\
    0 & 1 & 0 \\
    -\sin \beta & 0 & \cos \beta
    \end{bmatrix}
    \]
    
    \item Rotation by angle \(\alpha\) about the \(x\)-axis \cite{Dabiri2024BalloonUAV}:
    \[
    R_x(\alpha) = \begin{bmatrix}
    1 & 0 & 0 \\
    0 & \cos \alpha & -\sin \alpha \\
    0 & \sin \alpha & \cos \alpha
    \end{bmatrix}
    \]
\end{itemize}

The final transformation of the point \((x, y, z)\) to \((x', y', z')\) is given by the matrix multiplication \cite{Dabiri2024BalloonUAV}:
\[
\begin{bmatrix}
x' \\
y' \\
z'
\end{bmatrix}
= R_x(\alpha) R_y(\beta) R_z(\gamma)
\begin{bmatrix}
x \\
y \\
z
\end{bmatrix}
\]
This transformation allows us to model how the antenna pattern is affected by the pendulum motion and rotations, as illustrated in Fig. \ref{ballon2}. This coordinate transformation helps us calculate any oscillations and coordinate changes in the antenna pattern, enabling real-time updates to the network coverage. However, more detailed behavioral analysis and precise modeling remain open research areas, requiring further investigation and advanced modeling techniques.

\begin{figure*}
	\centering
	\subfloat[] {\includegraphics[width=4.4 in]{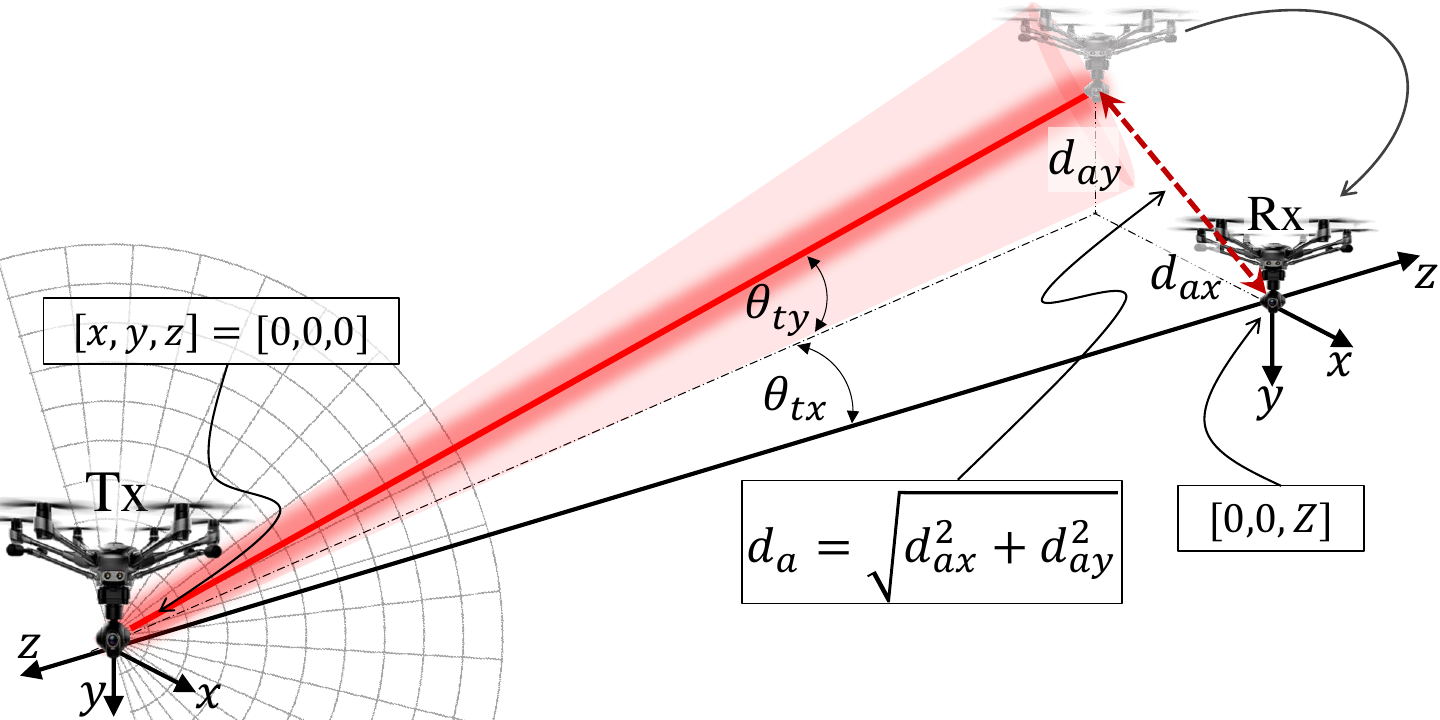}
		\label{ms1}
	}
	\hfill
        \subfloat[] {\includegraphics[width=2.0 in]{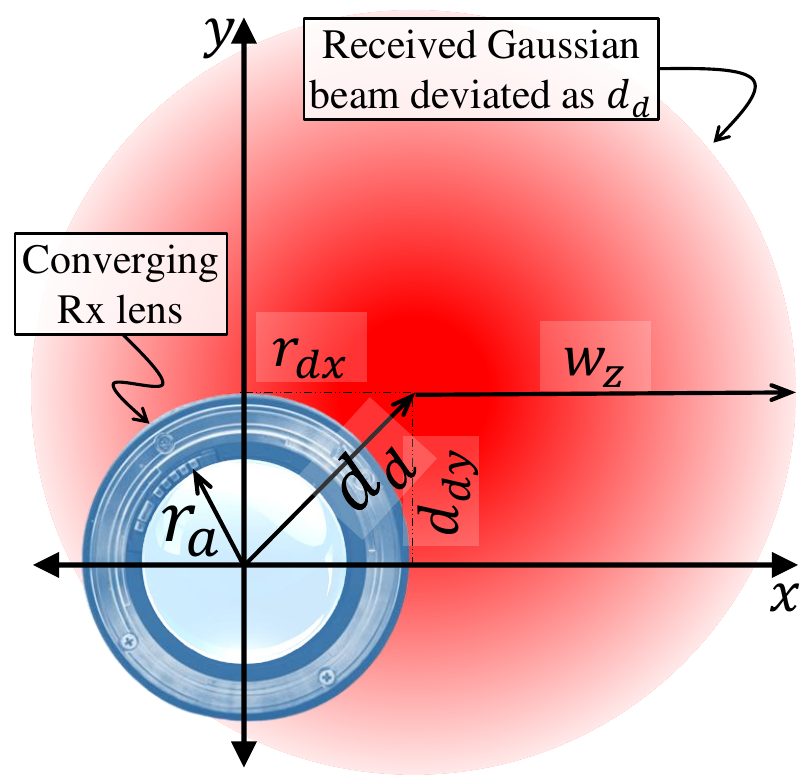}
		\label{ms2}
	}
	\hfill
	\subfloat[] {\includegraphics[width=4 in]{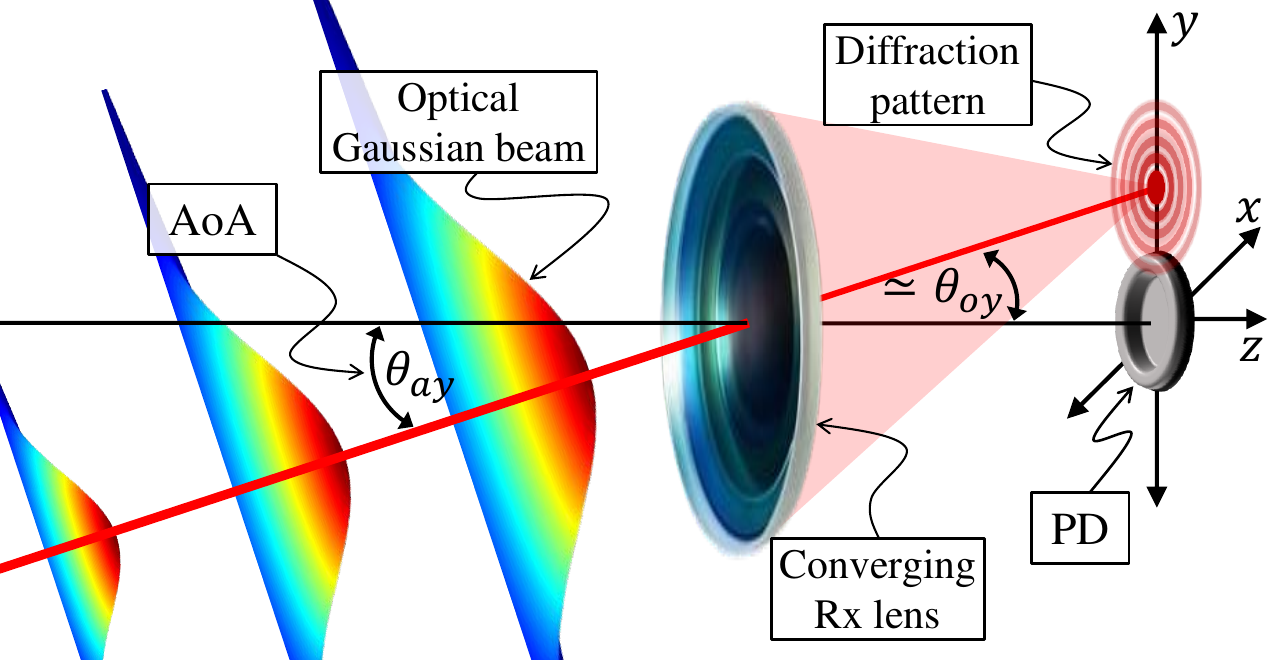}
		\label{ms3}
	}
	\hfill
	\subfloat[] {\includegraphics[width=2.4 in]{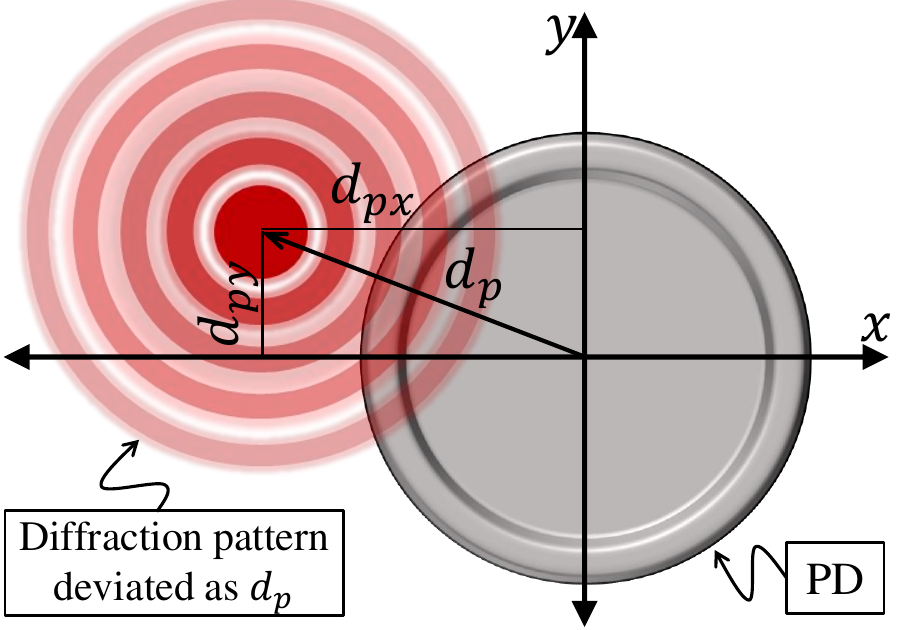}
		\label{ms4}
	}
	\hfill
	\caption{Illustration of pointing-error mechanisms in mobile FSO links: (a) transmitter angular deviations causing lateral displacement of the Gaussian beam at the receiver plane, (b) resulting geometrical loss due to beam miscentering at the receiver aperture, (c) angle-of-arrival (AoA) fluctuations producing a shifted Airy diffraction pattern at the focal plane of the receiving lens, and (d) corresponding received power degradation due to displacement of the Airy spot relative to the photodetector active area \cite{Dabiri2023THzFSO}.}
	\label{ms}
\end{figure*}
\section{Pointing Errors Modeling}
In this section, we focus on the fundamental modeling of pointing errors for both FSO links and directional THz/mmWave links. 

\subsection{FSO Pointing Error Modeling}
For FSO systems, pointing error is modeled as:
\begin{align}
    h_{p} = h_{pg} \times h_{pa},
\end{align}
where $h_{pg}$ is the geometrical loss due to pointing errors, and $h_{pa}$ is the link loss due to the AoA fluctuations.

\subsubsection{Geometrical Loss Due to Pointing Errors}
For modeling the geometrical loss caused by pointing error, it is essential to model the optical intensity distribution.
The normalized intensity distribution of an ideal single-mode Gaussian beam can be expressed as follows
\cite{saleh2019fundamentals}:
\begin{equation}
I(r, Z) = \frac{2P_t}{\pi w_z^2(Z)} \exp\left(-\frac{2r^2}{w_z^2(Z)}\right)
\end{equation}
where \(P_t\) represents the optical power of the beam, and $r$ denotes the radial position, and $Z$ is the link length. For an ideal Gaussian beam at a distance $Z$, the spot size $w_z(Z)$ is described by the following equation:
\begin{align}
w_z(Z) &= w_0 \sqrt{1 + \left( \frac{Z}{\pi w_0^2} \right)^2} 
\end{align}
where \(w_0\) is the beam waist and $\lambda$ is the wavelength of the laser. 
The beam divergence angle $\theta_\text{div}$ is defined as:
\begin{equation}
\theta_\text{div} = \lim_{Z \to \infty} \tan^{-1} \left( \frac{w_z(Z)}{Z} \right).
\end{equation}
For small values of $\theta_\text{div}$, we have:
\begin{equation}
\theta_\text{div} \approx \frac{\lambda}{\pi w_0}.
\end{equation}

For optical links, fluctuations in the transmitter cause a geometrical loss in the receiver. For better understanding, the effect of point error is depicted in Figs. \ref{ms1} and \ref{ms2}.
For a link length $Z$, as shown in Figs. \ref{ms1}, the angular deviation in the transmitter as $\theta_t=[\theta_{tx},\theta_{ty}]$ causes the received beam center to deviate from the center of the receiver lens as $d_a = \sqrt{d_{ax}^2 + d_{ay}^2}$, where
\begin{align}
	\label{cv1}
	d_{ax} = Z \tan\left( \theta_{tx} \right), ~~~\& ~~~
	d_{ay} = Z \tan\left( \theta_{ty} \right).
\end{align}
Therefore, for a circular receiver lens, the geometrical pointing loss $h_{pg}$ is defined as the ratio of the collected power by the receiver aperture to the total distributed power, which is modeled as follows:
\cite{farid2007outage}:
\begin{align}
	\label{df3}
	& h_{pg} = \frac{2}{\pi w_{z}^2} \nonumber \\ &\times \int_{-r_a}^{r_a}  \int_{-\sqrt{r_a^{2}-y^2}}^{\sqrt{r_a^{2}-y^2}}   e^{-2\frac{(x+Z \tan\left( \theta_{tx} \right))^{2}+(y+Z \tan\left( \theta_{ty} \right))^2}{w_z^2}} dx dy,
\end{align}
where $r_a$ is the receiver lens radius.
It should be noted that $h_{pg}$, as modeled in \eqref{df3}, represents an instantaneous factor of geometrical pointing loss, which itself is a function of the random variables $ \theta_{tx} $ and $ \theta_{ty} $.

A great deal of work has been done to derive the relation \eqref{df3}. In \cite{farid2007outage}, the authors show that if $w_z>>r_a$, then \eqref{df3} is simplified as follows:
\begin{align}
	\label{df4}
	h_{pg} &= A_0 \exp\left(- \frac{2 r_d^2}{w_{z_\text{eq}}} \right),
\end{align}
where 
\begin{align}
	\label{df5}
	w_{z_\text{eq}}^2 = w_z^2 \frac{\pi \text{erf}(\nu)}{2\nu \exp(-\nu^2)}, ~~\&~~
	A_0 = [\text{erf}(\nu)]^2,
\end{align}
and, $\nu=(\sqrt{\pi}r_a)(\sqrt{2}w_z)$.

\subsubsection{Link Loss Due to AoA Fluctuation}
In many references related to terrestrial FSO links, geometrical pointing error $h_{pg}$ is commonly considered the main factor influencing pointing loss $h_p$. It is often assumed that the optical signal is captured by the lens, which then focuses it onto the photodetector at the focal point, and that all the focused power is entirely received by the photodetector, i.e., it is assumed that $h_{pa} \simeq 1$. While this assumption holds true for stable fixed links, it cannot be directly applied to mobile links for three critical reasons:
\begin{itemize}
    \item \textit{First Reason:} Due to the background noise caused by sunlight during the day, the receiver's FoV is typically designed to be narrow to minimize interference. The FoV can be defined based on the radius of the lens ($r_a$) and the radius of the photodetector or fiber core ($r_d$) using the following relationship:
    $$ \theta_\text{FoV} \approx \frac{r_d}{r_a}. $$
    According to this relation, to achieve a smaller FoV, the effective area of the photodetector must be reduced. 

    \item \textit{Second Reason:} The electrical bandwidth of the photodetector is typically inversely proportional to its effective area. In order to achieve data rates on the order of tens of Gbps, the effective area of the photodetector must be reduced to the millimeter scale. This introduces further limitations in terms of optical power collection, especially in mobile applications.

    \item \textit{Third Reason:} In mobile systems, any angular error in the AoA of the incoming link, caused by transmitter-receiver misalignment, beam wandering during propagation, or platform instability, results in the fluctuation of the optical spot at the focal point. This causes the beam to drift from the small photodetector or fiber core, leading to link loss due to AoA fluctuation and significant degradation in link performance.
\end{itemize}

Now, we focus on modeling $h_{pa}$. Let $\theta_{ax}$ and $\theta_{ay}$ represent the AoA angles along the $x$ and $y$ axes, respectively. A common assumption is that the optical beam can be approximated as a plane wave at the receiver aperture, which holds true for two important reasons:
\begin{itemize}
    \item \textit{First Reason:} To mitigate mobility effects and reduce the negative impact of $h_{pg}$, mobile FSO links typically use Gaussian lasers with larger divergence angles at the transmitter, resulting in larger $w_z$ at the receiver, often on the order of several meters.
    \item \textit{Second Reason:} Unlike terrestrial links, mobile FSO systems, particularly those involving UAVs, are usually designed with smaller optical payloads due to weight and power consumption limitations. As a result, the receiver lens aperture is typically chosen to have a radius on the order of 2 to 4 cm, which is significantly smaller than the $w_z$. Consequently, variations in the Gaussian power distribution over the lens surface can often be ignored, making the constant approximation valid in most cases.
\end{itemize}

Under the constant (plane wave) assumption, the optical intensity distribution at the focal point of the lens follows an Airy pattern, as shown in Figs. \ref{ms3} and \ref{ms4}. The mathematical expression for the normalized intensity distribution of an Airy pattern at a radial distance $r$ from the center is given by \cite{saleh2019fundamentals}:
\begin{align} 
I(r) = I_0 \left( \frac{2 \mathbb{J}_1\left(k \cdot r\right)}{k \cdot r} \right)^2, 
\end{align}
where $I_0$ is the peak intensity, $\mathbb{J}_1(\cdot)$ is the Bessel function of the first kind, $r$ is the radial distance from the center, and $k = \frac{2 \pi}{\lambda d_f}$ is a constant dependent on the wavelength $\lambda$ and focal length $d_f$ of the lens.

Given AoA fluctuations represented by $\theta_{ax}$ and $\theta_{ay}$, the deviation of the Airy pattern’s center from the focal point’s center in the $x$ and $y$ directions can be determined as:
\begin{align}
d_{px} = d_f \cdot \tan(\theta_{ax}), \quad d_{py} = d_f \cdot \tan(\theta_{ay}).
\end{align}
As shown in Fig. \ref{ms4}, the values $d_{px}$ and $d_{py}$ indicate the shift of the Airy pattern’s center along the $x$ and $y$ axes, respectively. 
Finally, by integrating over the surface of the photodetector, the instantaneous coefficient $h_{pa}$ can be obtained as follows:
\begin{align} \label{hpa1}
  h_{pa} = I_0 \int_{\mathcal{A}_\text{phot}}  \left( \frac{2 \mathbb{J}_1\left(k \cdot \mathbb{B}(\theta_{ax},\theta_{ay},x,y) \right)}{k \cdot \mathbb{B}(\theta_{ax},\theta_{ay},x,y)} \right)^2
  	 \, dx \, dy,	 
\end{align}
where $\mathcal{A}_\text{phot}$ is the effective area of photodetector, and 
\begin{align}
&\mathbb{B}(\theta_{ax},\theta_{ay},x,y)= \nonumber \\
&~~~\sqrt{(x-d_f \cdot \tan(\theta_{ax}))^2 + (y-d_f \cdot \tan(\theta_{ay}))^2}.
\end{align}

It should be noted that the $h_{pa}$ modeled in \eqref{hpa1} is only an instantaneous coefficient and is a function of the random variables $\theta_{ax}$ and $\theta_{ay}$, as well as the design parameters such as the lens radius and photodetector radius. These parameters should be optimally designed to create a trade-off between system performance, overcoming $h_{pa}$, reducing background noise, and increasing transmission rates. 


\subsection{THz Pointing Error Modeling}
Recently, with the advancement of technology in high-frequency mmWave and THz communications, we have witnessed the use of directional mmWave/THz communications for mobile systems. Precise pointing error modeling is crucial for these communications for several important reasons:
\begin{itemize}
    \item First, high-frequency communications, especially THz communications, suffer from significant attenuation. To overcome this attenuation and establish longer free-space links, it is necessary to use directional antennas.
    
    \item Due to the smaller antenna dimensions at higher frequencies, it becomes feasible to utilize higher-gain antennas with smaller sizes. For example, consider an 10x10 antenna array. For a 10x10 antenna array, the theoretical gain is approximately 20 dBi. At a frequency of 3 GHz, the size of the array would be around $50 \times 50$ cm$^2$ ($\simeq 10\times \lambda/2$). In contrast, at 150 GHz, the size of the same array would be significantly smaller, around $1\times 1$ cm$^2$. This drastic reduction in size at higher frequencies is due to the much shorter wavelength, making it possible to achieve high gain with a compact form factor at THz frequencies.

    \item The smaller dimensions of these high-frequency antennas make them suitable for installation on mobile platforms, particularly UAVs, which are constrained by weight and size. THz-based UAV systems can even overcome NLoS limitations by providing a long-range LoS path between two transmitters through precise 3D positioning in space.

    \item Moreover, in next-generation communication systems, the demand for significantly higher data rates is driven by the proliferation of data-hungry applications, such as immersive augmented reality, ultra-high-definition video streaming, and massive IoT deployments. To address these growing demands, further enhancement of antenna gain becomes critical, as it enables improved spatial multiplexing capabilities. This, in turn, facilitates a higher number of frequency reuses within 3D space, a crucial factor for ensuring efficient spectrum utilization.
    
    \item The increase in antenna gain leads to narrower beamwidths and greater sensitivity to angular errors, even on the order of less than one degree. As the envisioned applications for next-generation mobile communication systems based on high-frequency THz links expand, pointing error becomes the dominant fading factor. Therefore, its accurate modeling and analysis of directional THz antennas are crucial for the optimal design of these systems.
\end{itemize}

Below, the modeling of pointing error for THz links is examined across major scenarios.

\subsubsection{Optical-based Pointing Model}
Most of the primary works in the literature of THz, directly use the optical pointing error model provided in \eqref{df4} for directional mmWave/THz links \cite{boulogeorgos2019analytical,singya2022hybrid}.

\subsubsection{General Pointing Error Model}
However, the optical-based pointing error model is not always valid for THz communications. As illustrated in Fig. \ref{ms}, the optical pointing error model includes two distinct factors of pointing error. In fact, the model provided in \eqref{df4} represents the geometrical pointing error and is solely a function of the transmitter’s angular deviations, $\theta_t = [\theta_{tx}, \theta_{ty}]$, and does not account for fluctuations in the receiver’s angle. However, in many THz scenarios, this assumption does not hold. In THz communication models, the overall pointing error is derived from the product of the misaligned gain of the transmitter antenna and the misaligned gain of the receiver antenna, as follows:
\begin{align}
    h_p = G_t(\theta_{tx}, \theta_{ty}) \times G_r(\theta_{rx}, \theta_{ry})
\end{align}
where $G_t(\theta_{tx}, \theta_{ty})$ and $G_r(\theta_{rx}, \theta_{ry})$ represent the gain of the misaligned transmitter and receiver antennas, respectively.

In some cases, the pointing error is modeled in terms of power rather than gain. Since antenna gain is related to the square of the electric field (proportional to power), the combined effect of transmitter and receiver misalignment is sometimes expressed under a square root, as shown below:
\begin{align} \label{gain1}
    h_p = \sqrt{G_t(\theta_{tx}, \theta_{ty}) \times G_r(\theta_{rx}, \theta_{ry})}.
\end{align}
The above relationship is a complex function of the random variables $\theta_t = (\theta_{tx}, \theta_{ty})$ and $\theta_r = (\theta_{rx}, \theta_{ry})$. For an antenna array, the gain relationship in \eqref{gain1} can be expressed as follows \cite[eqs. (6.89) and (6.91)]{balanis2016antenna}:
\begin{align}
	\label{gain2}
	h_{p} &= \sqrt{G_0(N_t) G_0(N_r)}
	\left( \frac{\sin\left(\frac{N_{tx} (k d_{x} \sin(\theta_t)\cos(\phi_t)+\beta_{x})}{2}\right)} 
	{N_{tx}\sin\left(\frac{k d_{x} \sin(\theta_t)\cos(\phi_t)+\beta_{x}}{2}\right)}
	\right. \nonumber \\
	&~~~\times \left. \frac{\sin\left(\frac{N_{ty} (k d_{y} \sin(\theta_t)\sin(\phi_t)+\beta_{y})}{2}\right)} 
	{N_{ty}\sin\left(\frac{k d_{y} \sin(\theta_t)\sin(\phi_t)+\beta_{y}}{2}\right)}\right) \nonumber \\
	&~~~\times \left( \frac{\sin\left(\frac{N_{rx} (k d_{x} \sin(\theta_r)\cos(\phi_r)+\beta_{x})}{2}\right)} 
	{N_{rx}\sin\left(\frac{k d_{x} \sin(\theta_r)\cos(\phi_r)+\beta_{x}}{2}\right)}
	\right. \nonumber \\
	&~~~\times \left. \frac{\sin\left(\frac{N_{ry} (k d_{y} \sin(\theta_r)\sin(\phi_r)+\beta_{y})}{2}\right)} 
	{N_{ry}\sin\left(\frac{k d_{y} \sin(\theta_r)\sin(\phi_r)+\beta_{y}}{2}\right)}\right).
\end{align}
where $G_0$ represents the peak gain, $N_{tx}$ and $N_{ty}$ represent the number of antenna elements along the $x$ and $y$ axes of the transmitter array, respectively, while $N_{rx}$ and $N_{ry}$ denote the number of antenna elements along the $x$ and $y$ axes of the receiver array. 
In \eqref{gain2}, for $q\in\{t,r\}$, the parameters $\theta_q$ and $\phi_q$ can be defined as  functions of random variables $\theta_{q_x}$ and $\theta_{q_y}$ as follows:
\begin{align}
	\label{f_2}
	\theta_q  &= \tan^{-1}\left(\sqrt{\tan^2(\theta_{q_x})+\tan^2(\theta_{q_y})}\right), \nonumber\\
	\phi_q    &=\tan^{-1}\left({\tan(\theta_{q_y})}\big/{\tan(\theta_{q_x})}\right).
\end{align} 
Moreover, $\beta_{x}$ and $\beta_{y}$ are progressive phase shift between the elements along the  $x$ and $y$ axes, respectively. 
More details on the elements and array radiation pattern is provided in \cite{niu2015survey,balanis2016antenna}.

Given the complexity of the relationship in \eqref{gain2}, many studies have attempted to simplify this equation for different scenarios to facilitate the analysis of systems under alignment errors. 

\subsubsection{Two-Sector Model}
One commonly used approximation is the two-sector antenna pattern model, where the directional antenna pattern is simplified by dividing the radiation pattern into two regions: the main lobe and the side lobes. This model allows the system to be described using two distinct gains, one for the main lobe and another for the side lobes, as follows:
\begin{align}
    G(\theta) = 
    \begin{cases}
    G_\text{main}, & \text{if } |\theta| \leq \theta_\text{m}, \\
    G_\text{side}, & \text{if } |\theta| > \theta_\text{m},
    \end{cases}
\end{align}
where $G_\text{main}$ represents the gain of the main lobe, $G_\text{side}$ is the gain of the side lobes, and $\theta_\text{m}$ is the angular width of the main lobe. This model provides a simplified yet effective means of capturing the essential characteristics of the directional antenna pattern while accounting for the effects of misalignment and side lobe interference. However, it is important to note that this is not an exact model, and analyses based on this approximation can exhibit significant deviations from real-world values, especially in scenarios with more complex radiation patterns or where side lobe effects are more pronounced.

\begin{figure}
	\begin{center}
		\includegraphics[width=3.35 in]{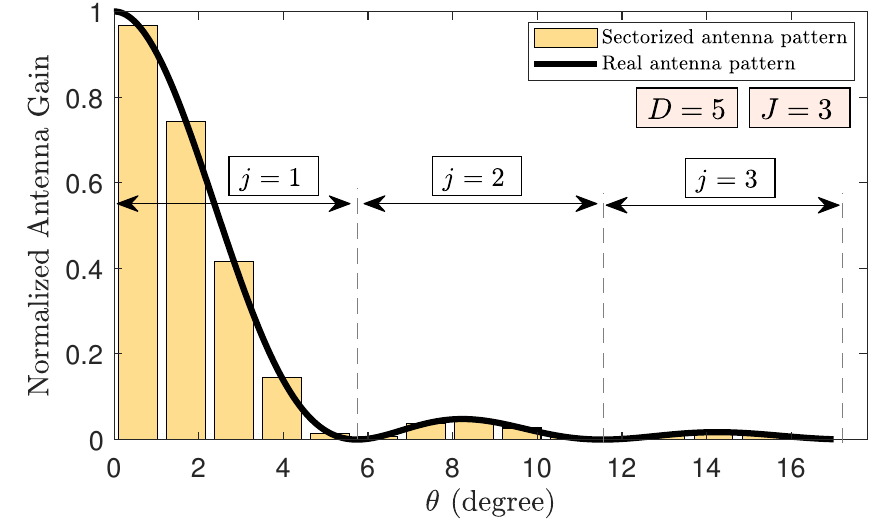}
		\caption{A graphical comparison of a real antenna pattern with a sectorized antenna pattern, characterized by $J=3$ (number of lobes), $D=5$ (number of sectors per lobe), and $J_u=0$. }
		\label{dis111}
	\end{center}
\end{figure}
\subsubsection{Multi-Sector Model}
For a more accurate representation of the antenna radiation pattern, the multi-sector antenna pattern model divides the antenna pattern into multiple angular sectors, each associated with a distinct gain. This model provides a finer approximation of the directional pattern compared to the two-sector model, allowing for better representation of side lobes and more gradual transitions between the main lobe and side lobes.

Following the results in \cite{dabiri2024non,dabiri20203d}, for a square antenna array with $N_{tx} = N_{ty} = N_t$ and $N_{rx} = N_{ry} = N_r$, the relationship in \eqref{gain2} can be approximated as follows:
\begin{align}
	\label{as6}
	h_{p} = \sqrt{G_0(N_t) G_0(N_r)} h_{pgt} h_{pgr},
\end{align}
where
\begin{align}
	\label{as7}
	&h_{pt} \simeq \sum_{j_u=J_\text{u,min} }^{J_\text{u,max}}  \left| \frac{\sin\left(\frac{N_t k d_{x} 
			\sin\left(  \frac{j_u}{D N_t}  \right)
		}{2}\right)} 
	{N_t\sin\left(\frac{k d_{x} 
			\sin\left(   \frac{j_u}{D N_t} \right)
		}{2}\right)}    \right| \nonumber \\
	&~~~~~~~\times\left[ \mathbb{U}\left( \Theta_{ui}  - \frac{j_u}{D N_t}  \right)
	- \mathbb{U}\left( \Theta_{ui}  - \frac{j_u+1}{D N_t}  \right)  \right], 
\end{align}
\begin{align}	
	\label{ass7}
	&h_{pr} \simeq  \sum_{j_g=J_\text{g,min} }^{J_\text{g,max}} \left|    \frac{\sin\left(\frac{N_r \left(k d_{x} 
			\sin\left(  \frac{j_g}{D N_r}  \right)
			\right)}{2}\right)} 
	{N_r\sin\left(\frac{k d_{x} 
			\sin\left(   \frac{j_g}{D N_r}  \right)
		}{2}\right)}  \right| \nonumber \\
	&~~~~~~~\times \left[ \mathbb{U}\left( \Theta_{gi} - \frac{j_g}{D N_r}  \right)
	- \mathbb{U}\left( \Theta_{gi} - \frac{j_g+1}{D N_r}  \right) \right],
\end{align} 
where $\mathbb{U}(x) = 
\begin{cases}
	1,~x>0\\
	0,~x\leq0
\end{cases} $, and
\begin{align}
	\begin{cases}
		J_u = \left\lfloor N_t \sqrt{\theta_{ix}^2 + \theta_{iy}^2}  \right\rfloor, ~~
		J_g = \left\lfloor N_r \sqrt{\mu_{ix}^2 + \mu_{iy}^2}  \right\rfloor,  \\
		J_\text{u,min} = \max\{0,(J_u-J)D\}, ~~ 
		J_\text{u,max} = {(J_u + J)D-1}, \\
		J_\text{g,min} = \max\{0,(J_g-J)D\}, ~~ 
		J_\text{g,max} = {(J_g + J)D-1}.
	\end{cases} \nonumber
\end{align}
The parameters $J$ and $D$ are directly related to the accuracy of the approximation in \eqref{as7} and \eqref{ass7}. Specifically, $J$ denotes the number of lobes considered around the $J_u$th lobe, while $D$ represents the number of sectors within each lobe. For a clearer understanding, Fig. \ref{dis111} illustrates the sectorization of the antenna pattern for the case of $J_u=0$ (which corresponds to the main lobe for $\mu_{ix}=\mu_{iy}=0$), $J=3$, and $D=5$. In this illustration, $J=3$ indicates the main lobe along with two adjacent side lobes, each divided into multiple sectors, represented by $D=5$. Note that the analysis of the relationships \eqref{as7} and \eqref{ass7} is quite straightforward, as they are merely a summation of simple coefficients.

\begin{figure}
	\centering
	\subfloat[] {\includegraphics[width=3.3 in]{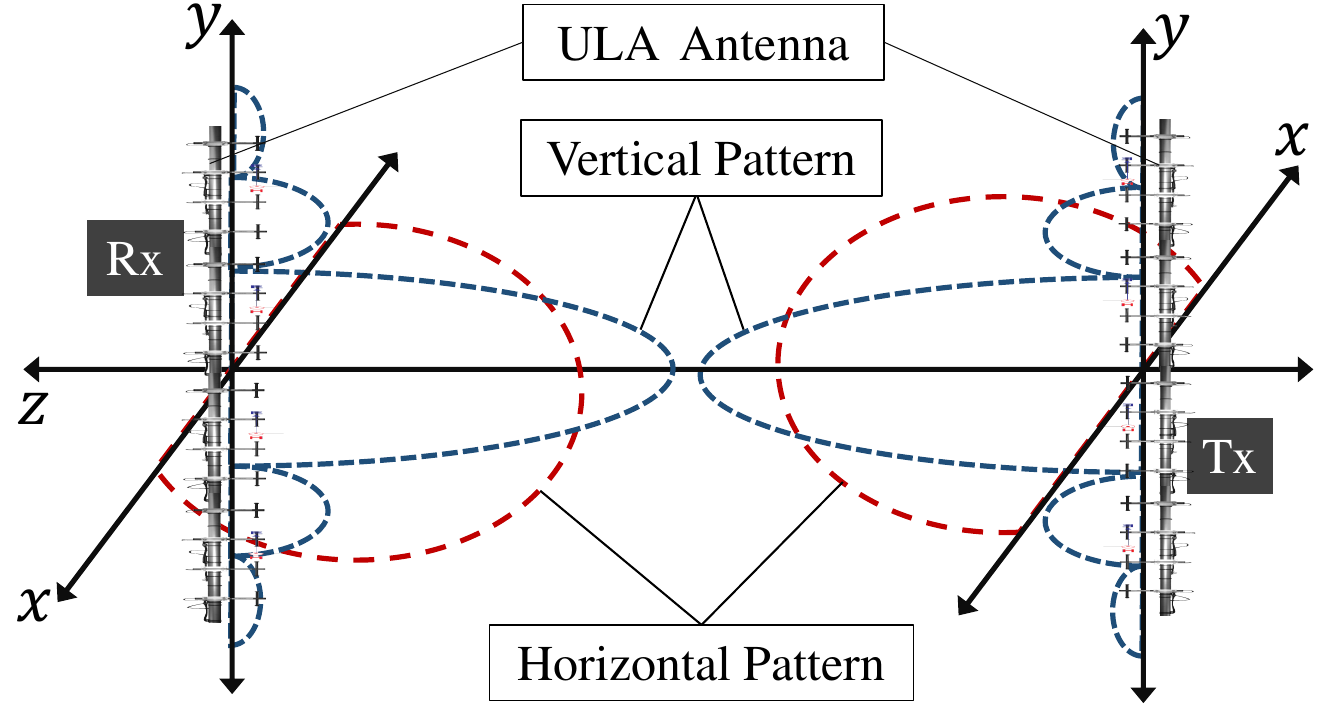}
		\label{ns1}
	}
	\hfill
        \subfloat[] {\includegraphics[width=3.3 in]{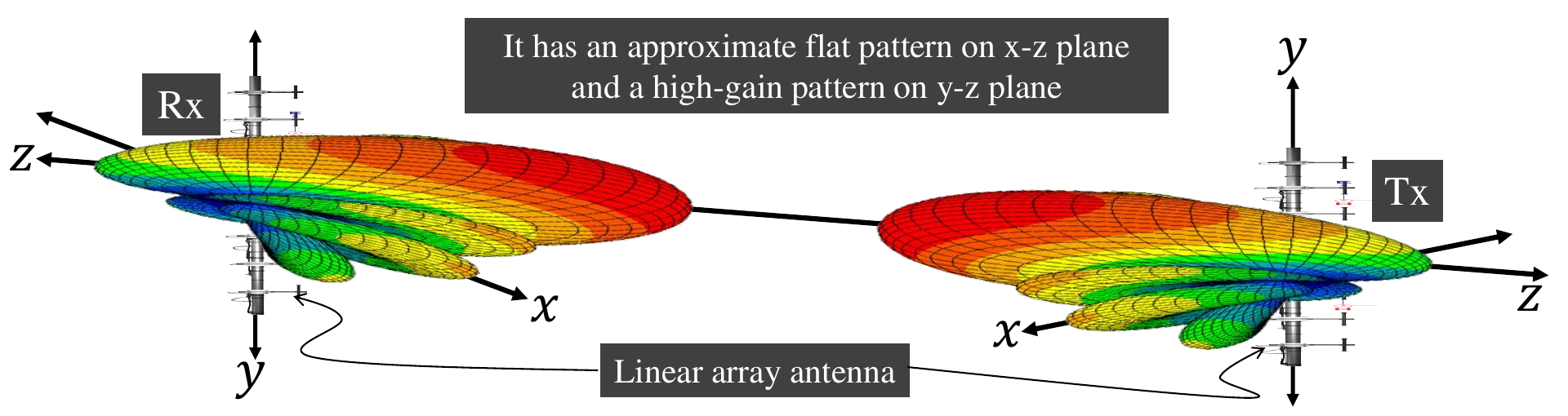}
		\label{ns2}
	}
	\hfill
	\caption{Radiation characteristics of a linear array antenna used in aerial platforms: (a) conceptual illustration of a ULA showing distinct vertical and horizontal radiation patterns between the transmitter (Tx) and receiver (Rx), and (b) three-dimensional radiation patterns highlighting a high-gain, narrow beam in the vertical (y–z) plane and an approximately omnidirectional pattern in the horizontal (x–z) plane, which enhances robustness against platform rotation and angular fluctuations \cite{dabiri2022general}.}
	\label{ns}
\end{figure}

\subsubsection{Multi-Sector Model for Linear Array Antennas}
Another category of antennas is the linear array antenna, which finds various applications. As depicted in Fig. \ref{ns}, linear array antennas are particularly well-suited for UAV communications due to the following two key reasons:
\begin{itemize}
    \item For lower frequencies, the dimensions of antennas typically become larger. Given the physical constraints of UAVs, mounting large arrays on UAVs reduces their maneuverability. In contrast, suspending a linear array antenna beneath the UAV not only provides favorable aerodynamic properties during flight but also allows for easy mechanical adjustment of the antenna's angle.
    
    \item As shown in the antenna patterns in Figs. \ref{ns1} and \ref{ns2}, the vertical pattern exhibits higher gain and a narrower beamwidth, which reduces interference with ground-based links. Meanwhile, the horizontal pattern is almost omnidirectional, making the UAV or balloon resilient to rotation.
\end{itemize}

The performance of linear array antennas under angular errors has been thoroughly analyzed in \cite{dabiri2020analytical}. In this case, the antenna pattern in the $x$-axis direction has a wider beamwidth and is relatively insensitive to pointing errors. Consequently, the pointing error in relationships \eqref{as7} and \eqref{ass7} becomes independent of fluctuations in $\theta_{tx}$ and $\theta_{ty}$. Therefore, we have $\theta_t \simeq \theta_{ty}$ and $\theta_r \simeq \theta_{ry}$ in this scenario.

\begin{figure}
	\centering
	\subfloat[] {\includegraphics[width=3.3 in]{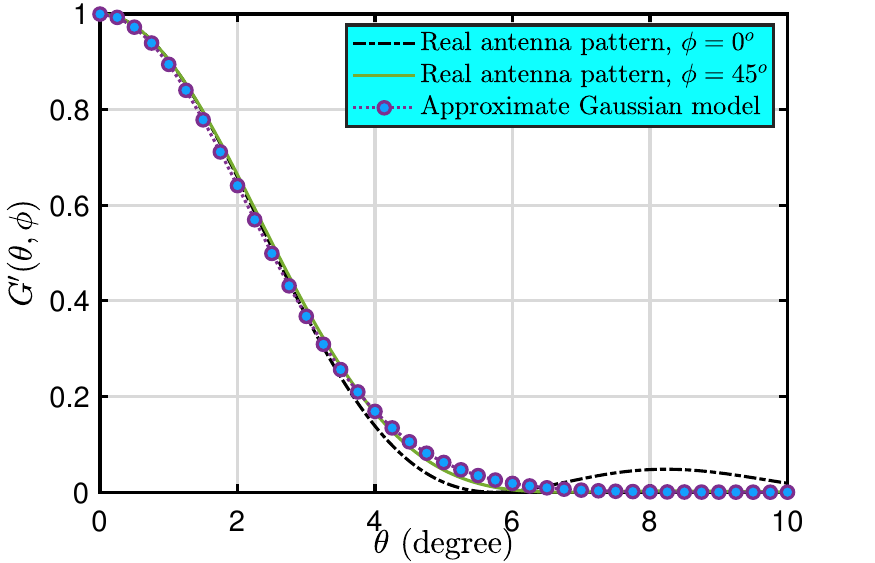}
		\label{m1}
	}
        \hfill
	\subfloat[] {\includegraphics[width=3.3 in]{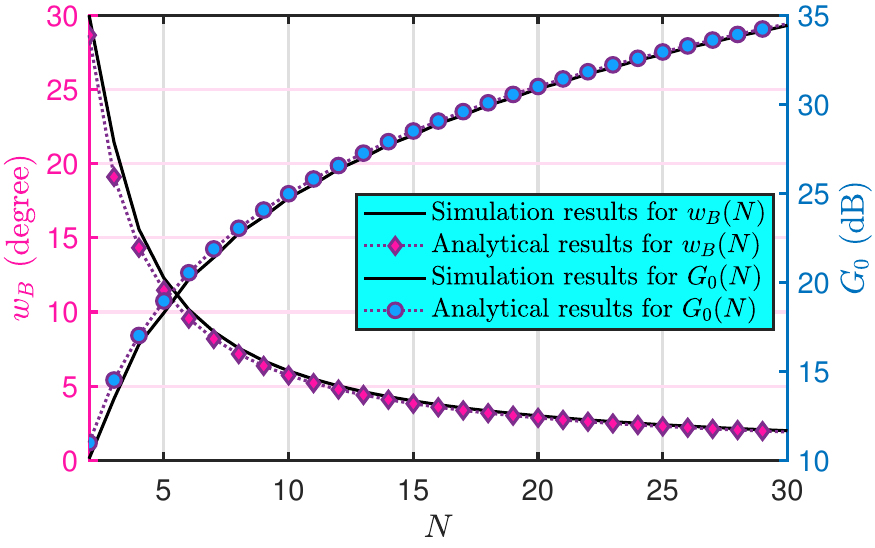}
		\label{m2}
	}
        \hfill
	\caption{(a) Comparison of Gaussian approximate pattern with actual antenna pattern of an array antenna for two different values of angle $\phi$.
		(b) Comparing the accuracy of approximate values for parameters $w_B$ and $G_0$ with the values obtained from numerical results \cite{dabiri2022pointing}.}
	\label{m3}
\end{figure}

\subsection{Pointing Error Modeling with a Gaussian Pattern}
A major category of directional mmWave/THz communications is used for point-to-point links. When the angular deviation is within a few degrees, the effect of side lobes can be reasonably neglected. As shown in Fig. \ref{m1}, the gain of the antenna pattern can be modeled by a Gaussian distribution, given by:

\begin{align}
	\label{b1}
	\resizebox{0.99\hsize}{!}{$
		G_q(N) = G_0(N) \exp\left(\! -\frac{\left( \tan^{-1}\left(\sqrt{\tan^2(\theta_{qx}) + \tan^2(\theta_{qy})}\right)\! \right)^2}{w^2_B(N)} \right),$} 
\end{align}
where \( q \in \{t,r\} \) represents the transmitter or receiver, and $w_B$ is the angular beamwidth (also called the beam divergence), which can be defined in several different ways.  
By substituting \eqref{b1} into \eqref{gain1}, the pointing error with the Gaussian approximation is modeled as follows:
\begin{align} \label{bx1}
	&h_p = \sqrt{ G_0(N_t) G_0(N_r) }  \nonumber \\
	&\exp\left( -\frac{\left( \tan^{-1}\left(\sqrt{\tan^2(\theta_{tx}) + \tan^2(\theta_{ty})}\right) \right)^2}{2w^2_B(N_t)} \right. \nonumber \\
	& ~~~~~~~ \left. -\frac{\left( \tan^{-1}\left(\sqrt{\tan^2(\theta_{rx}) + \tan^2(\theta_{ry})}\right) \right)^2}{2w^2_B(N_r)} \right).
\end{align}
For small values of $\theta_{tx}$, $\theta_{ty}$, $\theta_{rx}$, and $\theta_{ry}$, \eqref{bx1} simplifies as follows:
\begin{align} \label{bx2}
	&h_p = \sqrt{ G_0(N_t) G_0(N_r) }  
	\exp\left( -\frac{ \theta_{tx}^2 + \theta_{ty}^2  }{2w^2_B(N_t)} 
	 -\frac{\theta_{rx}^2 + \theta_{ry}^2}{2w^2_B(N_r)} \right).
\end{align}

It has been observed that calculating $w_B$ based on the $1/e$ criterion provides a better approximation, and it can be obtained as \cite{balanis2016antenna}:
\begin{align}
	\label{s1}
	~~~G'_q(w_B, \phi_q)  - e^{-1} = 0.
\end{align}

A comprehensive search in \cite{dabiri2022pointing} has shown that for a standard array antenna, $w_B$ can be accurately approximated as $w_B(N) = \frac{B}{N}$, where $B = 1.061$. The validity of this approximation is demonstrated in Fig. \ref{m1}. Moreover, based on the results in \cite{balanis2016antenna}, we have $G_0(N) < \pi N^2$ for array antennas. As illustrated in Fig. \ref{m2}, for a standard array antenna, $G_0(N)$ can be well approximated by $G_0(N) \simeq \pi N^2$. In Fig. \ref{m1}, the validity of the Gaussian main-lobe approximation for a standard array antenna is shown for different values of $\phi$. It is observed that changes in $\phi$ only affect the side lobes.

\section{Pointing Error Distribution Modeling}
In the previous section, we modeled pointing errors for both optical and 
THz links. Since these errors are inherently random, analyzing and studying key system metrics such as channel capacity, outage probability, and bit error rate (BER) requires an accurate probability density function (PDF) and cumulative distribution function (CDF) of the pointing errors. Consequently, extensive research has been conducted in the literature to model the PDFs and CDFs for both types of pointing errors. In this section, we provide a comprehensive overview of pointing error distribution models, focusing on two primary categories: optical pointing errors and mmWave/THz pointing errors.

\subsection{ Overview of Optical Pointing Errors Distributions  }

\subsubsection{Conventional FSO Links}
The foundational model for analyzing the PDF of pointing errors in optical communication links was introduced by Farid and Hranilovic in \cite{farid2007outage}. They developed a statistical model that characterizes pointing errors by accounting for factors like beam width, detector size, and jitter variance. Their work assumes that the radial displacement of the beam spot follows a Rayleigh distribution, which forms the basis for deriving the PDF of the collected power fraction at the receiver. 
Using the results of \cite{farid2007outage}, various studies have extended the foundational distribution model of pointing errors to improve accuracy under different conditions. For instance, in \cite{sandalidis2008ber}, it aimed to enhance the distribution model by incorporating the combined effects of pointing errors and severe atmospheric turbulence. This study provided a closed-form statistical model that covers both the PDF and other key system performance metrics.
The studies in \cite{borah2009pointing,huang2017free,trung2014pointing,varotsos2018df,Ghanbari2023MISOFSO} introduced the additional improvements by proposing a statistical model that accommodates varying beam width adjustments under turbulent conditions. These models consider the beam divergence's influence on pointing error distribution, offering a more flexible and realistic approach for FSO systems with adjustable parameters.
Another refinement was presented by \cite{ansari2015performance,balaji2018performance}, which proposed the Malaga distribution. This model further enhances the representation of pointing errors in FSO channels, especially under moderate-to-strong turbulence. The Malaga distribution is a generalized form that subsumes several earlier models, thus allowing for more comprehensive performance evaluation in different environmental settings.

The article \cite{yang2014free} presents a more comprehensive distribution model for pointing errors by considering nonzero boresight displacements, making it suitable for terrestrial FSO links impacted by building sway and other dynamic conditions. This model goes beyond the Gaussian assumption by introducing parameters that account for both fixed displacement (boresight) and random jitter, resulting in a versatile probability density function (PDF) that adapts to various atmospheric and structural effects in urban environments.
The reference \cite{boluda2016novel} introduces an advanced approximation to the Beckmann distribution specifically aimed at modeling pointing errors in terrestrial FSO links. This approximation is versatile, incorporating parameters such as beam width and different jitter values for both horizontal and elevation displacements, along with nonzero boresight errors. It allows the distribution to be adjusted to various scenarios, whether dominated by atmospheric turbulence or pointing errors. This model proves advantageous in complex analyses, where it accommodates additional random factors like atmospheric turbulence, making it suitable for detailed system performance metrics such as outage probability and BER in real-world FSO communication conditions.

Finally in \cite{bhatnagar2016performance,bhatnagar2015performance}, the authors provide a comprehensive model for FSO MIMO links under Gamma-Gamma fading conditions with pointing errors. Building upon previous models, such as those based on Beckmann and Rician distributions, this work generalizes the analysis to consider both atmospheric turbulence and pointing errors in a unified framework, which is particularly useful for terrestrial FSO links. The proposed model facilitates system performance analysis by leveraging power series representations of PDFs and CDFs, enabling efficient analysis of key metrics like BER and outage probability. Furthermore, the MIMO setup in this study offers an inherent relaxation to pointing error constraints, making the system more resilient to misalignments. The reviewed references constitute foundational works in the development of pointing error distribution modeling, serving as essential bases for subsequent analyses and the design of optical links in future studies.

\subsubsection{UAV-based FSO Links}
Recently, with the limitations of traditional FSO links, including the requirement for LoS connections, combined with advances in UAV technology enabling high maneuverability and precise 3D positioning amidst obstacles, there has been significant interest in UAV-based FSO networks. These networks are particularly useful in urban environments with tall structures, where UAVs can help establish LoS paths between transmitters and receivers.

The foundational work in this field is represented by \cite{dabiri2018channel}, which highlighted the distinct challenges faced in UAV-based mobile FSO links compared to terrestrial models.
In this study, it was shown that, unlike stationary terrestrial links, UAV-based FSO links are affected by severe pointing errors due to the spatial instability and oscillations of UAVs. The differences between pointing errors at the transmitter and receiver were analyzed, revealing that, unlike in terrestrial links, one cannot disregard the impact of angular fluctuations at the focal point of the receiver lens in UAV-based FSO systems. A general distribution for the pointing error in UAV-based FSO systems was then proposed, which accounts for fluctuations in the AoA.

To further enhance the modeling presented in \cite{dabiri2018channel}, subsequent studies have focused on developing more accurate pointing error distribution models by incorporating the unique mobility patterns of UAVs. 
In \cite{najafi2020statistical}, Najafi et al. present a statistical model for FSO fronthaul channels in UAV-based communications, focusing on geometric and misalignment losses due to UAV fluctuations. The paper develops models for pointing errors under independent Gaussian, correlated Gaussian, and uniform distributions, simulating these fluctuations to evaluate outage probability and ergodic rate. 
In \cite{dabiri2019tractable}, the authors introduce a tractable optical channel model designed for UAV-to-UAV FSO communication links. This model specifically addresses the pointing error distribution and provides simplified, closed-form solutions for UAV-based FSO links operating under varying atmospheric turbulence conditions. The model covers both weak and moderate-to-strong turbulence scenarios by employing log-normal and Gamma-Gamma distributions, respectively. 
For a more generalized case, the study \cite{dabiri2020channel} presents a comprehensive statistical model for UAV-based FSO links with nonzero boresight pointing errors. Unlike previous models, which typically assume zero boresight, this model incorporates the effects of angular fluctuations and positional displacements across the \( x \)-\( z \) and \( y \)-\( z \) planes.
In \cite{elamassie2023multi}, the authors extend the distribution modeling of pointing errors for multi-hop airborne FSO systems with relay selection, addressing outdated log-normal turbulence channels. This model contributes to understanding and mitigating pointing errors in complex relay-based FSO systems, which is particularly relevant for scenarios involving multiple UAV relays.
\cite{khallaf2021composite} introduces a composite fading model for MIMO FSO links between UAVs, incorporating pointing errors and atmospheric turbulence. This model is particularly effective for analyzing MIMO FSO systems with UAVs that experience random positional and orientation fluctuations, thus capturing realistic pointing errors and turbulence-induced fading effects for performance metrics like outage probability.
In the work by Bashir and Alouini \cite{bashir2022energy}, the authors focus on modeling the impact of pointing errors on both energy transfer and data transmission in FSO systems utilizing UAVs. They examine a simultaneous lightwave information and power transferscheme to optimize power allocation for UAV relays, taking into account channel conditions between the source, relay, and destination.

Another category of studies has focused on modeling the channel distribution function in the presence of pointing errors for longer-range stratospheric links, such as those involving HAPs or ground-to-satellite links. In \cite{safi2020analytical}, a closed-form channel model is developed for ground-to-HAP FSO communications, incorporating factors such as atmospheric turbulence, pointing error-induced losses, and angle-of-arrival fluctuations.
The work in \cite{le2024fso} extends the channel distribution modeling by considering pointing errors for multi-hop airborne FSO systems. It examines the combined effects of atmospheric turbulence, transceiver misalignment, and imperfect CSI on performance, focusing on the deployment of HAPs and UAVs for seamless multi-hop backhauling. 
In the study \cite{ghanbari2024outage}, the design of a FSO link between an UAV and a HAP is examined under generalized pointing errors and beam wandering. The authors focus on optimizing system parameters, such as the optical beamwidth at the UAV and its flight altitude, to minimize outage probability.
In \cite{xu2024outage}, Xu et al. focus on the AoA fluctuations and pointing errors in a UAV-assisted RF/FSO system designed for space-air-ground integrated networks. This work distinguishes itself by specifically modeling AoA fluctuations, a factor not addressed in previous models that typically concentrate on standard pointing errors alone. 

In \cite{ghanbari2022outage}, the outage performance for a mixed underwater-FSO communication link involving an underwater autonomous vehicle (AUV) and a UAV through a buoy relay is analyzed, specifically addressing the impact of pointing errors. The study examines the link's reliability across two environments, clear ocean and coastal sea, by exploring how distinct underwater conditions, such as absorption and scattering, affect the channel's outage probability. This model provides insights into designing robust optical underwater and aerial links where the path involves dual-hop communication between underwater and aerial mediums.

\subsubsection{Security and Jamming Analysis}
Pointing errors, beyond affecting the primary link performance, introduce detrimental effects on link security and vulnerability to jamming. As the severity of misalignment increases, the receiver's field of view must be widened to mitigate these errors, which consequently makes the system more susceptible to jamming and security breaches. This issue is particularly relevant for mobile links, such as those involving UAVs, where severe pointing errors are more probable due to instability. In \cite{saber2024security}, the authors have analyzed the security implications of integrated HAP-based FSO and UAV-enabled RF downlink systems, examining the impact of atmospheric turbulence, RF fading, and pointing errors on secrecy outage probability and positive secrecy capacity.
In the work by Alshaer et al. \cite{alshaer2021reliability}, the authors focus on the reliability and security of an entanglement-based QKD protocol within a dynamic ground-to-UAV FSO communication system. This study highlights the impact of AoA fluctuations and pointing errors on system performance, showing that larger FoV adjustments to counter these errors can increase vulnerability to security threats, especially photon number splitting attacks.

\cite{paul2019jamming} examines the impact of pointing errors on jamming within FSO systems. The authors demonstrate that increased pointing error severity necessitates a broader receiver FoV to maintain link reliability. However, this adaptation also heightens vulnerability to jamming, as a larger FoV increases the likelihood of jammer signals entering the detector. The study quantifies this trade-off, showing that the combination of pointing errors and a widened FoV significantly impacts the signal-to-jamming ratio (SJR), leading to performance degradation in terms of BER and outage probability.
In \cite{saxena2023analysis}, the authors examine the impact of pointing errors on jamming susceptibility in an RIS-assisted UAV-based dual-hop FSO communication system. The study highlights that pointing errors, combined with atmospheric turbulence and AoA fluctuations, substantially influence the system's average BER and outage probability, particularly in the presence of a malicious UAV jammer.

\subsection{ Overview of mmWave/THz Pointing Errors Distributions } 
Moving towards higher frequencies, such as millimeter-wave and terahertz bands, the antenna dimensions become smaller, and antenna arrays are utilized to overcome high propagation loss. This leads to increased antenna gain but also results in narrower beamwidths, making systems more sensitive to misalignment errors. Such pointing errors are especially critical for mobile links, including UAV-based links, where they contribute to signal fading. Given the random nature of these errors, accurate PDF and CDF modeling is essential for evaluating system performance, a focus of recent works in the literature. Among the foundational works on pointing error modeling for mmWave and THz links are studies by \cite{dabiri2020analytical,dabiri20203d}, where realistic antenna patterns were considered to derive the PDF and CDF for linear and planar array antennas, respectively. These works use a sector-based approach, dividing the antenna pattern into multiple sectors to capture gain variations, providing an efficient yet accurate representation of antenna patterns that supports a detailed analysis of pointing errors in mmWave UAV networks.

\subsubsection{For Point-to-Point Links}
Most directional mmWave and THz communications are constrained to point-to-point links. In this scenario, \cite{dabiri2022pointing} presented a more accurate and simplified PDF and CDF for modeling pointing errors, offering improvements over previous works. This model simplifies the analysis by deriving closed-form expressions for standard array antennas, focusing on the number of antenna elements and leveraging the Gaussian approximation to capture the main-lobe behavior of the antenna array under alignment fluctuations.
In \cite{badarneh2023channel}, the authors extended the model proposed in \cite{dabiri2022pointing} by incorporating the Meijer-G and H-functions. These functions, supported by most computational software such as Mathematica and MATLAB, facilitate a more streamlined analysis of directional THz communication systems under pointing errors. 
In \cite{dabiri2022general}, the authors developed a comprehensive model for pointing errors in high-frequency mmWave and THz communication links, accommodating various scenarios with different transmitter and receiver configurations. This work derives the PDF and CDF for pointing errors accounting for non-uniform vibrations in the yaw and pitch directions. A specific case featuring uniform linear array antennas for both transmitter and receiver was also examined. 
In \cite{badarneh2023general,bhardwaj2024generalized,bhardwaj2023exact}, the authors introduced several comprehensive frameworks for analyzing UAV-aided THz communication systems, incorporating generalized geometric loss. These models account for fluctuations in UAV positioning and orientation, as well as the non-orthogonality of the THz beam relative to the detector plane. The authors derive novel analytical expressions for the PDF and CDF of the instantaneous signal-to-noise ratio (SNR) under generalized fading and geometric loss conditions. 

In recent developments, Yadav and Mallik \cite{yadav2023terahertz} addressed pointing errors in THz links between unstable transceivers using the fluctuating two-ray (FTR) fading model, deriving closed-form performance metrics. Almeida et al. \cite{almeida2024cascaded} extended pointing error analysis to cascaded $\alpha-\mu$ fading channels for dual-hop and RIS-assisted THz systems. Jemaa et al. \cite{jemaa2024performance} examined pointing error impacts in outdoor THz links with mixture gamma fading, offering analytical insights for realistic channel conditions.

\subsubsection{For Relaying Links}
In recent work on extending pointing error modeling for long-distance and relay-assisted THz systems, Dabiri et al. \cite{dabiri2024enabling} focused on UAV-based aerial backhaul links for post-disaster scenarios, where pointing errors significantly impact link reliability due to UAV instability. Yadav and Mallik \cite{yadav2024performance} studied relay-assisted double-hop RF-THz hybrid systems, incorporating pointing error effects on performance metrics like outage probability and capacity in relay configurations. Chapala et al. \cite{chapala2024multiple} addressed multiple RIS-assisted multihop relaying in THz systems, deriving analytical expressions to quantify the impact of pointing errors over extended multihop THz links.

\subsection{Impact of Pointing Errors on performance of wireless system BER, capacity, and physical layer security}

\begin{figure}
    \centering
    \includegraphics[width=\linewidth]{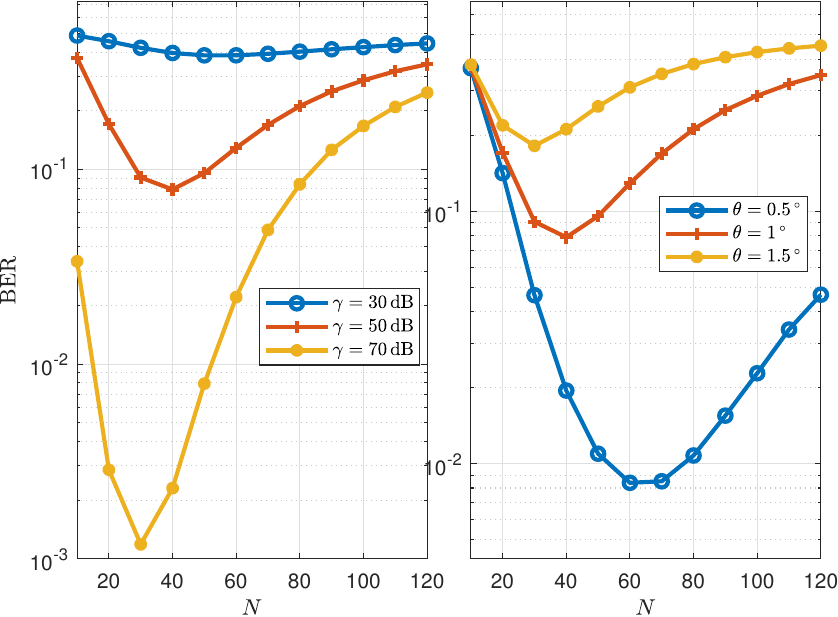}
    \caption{BER versus number of antennas $N$ under $\alpha$-$\mu$ fading and generalized pointing errors, with parameters $\alpha = 1.5$ and $\mu = 1.5$. Left: BER curves for different values of $\gamma$ at a fixed $\theta = 0.5^\circ$. Right: BER curves for different values of $\theta$ with a fixed $\gamma = 50$ dB.}
    \label{fig_BER}
\end{figure}

Pointing errors have a significant impact on system performance. Regarding the BER, the left subplot in Fig.~\ref{fig_BER} depicts the BER versus the number of antennas \( N \) system under different SNR conditions: 30 dB, 50 dB, and 70 dB. As illustrated, at higher SNRs (e.g., 70 dB), the system achieves significantly lower BERs across the range of antenna numbers compared to lower SNRs (e.g., 30 dB). This confirms the expected behavior that stronger received signals help mitigate the effects of fading and noise, leading to better error performance. On one hand, increasing $N$ enhances the spatial directivity of the transmitted beam, which theoretically improves the received SNR due to better focusing and higher gain. This leads to a reduction in BER. On the other hand, a more directive (i.e., narrower) beam becomes more sensitive to pointing errors and misalignment between the transmitter and receiver. This sensitivity degrades the BER performance beyond a certain point. As a result, the relationship between BER and $N$ becomes convex in nature: BER initially decreases with $N$, reaches an optimal point, and then starts increasing as pointing errors dominate.

Comparing the three curves for SNR = 30 dB, 50 dB, and 70 dB, we observe that higher SNR values consistently yield lower BER across all values of $N$, as expected due to the improved signal quality. However, the convex behavior is more pronounced at higher SNRs. This is because at high SNR, the system is more capable of leveraging the beamforming gain when alignment is near-optimal, but also more vulnerable to degradation due to misalignment as beamwidth narrows. Importantly, the optimal number of antennas corresponding to the minimum point on each curve decreases as the SNR increases. At lower SNR (e.g., 30 dB), a larger $N$ is needed to sufficiently improve the received power and mitigate noise. At higher SNR (e.g., 70 dB), the marginal benefit of adding more antennas is reduced due to already favorable signal conditions, and the impact of pointing error becomes the dominant factor. Therefore, a smaller optimal $N$ strikes a better balance between beamforming gain and robustness to misalignment.

To investigate the impact of the divergence angle of the antenna (i.e., $\theta$) on the BER performance, the right subplot in Fig.~\ref{fig_BER} shows the BER versus the number of antennas \( N \) for three values of $\theta$: $0.5^\circ$, $1^\circ$, and $1.5^\circ$. It is clear that increasing $\theta$ negatively impacts the BER because a larger divergence angle results in a wider beam, which reduces the optical power density at the receiver and increases susceptibility to background noise and atmospheric disturbances. This dispersion effect leads to lower received SNR and consequently higher BER. Also, we observe a similar convex relationship between the BER and $N$ for all $\theta$ curves. The location of the optimal \( N \) increases with decreased values of $\theta$ because narrower beams (smaller $\theta$) can benefit more from additional antennas without immediately suffering from significant pointing error penalties. In other words, a smaller divergence angle allows the system to leverage higher directivity longer before misalignment effects become dominant, thus shifting the optimal $N$ to higher values.

\begin{figure}
    \centering
    \includegraphics[width=\linewidth]{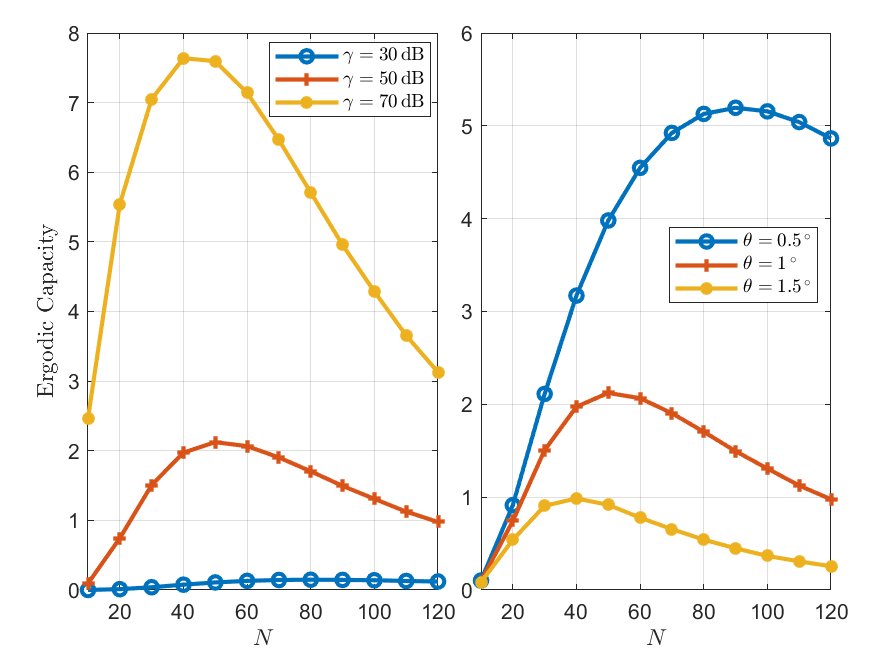}
    \caption{Ergodic capacity number of antennas versus $N$ under $\alpha$-$\mu$ fading and generalized pointing errors, with parameters $\alpha = 1.5$ and $\mu = 1.5$. Left: Ergodic capacity curves for different values of $\gamma$ at a fixed $\theta = 0.5^\circ$. Right: Ergodic capacity curves for different values of $\theta$ with a fixed $\gamma = 50$ dB.}
    \label{fig_EC}
\end{figure}

Fig.~\ref{fig_EC} examines the ergodic capacity performance under the same setups namely, as a function of the number of antennas $N$ for various SNR values (left subplot), and for different divergence angles $\theta$ (right subplot). Unlike BER, which demonstrates a convex relationship with $N$, the ergodic capacity shows a concave behavior. Initially, increasing $N$ improves the beamforming gain and hence the channel capacity, as the directivity of the beam increases and more power is focused at the receiver. However, beyond a certain point, further increasing $N$ narrows the beam excessively, making the system more susceptible to pointing errors. These misalignments reduce the effective received signal power and counteract the benefits of directivity, resulting in a saturation or even slight decline in capacity hence the observed concave trend.

Similarly, for varying divergence angles $\theta$, smaller angles lead to higher capacity due to better directivity and power concentration. However, they also introduce greater sensitivity to pointing errors. As a result, systems with smaller $\theta$ can achieve higher peak capacity, but only at an optimal $N$ that balances directivity and alignment robustness. For larger $\theta$, the peak capacity occurs at a lower $N$, reflecting the reduced benefit of adding antennas in more dispersed beam configurations.

\begin{figure}
    \centering
    \includegraphics[width=\linewidth]{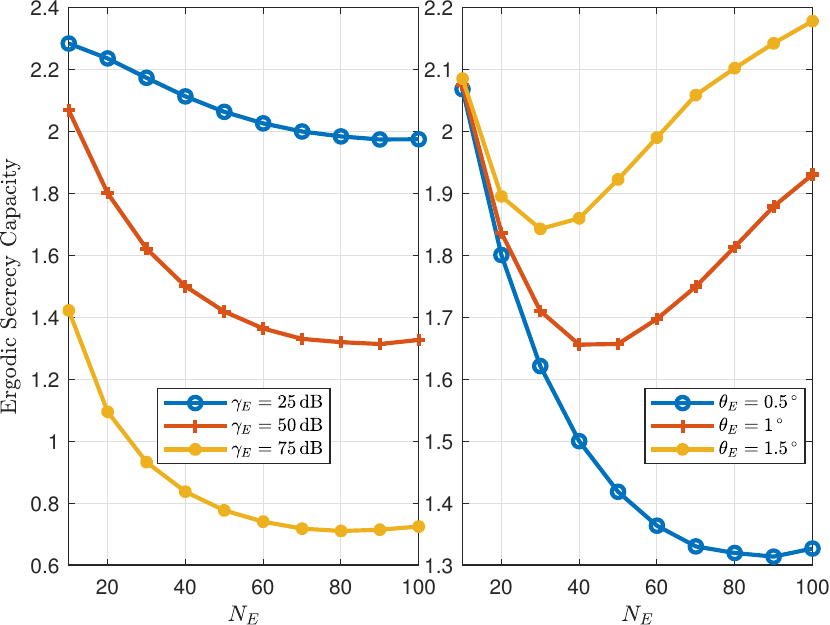}
    \caption{Ergodic secrecy capacity number of antennas versus $N_E$ under $\alpha$-$\mu$ fading and generalized pointing errors, with $\alpha = 0.5$ and $\mu = 0.5$ for both Bob's and Eve's channels. Bob operates with parameters $\theta_B = 0.5^\circ$, $\gamma_B = 50$ dB, and $N_B = 30$. Left: Secrecy capacity for various values of $\gamma_E$ at a fixed pointing error angle $\theta_E = 0.5^\circ$. Right: Secrecy capacity for various values of $\theta_E$ with a fixed $\gamma_E = 50$ dB.}
    \label{fig_SC}
\end{figure}

Fig.~\ref{fig_SC} illustrates the variation of ergodic secrecy capacity as a function of the number of antennas at the eavesdropper, denoted by \( N_E \). The left subplot considers different eavesdropper SNR levels (\( \gamma_E = 25 \) dB, 50 dB, and 75 dB), while the right subplot examines the impact of different divergence angles at the eavesdropper (\( \theta_E = 0.5^\circ \), \( 1^\circ \), and \( 1.5^\circ \)). In the left subplot, we observe that increasing the number of antennas \( N_E \) generally reduces the secrecy capacity. This trend is intuitive, as a larger \( N_E \) allows the eavesdropper to capture more spatial diversity, improving its ability to intercept the legitimate signal. Additionally, the secrecy capacity degrades further with increasing eavesdropper SNR \( \gamma_E \), since higher received power at the eavesdropper improves its decoding capability. For instance, the case with \( \gamma_E = 75 \) dB shows the lowest secrecy capacity across the entire \( N_E \) range.

Notably, all curves exhibit a \textit{convex} relationship with respect to \( N_E \): the secrecy capacity decreases initially, reaches a minimum, and then slightly increases as \( N_E \) becomes large. This convex behavior arises because while more antennas at the eavesdropper initially enhance its reception capability, at high \( N_E \), the sensitivity to misalignment may mitigate the added benefit, resulting in a modest recovery in secrecy performance.

In the right subplot, the impact of the divergence angle \( \theta_E \) is analyzed. A smaller divergence angle (e.g., \( \theta_E = 0.5^\circ \)) results in lower secrecy capacity at high \( N_E \), because narrow beams are more focused and can be efficiently captured by an aligned eavesdropper. In contrast, larger divergence angles (e.g., \( \theta_E = 1.5^\circ \)) lead to better secrecy performance overall, especially for large \( N_E \), due to the increased system susceptibility to pointing errors, which hinders the eavesdropper’s ability to collect the signal effectively. Similar to the SNR analysis, the curves for different \( \theta_E \) values also display a \textit{convex} profile with respect to \( N_E \). For smaller \( \theta_E \), the optimal (i.e., minimum) secrecy capacity occurs at lower \( N_E \), while for larger \( \theta_E \), the minimum is shifted to higher \( N_E \), and the capacity degradation is less severe. This again highlights the trade-off between beam directivity and robustness against eavesdropping.

\section{Mitigation Techniques for Pointing Errors in Optical Systems}
Various techniques have been proposed to mitigate pointing errors, each tailored to specific use cases and system requirements. This section provides an overview of these techniques, highlighting their advantages and limitations.

\subsection{Gimbals for Pointing Error Mitigation}

One of the most effective solutions for mitigating pointing errors is the use of \textit{gimbals}, which mechanically adjust the alignment of communication systems by compensating for platform movements or angular deviations. Gimbals provide stability in scenarios where the transmitter or receiver experiences motion, making them suitable for ground-based optical communication systems, such as ground-to-UAV or ground-to-satellite links. While effective in stationary or semi-stationary ground-based systems, their limitations in mobile platforms need careful consideration \cite{Peng2021GimbalPointing}:
\begin{itemize}
    \item \textbf{Weight and Power Consumption}: Although gimbals are commonly used in ground-based optical transmitters, their weight and power demands make them unsuitable for mobile platforms like UAVs or LEO satellites, where minimizing load and power consumption is critical.
    \item \textbf{Slower Response Times}: Mechanical gimbals tend to have slower response times compared to more advanced solutions, which limits their application in high-speed dynamic environments where rapid adjustments are needed.
\end{itemize}

For longer communication links, such as ground-to-satellite, achieving the necessary precision becomes much more challenging. One key limitation of gimbals is that their angular accuracy is lower compared to advanced optical devices. While gimbals can provide sufficient precision for shorter links, such as ground-to-UAV communications on the order of a few kilometers, they are inadequate for much longer links, like ground-to-satellite communications, which span several thousand kilometers. To better understand this limitation, consider a gimbal with an angular accuracy of 1 milliradian. For a 100 km link, the beam center would deviate by about 100 meters, increasing to 1000 meters for a 1000 km link and 10,000 meters for a 10,000 km link. Such large deviations are unacceptable in optical communication systems, where the beam width is relatively small. Therefore, gimbals alone cannot provide the required precision for long distance optical links, and additional techniques are necessary. In such cases, more advanced solutions like fast steering mirrors (FSMs) are combined with gimbals to enhance both the accuracy and the response time, allowing for precise beam alignment even over long distances \cite{Guo2025FSMCompensation}.

\begin{figure}
	\begin{center}
		\includegraphics[width=3.3 in]{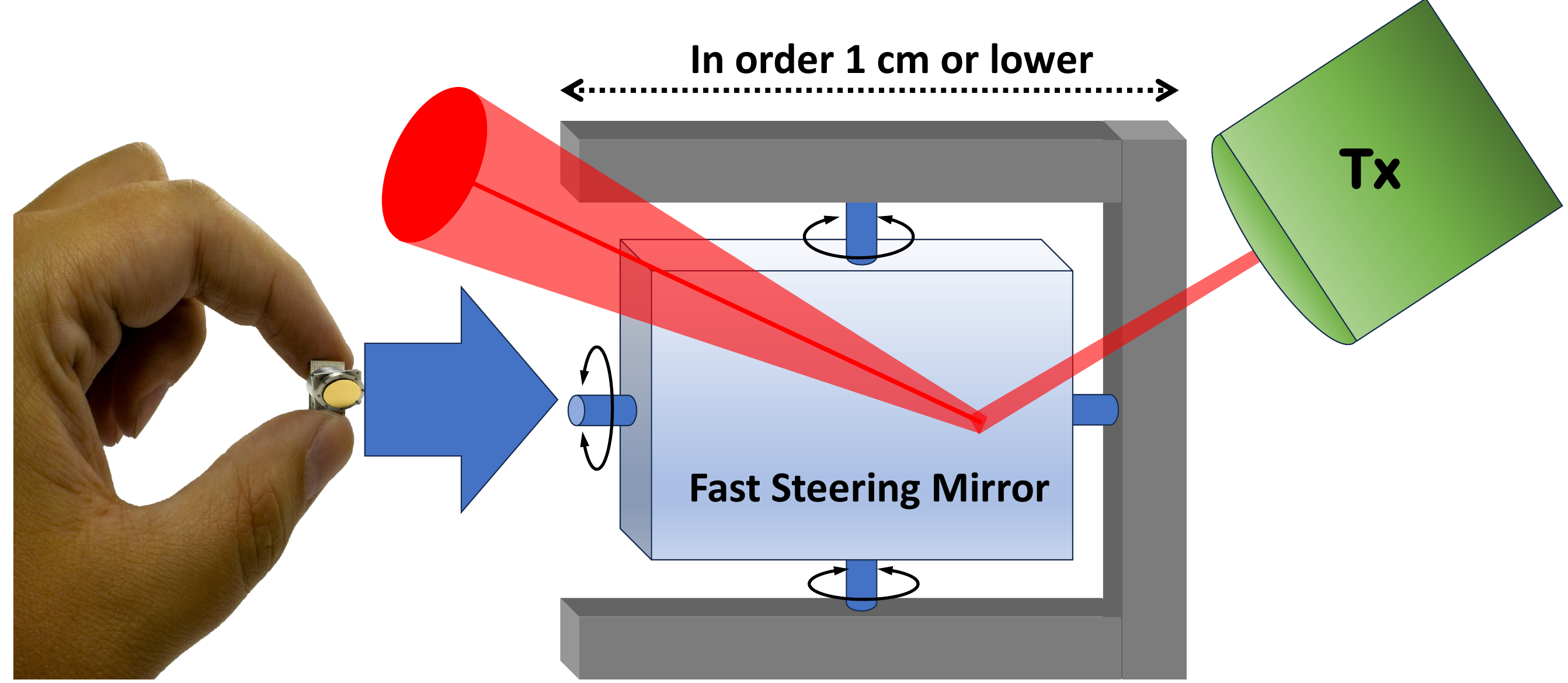}
		\caption{Illustration of a FSM system. The FSM rapidly adjusts the outgoing beam's angle using either electromagnetic or electrostatic actuation, offering precise angular corrections. This figure demonstrates how FSMs fine-tune the beam's direction to maintain alignment with the target in high-precision communication links.}
		\label{dis1}
	\end{center}
\end{figure}

\subsection{Fast-Steering Mirrors (FSMs)}
\textit{FSMs} are advanced optical devices used to make rapid and precise adjustments to the direction of a communication beam. As shown in Fig. \ref{dis1} they play a crucial role in systems where even minor pointing errors can degrade signal quality. Due to their small size and lightweight design, FSMs are widely used in mobile platforms such as LEO satellites, where fast and fine-tuned adjustments are required \cite{Guo2025FSMCompensation}.

\subsubsection{~Advantages and Limitations of FSMs}
FSMs offer several advantages over mechanical systems like gimbals, but they also have limitations that should be considered:
\begin{itemize}
    \item \textbf{Higher Angular Accuracy}: FSMs can achieve angular adjustments with precision on the order of micro-radians, making them ideal for high-precision applications such as ground-to-satellite links or UAV communications.
    
    \item \textbf{Faster Response Time}: FSMs can respond to angular fluctuations significantly faster than mechanical gimbals, typically in milliseconds or less, which makes them ideal for fast-moving platforms where quick adjustments are necessary.
    
    \item \textbf{Limited FoV}: Although FSMs are highly effective at making fine adjustments, their range of motion is limited. If the pointing error exceeds this range, FSMs cannot independently recover the target. 
    
    \item \textbf{Dependency on Larger Systems}: Due to their limited angular range, FSMs are often combined with gimbals or other systems that handle large-scale pointing. Once the beam is roughly aligned by the gimbal, FSMs take over for fine corrections, ensuring precise alignment over long distances.
\end{itemize}

\subsubsection{Actuation Mechanisms}
FSMs use different types of actuation mechanisms to adjust the mirror’s orientation, allowing for precise and rapid beam alignment. The two most common types of actuation mechanisms are:
\begin{itemize}
    \item \textbf{Electromagnetic Actuation}: This mechanism uses magnetic fields to adjust the mirror's position. Electromagnetic actuators provide high precision and rapid response times, making them suitable for larger FSM systems where both speed and accuracy are critical.
    
    \item \textbf{Electrostatic Actuation}: Electrostatic actuators use voltage to induce force and adjust the mirror's orientation. They are commonly used in microelectromechanical systems (MEMS) due to their compact design and low power consumption. However, they offer a smaller range of motion and lower force compared to electromagnetic systems, making them more suitable for lightweight mirrors.
    
\end{itemize}

\subsubsection{~Practical Use of FSMs in Communication Systems}
FSMs are most effective when used as part of a hybrid system. For instance, in long-distance communication links like ground-to-satellite connections, FSMs are responsible for fine-tuning the beam alignment after initial acquisition by gimbals or beacon-based systems. While the gimbal handles coarse adjustments to bring the beam within range, the FSM ensures continuous realignment to maintain signal integrity and prevent loss due to minor fluctuations in platform movement or environmental factors.

\begin{figure}
	\begin{center}
		\includegraphics[width=3.4 in]{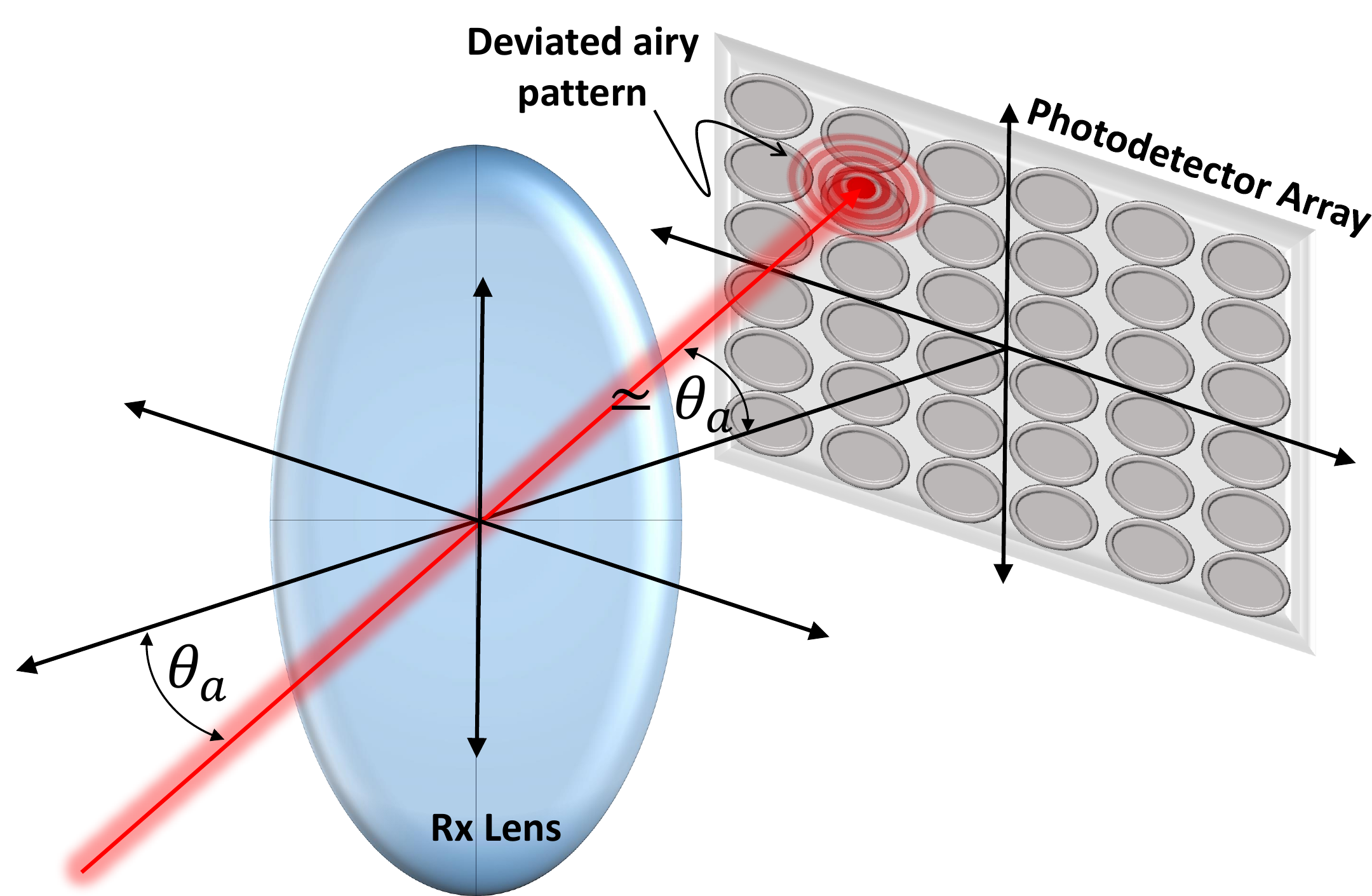}
		\caption{A simple illustration of a photodetector array system used for detecting the AoA of incoming signals. The system utilizes multiple photodetectors to capture variations in the intensity and direction of the received optical beam, enabling accurate alignment adjustments in real-time.}
		\label{AoA_est}
	\end{center}
\end{figure}
\subsection{Photodetector Arrays for Pointing Error Mitigation}
An effective technique for mitigating pointing errors, particularly in dynamic environments, is the use of \textit{photodetector arrays}. These systems can accurately estimate the AoA of incoming signals, enabling real-time adjustments to the receiver's alignment and maintaining proper communication with the transmitter.

As shown in Fig. \ref{AoA_est} photodetector arrays, composed of multiple photodetectors arranged across an array, can measure spatial variations in the intensity of the incoming signal. This spatial information is processed to compute the AoA, allowing the system to make precise real-time adjustments to the receiver's orientation. This method is particularly useful in CubeSat-assisted FSO communication systems, where accurate alignment is critical despite the limited size, weight, and power (SWaP) constraints \cite{safi2024cubesat}. These arrays offer a reliable and efficient way to track the direction of incoming beams in dynamic environments, helping maintain stable communication links in mobile platforms. However, the added complexity of the system, including the need for continuous signal processing to estimate AoA, can increase the computational demands on the receiver \cite{safi2024cubesat}.

\subsection{Beam Divergence Control for Mitigating Pointing Errors}
One of the key factors in mitigating pointing errors, particularly in dynamic environments, is the control of the \textit{beam divergence angle}. The beam divergence angle ($\theta_\text{div}$) determines the spread of the beam as it propagates over a distance $Z$. As the beam travels farther, the spot size ($w_z(Z)$) increases, which can either help compensate for angular misalignments or reduce the received signal power, depending on the chosen divergence angle. Thus, controlling $\theta_\text{div}$ plays a critical role in balancing pointing accuracy and signal strength.

\subsubsection{Relationship Between Beam Divergence and Beam Width}
The beam width $w_z(Z)$ is directly related to the beam divergence angle $\theta_\text{div}$, as given by the following equation:
\begin{align}
w_z(Z) = w_0 \sqrt{1 + \left( \frac{Z \theta_\text{div}}{w_0} \right)^2}
\end{align}
where $w_0$ is the initial beam waist at the transmitter, and $Z$ is the link length. As the beam divergence angle increases, the beam width $w_z(Z)$ grows, which can help mitigate pointing errors by allowing the beam to cover a larger area at the receiver. However, a larger beam divergence angle also reduces the beam’s intensity, leading to a lower SNR at the receiver.

\subsubsection{Adaptive Beam Divergence Control}
In optical communication systems, adaptive lenses can be used to modify the divergence angle in real-time. By changing the curvature or focal length of the lens, the system can adjust $\theta_\text{div}$ to increase or decrease the beam’s spread. This method is particularly effective in situations where the link distance $Z$ varies rapidly, such as in UAV or satellite communication systems.
    
\subsubsection{Trade-off Between Divergence Angle and Signal Power}
As mentioned, the choice of divergence angle is a balancing act between reducing pointing error and maintaining signal power. In cases where environmental conditions worsen, such as increasing wind speed leading to platform instability, the optimal strategy is to increase $\theta_\text{div}$ to account for larger angular deviations. However, widening the beam too much will result in a significant reduction in received signal power. Therefore, the system must find the optimal divergence angle that minimizes the effects of pointing error while maintaining acceptable signal strength.

This trade-off becomes particularly important in long-distance communication links, such as ground-to-satellite or UAV-based communication systems. As the link distance $Z$ increases, the divergence angle needs to be carefully controlled to ensure the beam remains aligned with the receiver without losing too much signal power. In adaptive systems, this balance is achieved through real-time monitoring and adjustments, ensuring that the communication link remains stable even under fluctuating environmental conditions.

\begin{figure}
	\begin{center}
		\includegraphics[width=3.4 in]{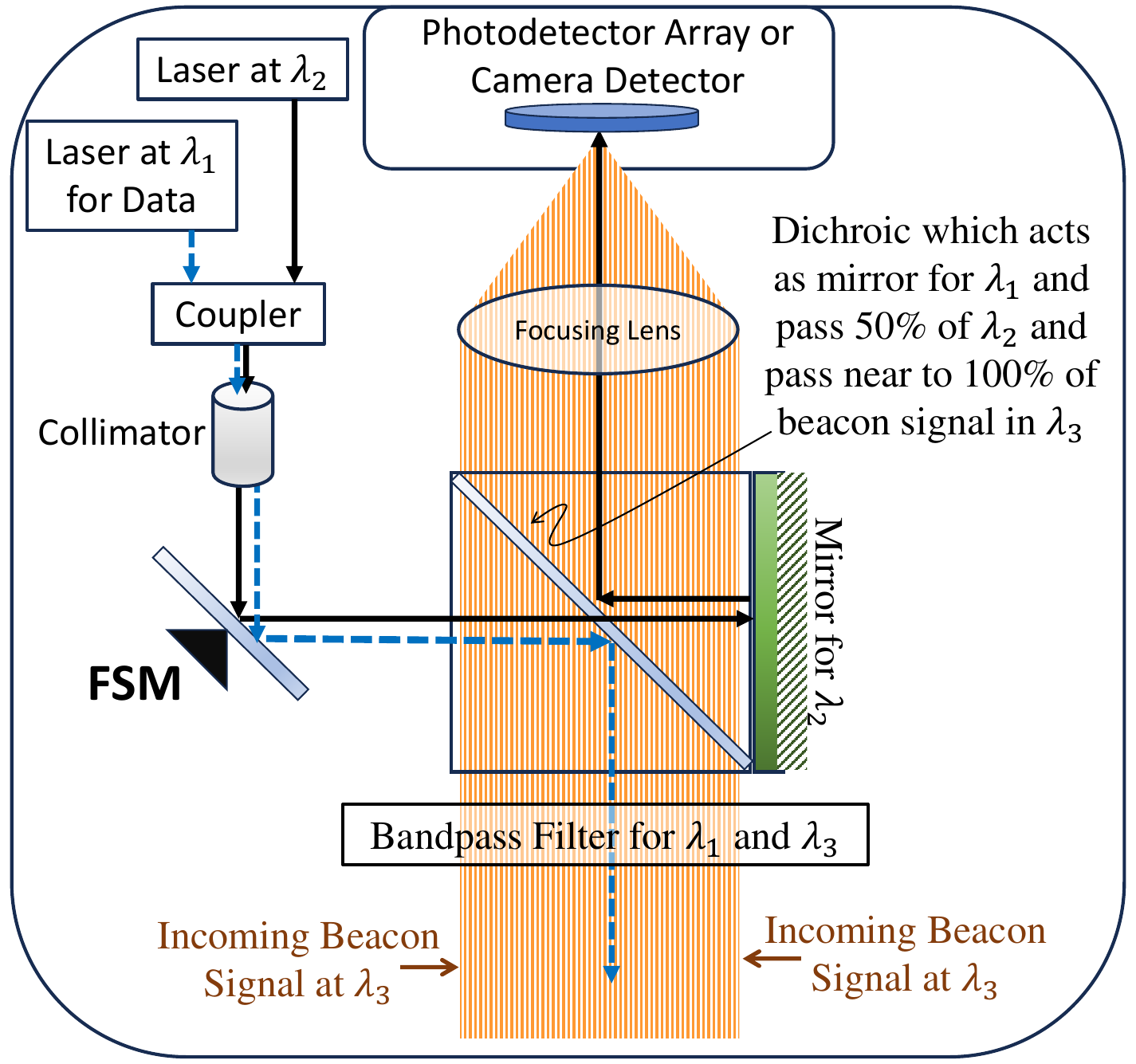}
		\caption{A beacon-based pointing correction system for asymmetric links, such as ground-to-satellite communication. The beacon signal helps align the outgoing data beam by comparing the incoming beacon signal with the transmitted data beam’s direction.}
		\label{beac1}
	\end{center}
\end{figure}
\subsection{Beacon-Based Systems for Asymmetric Links}
In asymmetric communication links, such as ground-to-UAV or ground-to-satellite links, \textit{beacon-based systems} are a practical solution for mitigating pointing errors. In these systems, a beacon signal is transmitted from one side (typically the ground station) to assist in aligning the communication beam on the satellite or UAV. Beacon systems are particularly useful where the transmitter (e.g., satellite or UAV) has limited power or computational resources to accurately track the receiver’s position. The main advantage of this technique is that the ground station can handle the majority of the complexity and computation, reducing the burden on the airborne or spaceborne transmitter. For a clearer understanding, a beacon-assisted satellite communication system is illustrated in Fig. \ref{beac1}. Here’s how the system works:

1. The ground station transmits a beacon signal at wavelength $\lambda_3$ toward the satellite. This signal is directed onto a photodetector array located on the satellite's aperture.

2. As discussed earlier, any misalignment between the satellite’s aperture and the incoming beacon signal will result in the beam being displaced on the photodetector array as shown in Fig. \ref{AoA_est}.

3. The satellite’s transmitter generates the data signal at wavelength $\lambda_1$, while a control signal at wavelength $\lambda_2$ is used to adjust the alignment of the outgoing data beam. Both signals are coupled into the same optical fiber path.

4. The combined signals (data and control) are directed to a FSM to control the output beam's direction.

5. A dichroic mirror (beam splitter) reflects most of the $\lambda_1$ signal toward the ground station, acting as a mirror. It also allows part of the control signal $\lambda_2$ to pass through.

6. The portion of the $\lambda_2$ signal that passes through the dichroic mirror is reflected by another mirror and re-directed back toward the dichroic mirror.

7. Now, the dichroic mirror acts as a mirror for the $\lambda_2$ beam, and this time it directs the $\lambda_2$ signal toward the photodetector array.

8. If the transmitter’s angle, controlled by the FSM, is perfectly aligned with the ground station without any error, the beam spots of $\lambda_2$ and the beacon signal $\lambda_3$ should overlap exactly on the photodetector array.

9. Any misalignment between the outgoing data beam ($\lambda_2$) and the beacon signal ($\lambda_3$) is detected by comparing their spot positions on the photodetector array.

10. The FSM then adjusts the beam direction until the spots of the $\lambda_2$ and $\lambda_3$ signals are perfectly aligned on the photodetector array, ensuring that the outgoing data signal is accurately pointed toward the ground station.

\begin{figure}
	\begin{center}
		\includegraphics[width=3.4 in]{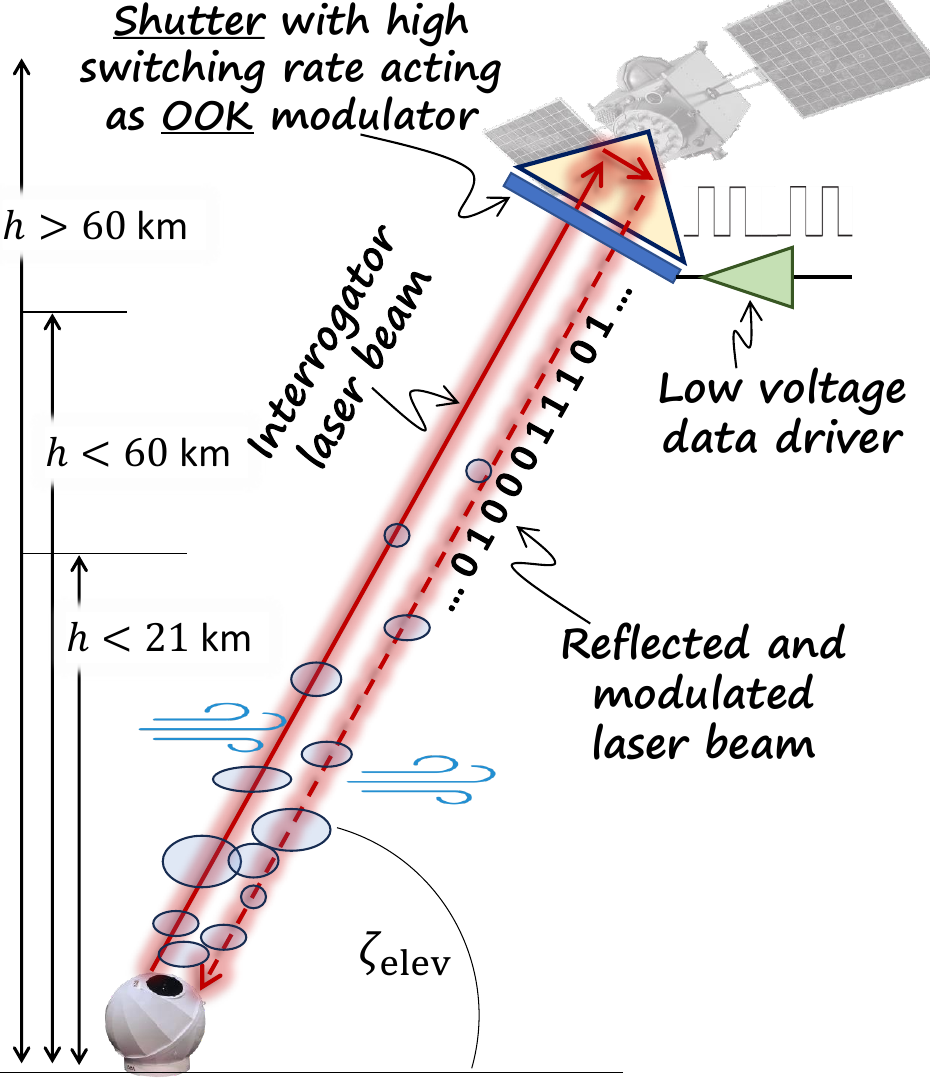}
		\caption{Illustration of the structure of MRR system to creat a satellite-to-ground link. The interrogator signal is transmitted from the ground station, reflected by the MRR, and modulated by an electro-optic shutter to encode data before being sent back to the ground station.}
		\label{mrr1}
	\end{center}
\end{figure}
\subsection{Modulating Cube Retroreflectors (MRRs)}
\textit{MRRs} provide an innovative solution for asymmetric links, especially in low-power systems such as UAVs or LEO satellites. MRRs are passive devices consisting of three perpendicular mirrors that directly reflect the signal back to the ground station, regardless of alignment errors or momentary fluctuations in the UAV or satellite's position. Unlike beacon-based systems that only assist in alignment, MRRs enable the ground station to supply the transmission power for the satellite or UAV, making them ideal for low-power platforms \cite{Dabiri2025LWC_AAVFSO,Dabiri2025TCOMM_MRTracking}.

\subsubsection{How MRR Systems Work}
The MRR system, as illustrated in Fig. \ref{mrr1}, operates as follows:

1. The ground station transmits an interrogator signal towards the cube retroreflector installed on the UAV or satellite. 

2. The cube retroreflector, which consists of three perpendicular mirrors, reflects the incoming laser beam directly back to the ground station. Due to the perpendicular arrangement of the mirrors, the reflected beam is always directed back to its ground source, independent of the UAV or satellite's real-time misalignmnets.

3. An electro-optic shutter (modulator) is placed in front of the MRR. The shutter opens or closes based on the data being transmitted, typically using On-Off Keying (OOK) modulation, to either block or allow the passage of the signal. This modulates the reflected signal with the data.

4. The data signal, modulated by the electro-optic shutter, is reflected back to the ground station. 

5. Finally, the modulated signal is detected at the ground station, where the encoded data is recovered. Without the need for any transmitted power or alignment devices like gimbals, this downlink from the UAV or satellite to the ground station is established. In fact, both the alignment and power provision tasks are handled by the ground station, which makes it suitable for many scenarios and applications.

\subsubsection{Challenges in MRR System Design}

The design of MRR systems presents several challenges, particularly when optimizing for high data rates and efficient power usage:

\begin{itemize}
 \item \textbf{Bandwidth vs. Aperture Size:} The bandwidth of the MRR’s modulator is inversely related to its effective aperture size. For modulation rates in the tens of gigahertz, the modulator must have an aperture area smaller than 1 cm$^2$. However, smaller aperture sizes lead to lower signal reception due to reduced capture area.

 \item \textbf{Received Signal Power:} Since the MRR system relies on a round-trip path (ground station to satellite and back), the received signal power is significantly reduced. The smaller the effective aperture, the less power is reflected back, leading to a weak signal at the ground station.

 \item \textbf{Narrow Beam Divergence:} To compensate for the low power return, the system must use a narrower beam divergence. While this helps focus more power on the retroreflector, it also increases sensitivity to geometric pointing errors. If the beam is even slightly misaligned, a significant portion of the power will be lost, and the link could degrade or fail.

 \item \textbf{Precision Design:} Due to the narrow beam divergence and sensitivity to misalignment, the design of MRR systems requires high precision. Even small pointing errors or environmental disturbances can significantly impact performance, making the alignment process critical for maintaining a stable communication link.
 \end{itemize}

\subsubsection{~Technologies Used in MRR Modulators}
The selection of modulation technology in MRR systems is largely determined by the specific needs of the application, such as communication speed, power efficiency, environmental robustness, and system miniaturization. Recent advancements in materials science and microfabrication have broadened the scope of these devices, making them suitable for a wide range of cutting-edge optical communication applications. Some common modulator technologies include: 
\begin{itemize}
    \item \textbf{Liquid Crystal Modulators}: These modulators use liquid crystals to modulate the light signal. While they offer lower switching speeds compared to electro-optic devices, they are advantageous for applications where low power consumption is critical.

    \item \textbf{Electro-Optic Modulators (EOMs)}: These modulators use electric fields to modify the optical properties of signals for high-speed data communications. EOMs provide very fast switching rates (up to GHz), though they require high voltages and more complex integration.
    
    \item \textbf{Multiple Quantum Well (MQW) Modulators}: Voltage is applied to quantum wells to affect light absorption, typically used in low-power military and space communication systems. These modulators support Gbit/s speeds but come with limitations in modulation range and wavelength precision.
    
    \item \textbf{Electro-Absorption Modulators (EAMs) with MQW}: Leveraging multiple quantum wells to modulate light absorption, EAMs are well-suited for UAV communications with Gbit/s data rates. They improve bandwidth and lower capacitance through a pixelated design, making them ideal for space-constrained platforms.

    \item \textbf{All-Optical Retro-Modulators (AORMs):} These modulators utilize high-refractive-index hemispheres along with ultrafast CuO nanocrystals to achieve terabit-per-second all-optical modulation. They are particularly suited for high-speed, multidirectional optical communications without the need for electrical inputs \cite{Dabiri2025LWC_AAVFSO,Dabiri2025TCOMM_MRTracking}.

\end{itemize}

\begin{figure*}
	\centering
	\subfloat[] {\includegraphics[width=5 in]{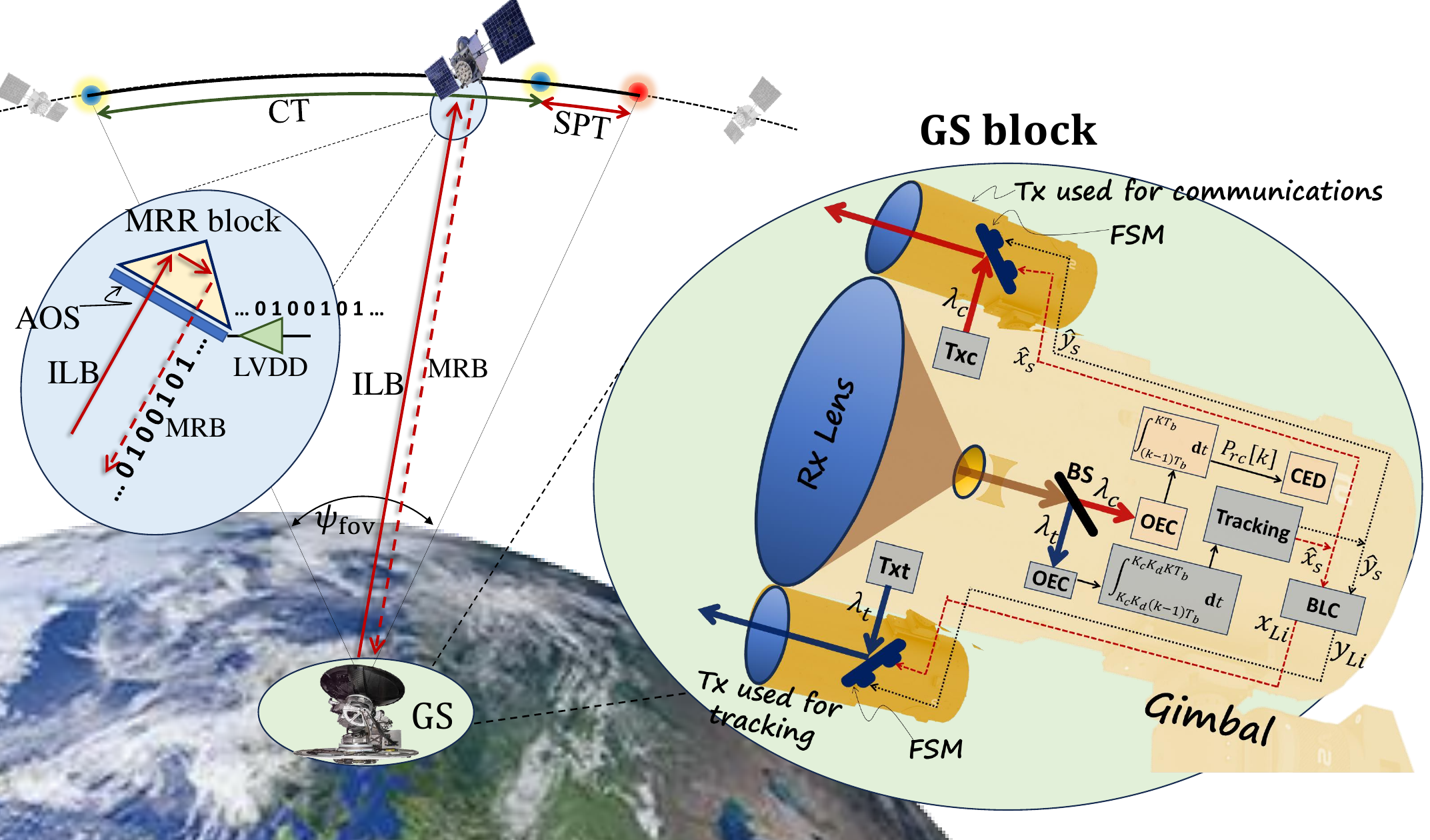}
		\label{bm1}
	}
        \hfill
	\subfloat[] {\includegraphics[width=2 in]{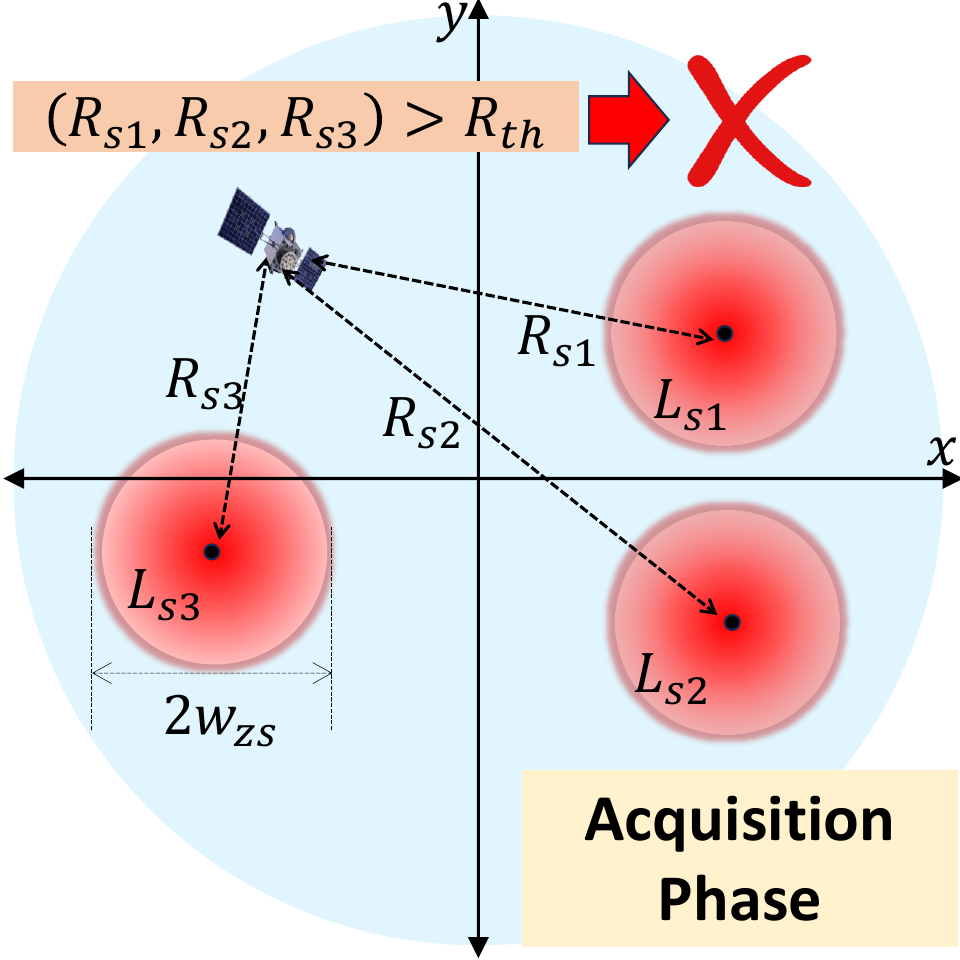}
		\label{bm2}
	}
        \hfill
        \subfloat[] {\includegraphics[width=2 in]{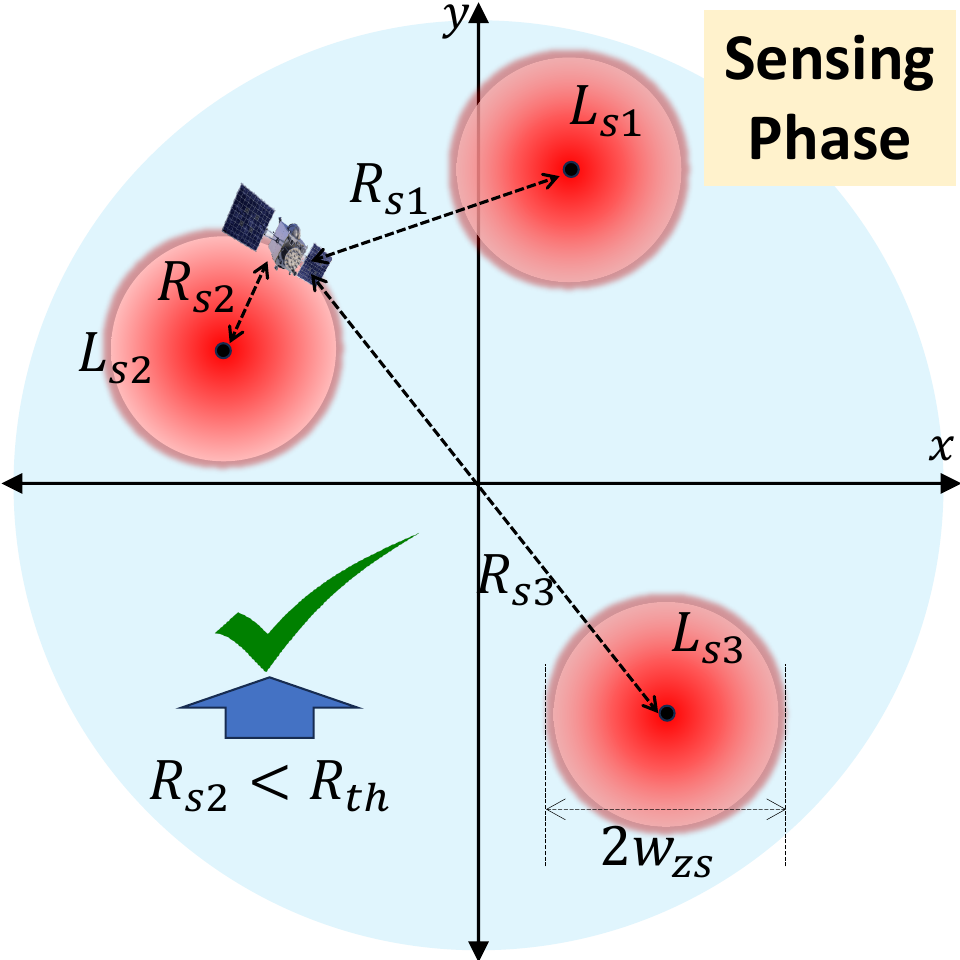}
		\label{bm3}
	}
        \hfill
        \subfloat[] {\includegraphics[width=2.1 in]{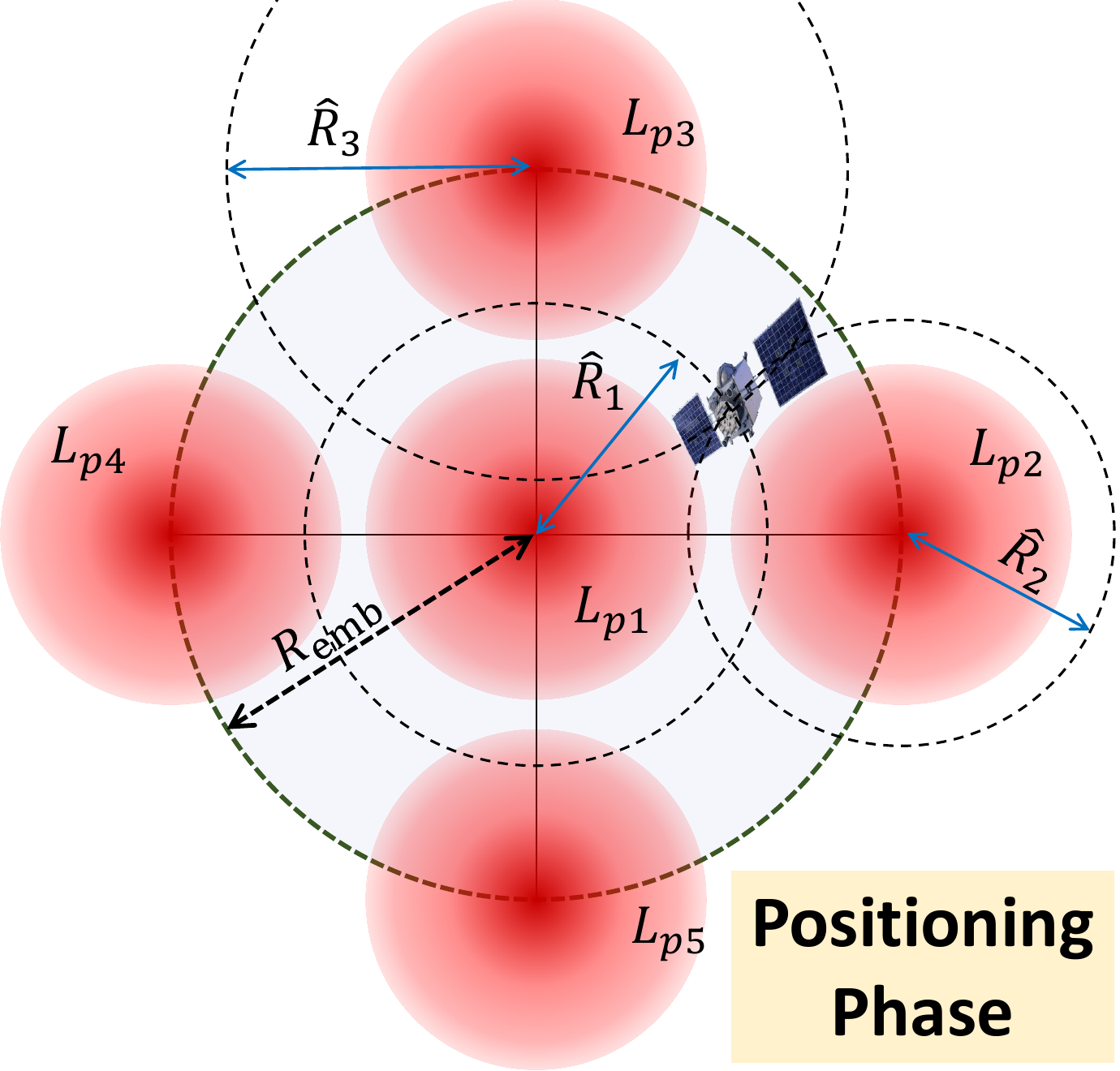}
		\label{bm4}
	}
        \hfill
        \subfloat[] {\includegraphics[width=2.5 in]{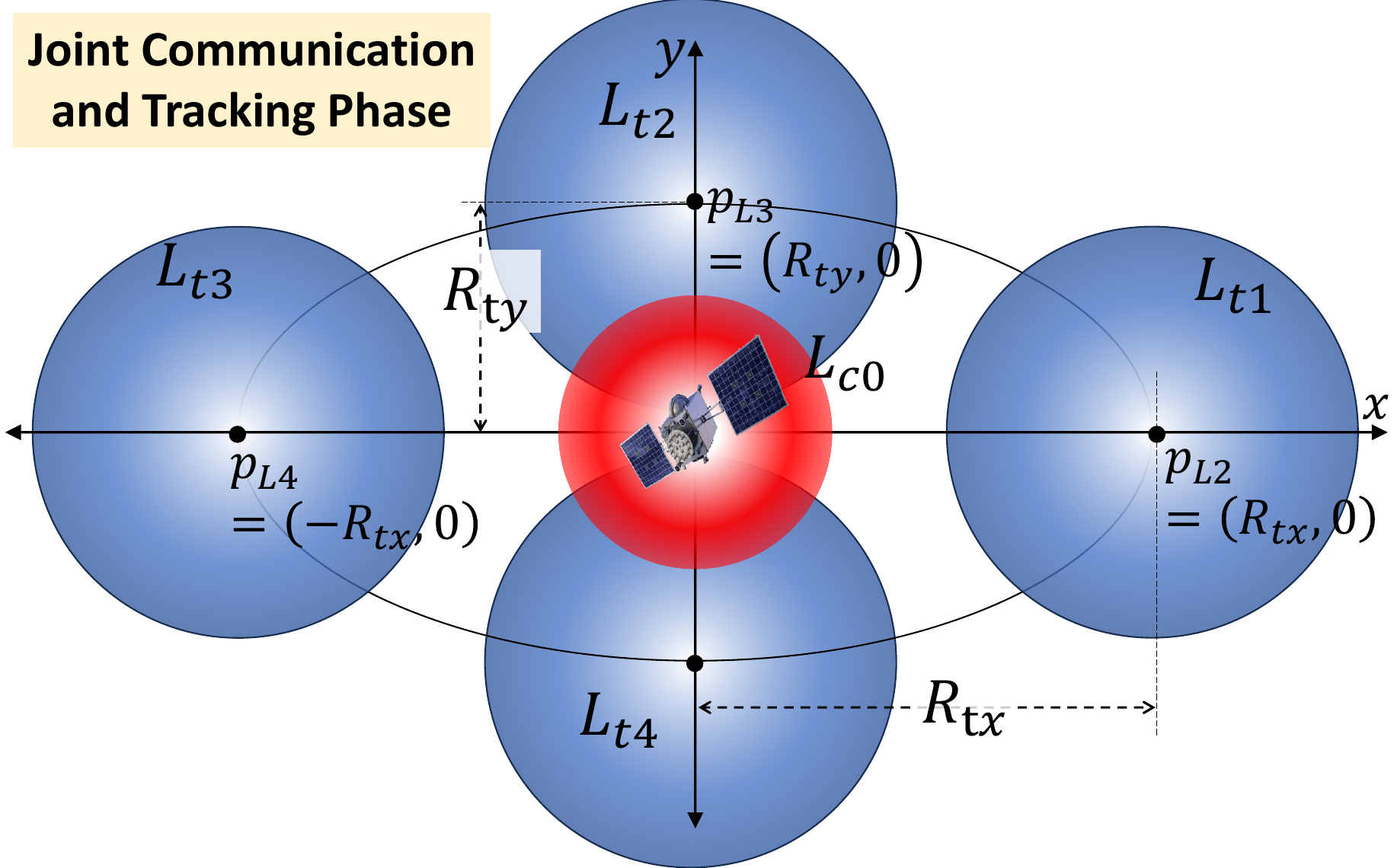}
		\label{bm5}
	}
        \hfill
	\caption{Illustration of the MRR-based communication system for LEO satellites. (a) The ground station setup with two laser transmitters and MRRs mounted on a gimbal. (b) Illustration of the acquisition phase, where the search algorithms explore a region with a high probability of satellite presence. (c) Illustration of the sensing phase. Once one of the beams in the acquisition phase gets close to the satellite, the reflected power from the MRR exceeds the threshold, allowing the ground station to sense the presence of the satellite. (d) Illustration of the positioning phase. At this stage, one beam is positioned at the obtained sensing point, and additional beams are placed around this point at different time intervals according to various algorithms to achieve a more accurate estimation of the satellite's position. (e) Continuous tracking during communication to maintain the link.
		Note that the four laser beams used for tracking are generated by the Txt at specific time intervals. 
	The blocks OEC, CED, BS, SPT, CT, ILB, and LVDD are
		optical-to-electrical converter, 
		channel estimation and detection, 
		beam splitter,
		sensing and positioning time, 
		communication time,
		interrogator laser beam, and
		low voltage data driver, respectively \cite{Dabiri2025TCOMM_MRTracking}.}
	\label{bm}
\end{figure*}
\subsection{~MRR System Design for Long-Distance Links}
One promising application for \textit{MRRs} is solving the downlink problem associated with small LEO satellites. Due to the constraints of size and power consumption in small LEO satellites, which are often around 10 cm in size, these systems typically suffer from limited transmission power and low data rates, reaching a maximum of only a few Mbps. However, with the growing demand for high-resolution monitoring and other data-intensive applications, higher data rates are required to support such missions.

The MRR technology offers a viable solution to overcome these challenges by offloading the power transmission burden to the ground station. As illustrated in Fig. \ref{mrr1}, the ground station can continuously track the LEO satellites as they enter its field of view and establish an MRR-based downlink using an interrogator signal. However, the design of such a system is significantly more complex than traditional satellite communication links due to several factors \cite{Dabiri2025LWC_AAVFSO,Dabiri2025TCOMM_MRTracking}:

\begin{itemize}

\item {\bf Beam Width Sensitivity}: 
    \begin{itemize}
        \item In conventional satellite communication links, the beam width is typically around 100 meters or more.
        \item This large beam width allows for maintaining the link despite pointing errors of just a few millimeters.
        \item However, in MRR systems, due to the small size of the MRR and the signal's round-trip nature (which doubles the path loss), geometric losses are significantly higher.
    \end{itemize}

\item {\bf Narrow Beam for Long-Distance Links}: 
    \begin{itemize}
        \item To offset these geometric losses, the beam width must be reduced, typically to about 10 meters.
        \item This focuses more power on the MRR but greatly increases the system's sensitivity to ground station pointing errors.
        \item Achieving the required alignment precision demands pointing accuracy in the microradian range, making the system highly susceptible to misalignments.
        \item Recent studies, such as \cite{Dabiri2025TCOMM_MRTracking} and \cite{dabiri2024modulating2}, have explored these challenges in detail.
    \end{itemize}

\item {\bf Acquisition and Sensing Phases}:
    \begin{itemize}
        \item The ground station setup for MRR-based communication is illustrated in Fig. \ref{bm1}, where two laser transmitters controlled by two MRRs are mounted on a gimbal platform.
        \item The process begins when a LEO satellite enters the field of view of the ground station, as shown in Fig. \ref{bm2}.
        \item The acquisition phase starts using various techniques (further research is required in this area).
        \item Once the laser beam approaches the satellite, the reflected power increases, allowing the satellite’s presence to be sensed, as illustrated in Fig. \ref{bm3}.
    \end{itemize}

\item {\bf Fine Positioning Phase}:
    \begin{itemize}
        \item Sensing the satellite is not sufficient to establish the link.
        \item A positioning phase must follow to achieve the high precision necessary for maintaining the link within a few meters, as shown in Fig. \ref{bm4}.
        \item Given the short window of time in which the satellite is visible, it is critical that the acquisition, sensing, and positioning phases are completed as quickly as possible.
        \item This is an ongoing challenge in the literature and requires further research.
    \end{itemize}

\item {\bf Tracking and Communication}:
    \begin{itemize}
        \item Once the satellite has been positioned, the ground station's alignment accuracy is sufficient to establish the communication link.
        \item However, due to the high velocity of the LEO satellite, if precise tracking is not continuously maintained during communication, the link will quickly be lost.
        \item The acquisition phases must then be repeated, as shown in Fig. \ref{bm5}.
        \item This dynamic tracking phase is essential for ensuring the continuous operation of the link.
    \end{itemize}
\end{itemize}

Finally, although the ground station's structure becomes more complex with the inclusion of multiple MRRs, it is important to note that the same ground station can continuously serve dozens of LEO satellites. These satellites are highly cost-effective and can provide high-speed, secure downlinks, offering a significant improvement in overall system performance.

Although the ground station structure becomes more intricate due to the larger lens dimensions, the use of precise and fast sensors for MRR, more complex processing algorithms, and the inclusion of faster response gimbals, it is important to highlight that the one ground station can continuously support multiple LEO satellites. These cost-effective satellites offer high-speed, secure downlinks, leading to a substantial improvement in overall system performance and cost.

\subsection{MIMO-Based Techniques for Pointing Error Mitigation}
\textit{Multiple-input multiple-output (MIMO)} techniques can also play a crucial role in mitigating pointing errors, especially in high-frequency communication systems like mmWave and THz. In MIMO systems, multiple antennas are used to transmit and receive signals, providing redundancy and improving the system's resilience to pointing errors.

\begin{itemize}
    \item \textbf{Spatial Diversity}: MIMO systems exploit spatial diversity by transmitting the same signal over multiple antennas, reducing the impact of pointing errors on overall system performance.
    \item \textbf{Beamforming}: Advanced beamforming techniques in MIMO systems can dynamically adjust the direction of the communication beam, compensating for angular deviations and ensuring that the transmitted signal remains aligned with the receiver.
\end{itemize}

MIMO-based solutions are highly effective in dynamic environments where the communication link is subject to constant movement. However, these techniques require advanced signal processing and can increase the system's complexity and power requirements.

\subsection{Conclusion}
Each of the techniques discussed offers unique advantages and trade-offs for mitigating pointing errors in communication systems. Gimbals and FSMs provide high precision but are limited by their weight and power consumption. Photodetector arrays and camera-based systems offer a high degree of accuracy in determining the angle of arrival but may increase system complexity. Adjusting the beam divergence angle provides a balance between robustness to misalignments and maintaining adequate signal strength. Beacon-based systems and MRRs offer efficient solutions for asymmetric links, particularly in low-power platforms like UAVs and satellites. MIMO-based techniques provide resilience and adaptability in dynamic environments, making them ideal for high-frequency communications.


\section{Case Studies and Applications}
\subsection{Reconfigurable Intelligent Surfaces}

RIS revolutionize wireless network design by facilitating programmable manipulation of the electromagnetic propagation environment \cite{8796365,Dabiri2025LWC_IRS}. Comprising densely arranged sub-wavelength passive elements with adjustable impedance, RIS can induce phase shifts on incident electromagnetic waves to achieve desired beamforming, reflection, or focusing effects \cite{9140329}. In contrast to conventional amplify and forward (AF) relays, RIS can operate without active signal amplification, providing a spectrally efficient, low power, and noise-free solution to extend coverage, improve spectral efficiency, and build ultra-massive multiple-input multiple-output (UM-MIMO) systems \cite{8811733, cui2014coding, 8741198}. With its ability to support intelligent, flexible, and sustainable wireless ecosystems, RIS will undoubtedly play a key role in 6G networks.

Recent research has significantly expanded the operational capabilities of RIS beyond basic beamforming applications. Modern RIS implementations now support multifunctional operation modes, including simultaneous reflection and absorption for interference management \cite{9473762}, dynamic holographic beam shaping \cite{9690474}, spectrum sharing systems \cite{9882367}, \cite{10557617}, energy harvesting systems \cite{alshawaqfeh2025performance}, and integrated sensing and communication (ISAC) systems \cite{10243495}.  Developments in RIS technology have enabled finer control at the sub-wavelength level, facilitating more agile beam-steering and spatial multiplexing \cite{bartoli2023spatial}. Additionally, efforts in machine learning-aided configuration algorithms have improved the feasibility of deploying RIS in highly dynamic scenarios \cite{10552229}, where traditional channel estimation and optimization approaches would otherwise be computationally prohibitive. Emerging research is also investigating RIS  integration with UAVs \cite{9703337}, satellite links \cite{10559954}, and vehicular networks \cite{chen2022reconfigurable}. 

When considering RIS in practical deployments, pointing error becomes a significant hindrance, particularly in highly directed, narrow-beam situations. Since RIS-assisted links depend on coherent phase alignment across all reflecting elements, they are considerably more prone to angular variations than conventional systems. The advantages of RIS beamforming can be significantly compromised by misalignments in the angle of incidence or reflection. These phase errors increase path loss and outage probability, reduce energy focusing capabilities, and deteriorate the effective channel rank \cite{9530530}. This effect becomes even more pronounced in RIS-assisted systems operating at mmWave or THz bands, where beamwidths are extremely narrow, resulting in sharp performance degradation with even minimal misalignment.

Several technical challenges emerge when attempting to mitigate the detrimental effects of pointing errors in RIS-assisted systems. Conventional passive beamforming approaches, which often assume perfect knowledge of RIS phases, become impractical under the stochastic nature of pointing error. Moreover, accurate and timely estimation of the effective channel in the presence of pointing errors is crucial, yet traditional channel estimation methods typically rely on extensive pilot overhead and static assumptions, rendering them insufficient in dynamic environments. These limitations necessitate the development of robust beamforming schemes that are tolerant to angular deviations, potentially leveraging stochastic optimization, worst-case design, or adaptive learning strategies.

From a system design perspective, RIS deployment strategies must also account for the nature of pointing error dynamics during the planning stage. This includes optimizing the RIS location, orientation, and surface geometry to reduce the likelihood of severe misalignment. The use of redundant RIS panels \cite{yuan2022reconfigurable}, distributed surfaces \cite{yang2021energy}, or hybrid active-passive architectures can further improve robustness. Machine learning-driven techniques may offer promising alternatives for real-time channel state inference under angular perturbations \cite{10361836}. Lastly, experimental validation and standardization efforts remain essential to accurately characterize pointing error effects in realistic conditions and to guide the design of reliable  RIS-assisted networks for future wireless systems.

\subsection{Non-Orthogonal Multiple Access (NOMA)}

NOMA represents a paradigm shift in multiple access methodologies, offering superior spectral efficiency relative to conventional orthogonal schemes \cite{7263349}, \cite{7676258}. In contrast to traditional approaches that mandate exclusive time frequency resource allocation, NOMA facilitates simultaneous resource sharing among multiple users through innovative power domain superposition coding and sophisticated successive interference cancellation (SIC) techniques. The architecture strategically allocates greater transmission power to disadvantaged users while employing SIC at receivers with favorable channel conditions, thereby optimizing system throughput, ensuring equitable service distribution, and extending reliable coverage to network edge users.These distinctive capabilities position NOMA as a transformative technology for modern wireless infrastructures particularly when synergistically combined with MIMO architectures and advanced beamforming techniques further strengthening its role in addressing the growing demands for high data rates, low latency, and ultra-reliable communications \cite{akbar2021noma}.

In practical deployments, pointing errors pose significant challenges to NOMA systems, particularly in those employing FSO \cite{islam2021impact} or mmWave communications \cite{xiao2017millimeter}, as they introduce random fluctuations in the received signal power. These variations critically undermine NOMA's fundamental power allocation strategy, which depends on maintaining precise power differences between users for effective SIC. As a result, three major performance degradations typically arise: (1) increased inter-user interference due to imperfect SIC, (2) elevated bit error rates caused by distorted signal decoding, and (3) a reduction in overall system capacity. The time-varying nature of pointing errors further exacerbates these issues, making it especially difficult to maintain consistent Quality of Service (QoS) in mobile scenarios or outdoor environments where alignment stability is inherently limited. To address the challenges posed by pointing errors in NOMA systems, several mitigation strategies have been explored, each targeting a different aspect of the performance degradation caused by angular misalignments and power fluctuations:

\begin{itemize}
    \item \textbf{Adaptive Power Allocation:} Adaptive power allocation techniques dynamically adjust the transmit power levels of users in real time based on instantaneous channel state information and estimates of pointing error severity \cite{dong2019adaptive}. By maintaining the required power disparity between strong and weak users—even under signal fluctuations—these algorithms help preserve the effectiveness of SIC, which is critical to NOMA’s core operation.

    \item \textbf{Beam Tracking and Alignment:} Beam tracking systems are particularly effective in high-frequency NOMA scenarios, such as mmWave or FSO links, where narrow beams are highly susceptible to misalignment \cite{liu2020robust}. These systems continuously monitor and adjust the transmitter and receiver orientations using feedback control, mechanical actuators, or electronically steerable arrays to counteract dynamic mispointing induced by environmental factors like wind, vibration, or user mobility.

    \item \textbf{Robust SIC Techniques:} Robust SIC algorithms enhance the receiver’s ability to decode superimposed signals despite imperfections in power separation caused by pointing errors \cite{iswarya2021survey}. These mechanisms leverage error correction, adaptive thresholds, or probabilistic decoding to reduce inter-user interference and improve decoding success rates under non-ideal channel conditions.

    \item \textbf{Hybrid Multiple Access Schemes:} Hybrid access schemes integrate NOMA with orthogonal multiple access (OMA) to provide architectural flexibility \cite{giang2020hybrid}, \cite{suganuma2019hybrid}. In severe pointing error scenarios, the system can temporarily switch to OMA-like resource allocation, preserving link reliability and preventing service degradation.

    \item \textbf{Machine Learning Approaches:} Machine learning-driven methods proactively predict and compensate for pointing errors based on historical data and environmental variables \cite{kara2021lightweight}, \cite{yang2021machine}. These models can dynamically tune parameters such as power coefficients, beam directions, or decoding orders, enabling intelligent, context-aware system behavior even in the presence of rapid alignment changes.
\end{itemize}

By integrating these strategies, NOMA systems can significantly enhance their resilience against pointing errors, ensuring reliable communication, improved spectral efficiency, and support for next-generation wireless goals like ultra reliable low-latency communications (URLLC) and massive connectivity. These advancements reduce latency and packet loss for mission critical applications while enabling massive connectivity in IoT ecosystems, with future innovations like artificial intelligence (AI)-powered predictive beam management and RIS-assisted NOMA poised to further strengthen performance in 6G networks, particularly in high-frequency terahertz communications where beam alignment is even more critical.

\subsection{OAM systems}
\begin{figure}
    \centering
    \includegraphics[width=0.5\textwidth]{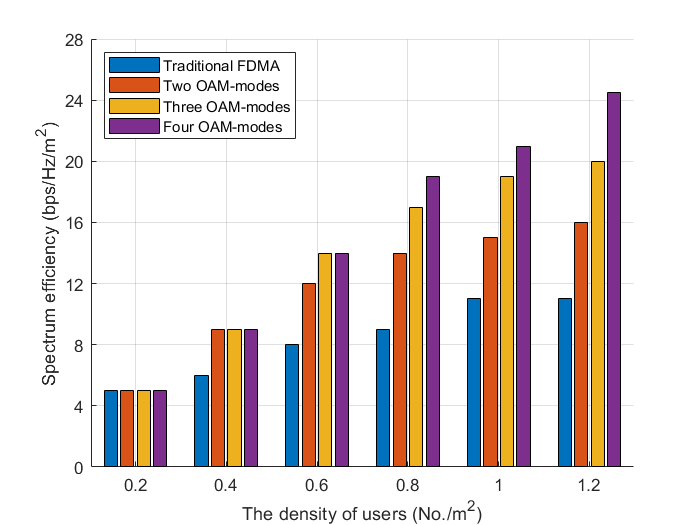}
    \caption{Comparison of OAM modes with FDMA system.}
    \label{fig:oam_fdma}
\end{figure}

As the demand for higher data rates and increased capacity grows in next-generation wireless communication, innovative technologies are required to overcome current limitations. OAM has emerged as a promising solution, offering a unique way to multiplex signals by utilizing the spatial phase structure of electromagnetic waves. Unlike conventional multiplexing techniques, OAM enables multiple independent data streams to coexist in the same frequency band by exploiting its helical phase distribution, expressed as $e^{i l \phi}$, where $\phi$ is the azimuthal angle and $l$ is the topological charge determining the number of twists per wavelength. This additional degree of freedom significantly enhances spectral efficiency and system capacity, making OAM a key enabler for future high-speed communication networks such as 6G, optical communication, and satellite links \cite{Yao2011OAM}. Traditional wireless communication systems, which use plane electromagnetic waves, often face signal interference when transmitting and receiving data at the same frequency, limiting network capacity and reliability. In contrast, OAM mitigates these interference issues by leveraging its inherently orthogonal modes, allowing multiple independent data streams to coexist without overlapping  \cite{Affan2021OAM6G,Noor2022OAMReview}. Fig.~\ref{fig:oam_fdma} illustrates that the spectrum efficiency of conventional frequency-division multiple access (FDMA) is consistently lower than that of systems employing multiple OAM modes, particularly as user density increases. Notably, increasing the number of OAM modes from two to four yields a marked improvement in spectrum efficiency, with the four-mode configuration demonstrating the highest performance at elevated user densities. Furthermore, the comparable efficiency observed at user densities of 1.0 and 1.2 users/m² underscores the effectiveness of OAM in managing increased user loads \cite{Cheng2019OAMWireless}.

Various beam types are capable of carrying OAM, each deriving its unique properties from solutions to fundamental wave equations. Laguerre-Gaussian (LG) beams, obtained as solutions to the paraxial wave equation in cylindrical coordinates. In contrast, Hermite-Gaussian (HG) beams, which originate from Cartesian-coordinate solutions of the paraxial wave equation, do not naturally possess OAM but can be converted into OAM modes via appropriate transformations. Ince-Gaussian beams, emerging from elliptical coordinate solutions to the paraxial wave equation, provide a natural bridge between the LG and HG beam families, thereby offering versatile OAM capabilities. Meanwhile, Bessel beams, as exact solutions to the Helmholtz equation, are renowned for their non-diffracting and self-healing properties while efficiently carrying OAM, and Mathieu beams, derived from the Helmholtz equation in elliptical cylindrical coordinates, exhibit complex spatial profiles that further enhance their capacity to support OAM  \cite{ref6,ref10,McGloin2005Bessel,GutierrezVega2000Mathieu}. One of the most frequently employed beams in orbital angular momentum–based communication is the LG beam, mainly due to its ring-shaped intensity distribution, inherent orthogonality among different topological charges, and ease of generation in cylindrical coordinates. Mathematically, the electric field for an LG beam can be expressed as \cite{Dabiri2024TechRxiv}:
\begin{equation}
\begin{split}
    u_{p,\ell}(r,\phi,z) &= \sqrt{\frac{2p!}{\pi (p + |\ell|)!}} \frac{1}{w(z)}
    \left(\frac{\sqrt{2} r}{w(z)}\right)^{|\ell|} \\
    &\quad \times L_p^{|\ell|} \left(\frac{2r^2}{w^2(z)}\right) e^{-i \ell \phi} \\
    &\quad \times \exp\left(-i k \frac{r^2}{2 R(z)} + i \zeta(z) \right)
\end{split}
\end{equation}

where $\zeta(z)$ denotes the Gouy phase. Furthermore, $L_p^{|\ell|}$ is the generalized Laguerre polynomial, $\ell$ is the topological charge (related to OAM), $p$ is the radial index, and $k = 2\pi / \lambda$ is the wave number. $\phi$ is the azimuthal angle in cylindrical coordinates, indicating how the field’s phase and amplitude vary around the beam’s axis.

\begin{figure*}[t]
\centering
\includegraphics[width=7in]{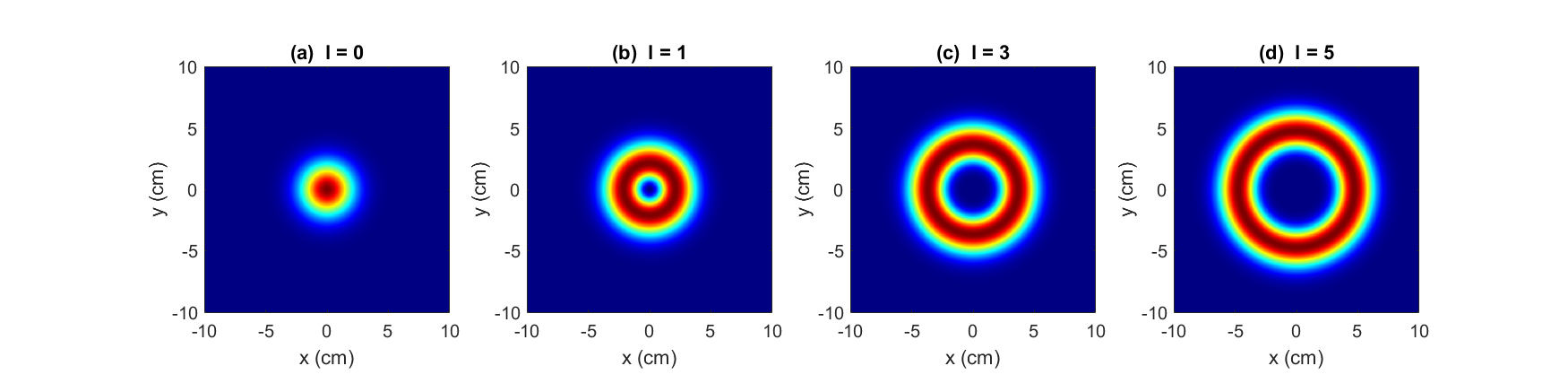}
\caption{Intensity patterns for different Laguerre-Gaussian modes in an inter-satellite link with $z = 1000$ km. Subfigures illustrate the power distribution for modes: (a) $l = 0$, (b) $l = 1$, (c) $l = 3$, and (d) $l = 5$. Larger modes, such as $l = 5$, exhibit a donut-shaped pattern with low intensity at the beam center, leading to poorer performance under small pointing errors but improved performance as pointing errors increase \cite{Dabiri2024TechRxiv}.}
\label{fig:LG_modes}
\end{figure*}

Fig.~\ref{fig:LG_modes} illustrates the intensity patterns of LG beams for different topological charges ($l = 0, 1, 3, 5$) in an inter-satellite link at a distance of 1000 km. Each subfigure shows the 2D power distribution across the beam cross-section. As the topological charge $l$ increases, the beam transitions from a nearly Gaussian shape ($l = 0$) to a “donut-like” ring structure. Notably, the ring’s radius grows larger with higher $l$, and the beam center exhibits a null intensity (i.e., the center is dark). This null arises due to the phase singularity at the center of higher-order LG beams. The larger donut radius can offer advantages under certain conditions (e.g., reducing the impact of minor pointing errors in free-space optical links).

 However, HG beams can also be utilized for free-space optical communication. While LG beams exhibit circular symmetry, HG beams have rectangular symmetry and can be generated in Cartesian coordinates. The General Electric field of a HG beam is given by \cite{Dabiri2024TechRxiv}:
 
\begin{equation}
\begin{split}
    u_{n,m}(x,y,z) &= \sqrt{\frac{2}{2^{n+m} \pi n! m! \omega_0^2}} 
    \frac{q(0)}{q(z)}
    \left[ \frac{q^*(z)}{q(z)} \right]^{\frac{n+m}{2}} \\
    &\quad \times H_n \left(\frac{\sqrt{2} x}{w(z)}\right) H_m \left(\frac{\sqrt{2} y}{w(z)}\right) \\
    &\quad \times \exp\left( -\frac{x^2 + y^2}{w^2(z)} \right) \\
    &\quad \times \exp\left( -i k \frac{x^2 + y^2}{2 R(z)} \right) \\
    &\quad \times \exp[i \psi_{n,m}(z)] \exp(-i k z).
\end{split}
\end{equation}

where

\begin{itemize}
    \item $n, m$ are the Hermite-Gaussian mode indices (integers $\geq 0$).
    \item $H_n(\cdot)$ and $H_m(\cdot)$ are Hermite polynomials of orders $n$ and $m$.
    \item $\omega_0$ is the waist radius at $z = 0$.
    \item $q(0)$ and $q(z)$ are the so-called \textit{complex beam parameters} at $z = 0$ and $z$, respectively (often used in paraxial wave solutions).
    \item $\psi_{n,m}(z)$ is the Gouy phase term specific to the $(n,m)$ mode.
\end{itemize}

Parameters such as $\omega(z)$, $R(z)$, and $k$ follow the usual definitions in Gaussian beam theory. Although HG beams can simplify certain Cartesian-coordinate analyses, LG beams often provide natural cylindrical symmetry for exploiting orbital angular momentum. Consequently, we base our derivations on the LG formulation.

On the transmit side, each mode $U_{p,\ell}(r,\phi,z)$ can be individually intensity modulated for instance, via OOK, M-ary PAM, or other IM techniques. Let $p_\ell$ represent the modulation amplitude (or power) associated with each mode $\ell$. Then, the transmitted multiplexed electric field becomes:

\begin{equation}
    u_{\text{MUX}}(r,\phi,z) = \sum_{\ell} p_{\ell} U_{p,\ell}(r,\phi,z),
\end{equation}

where each unmodulated basis function $U_{p,\ell}$ is scaled by $p_{\ell}$. The resulting composite field $u_{\text{MUX}}$ is launched into free space for subsequent demultiplexing and detection at the receiver.

Fig.~\ref{fig:beam_misalignment} illustrates how beam misalignment arises in an OAM-based free-space optical (FSO) link. 
The diagram depicts a lateral offset $\mathbf{r}_{ch} = (x_{ch}, y_{ch})$, which shifts the beam’s nominal center away from 
the receiver aperture’s origin $(x', y') = (0, 0)$. As a result, the received beam is expressed in two distinct coordinate sets:

\begin{enumerate}
    \item Cylindrical Coordinates $(r', \phi')$ from the receiver’s perspective, defining radial distance $r'$ and azimuthal angle $\phi'$.
    \item Cartesian Coordinates $(x', y')$, similarly used by the aperture plane.
\end{enumerate}

In the absence of pointing errors, the transmitted OAM field $u_{\text{MUX}}(r, \phi, z)$ at distance $z$ would directly match $(r', \phi')$. 
That is, $(r', \phi') = (r, \phi)$. However, misalignment causes a transformation such that

\begin{equation}
    x_F = x' - x_{ch}, \quad y_F = y' - y_{ch},
\end{equation}

with

\begin{equation}
    r' = \sqrt{x_F^2 + y_F^2}, \quad \phi_F = \tan^{-1} \left( \frac{y_F}{x_F} \right).
\end{equation}
Hence the beam’s apparent center moves to $\mathbf{F} = (x_F, y_F)$, possibly causing partial ring cutoff or radial distortion.

To model the receiver in such a scenario, let

\begin{equation}
    u_{\text{MUX}}(r', \phi', z)
\end{equation}

denote the beam as seen at the aperture input (now measured in the $(r', \phi')$ coordinates). The aperture, of radius $r_a$, only allows signals where $r' \leq r_a$. As a result, the output field from the aperture plane becomes

\begin{equation}
u_{\mathrm{aperture}}(r', \phi', z) =
\begin{cases}
    u_{\mathrm{MUX}}(r', \phi', z), & r' \leq r_a, \\
    0, & \text{otherwise}.
\end{cases}
\label{eq:aperture}
\end{equation}

This truncated output $u_{\text{aperture}}(r', \phi', z)$ is then divided among $N_m$ demultiplexing components, with each component filtering out a particular LG mode (or whichever OAM basis is used). The filter is designed to isolate the mode’s unique phase and intensity pattern, thereby minimizing cross-talk among different $\ell$ channels. In the presence of nonzero $\mathbf{r}_{ch}$, the ring-like structure may be off-center, causing more crosstalk and power loss.

The filtering (or mode extraction) process for each OAM index $\ell'$ involves computing the overlap between the aperture output field $u_{\text{aperture}}(r', \phi', z)$ and a phase-matching exponential that isolates $\ell'$. Assuming each mode $\ell'$ has a modulation amplitude $p_{\ell'}$, the filter’s output $S_{p,\ell'}(r', z)$ can be written as \cite{Dabiri2024TechRxiv}:

\begin{equation}
S_{p,\ell'}(r', z) = \int_0^{2\pi} p_{\ell'} 
\frac{u_{\mathrm{aperture}}(r', \phi', z)}{\sqrt{2\pi N_m}}
\exp(-i \ell' \phi') \, d\phi',
\label{eq:modal_projection}
\end{equation}

where $\exp(-i \ell' \phi')$ serves as the phase-matching function for the desired OAM index $\ell'$. In practice, $N_m$ can be the number of modes or an appropriate normalization constant. The factor $p_{\ell'}$ corresponds to the amplitude (or power) assigned to mode $\ell'$ during transmission.

\begin{figure}
    \centering
    \includegraphics[width=0.5\textwidth]{3.png}
    \caption{Illustration of beam misalignment in OAM-based FSO systems. The figure shows the misalignment represented by $\mathbf{r}_{ch} = (x_{ch}, y_{ch})$. Cylindrical coordinates $(r', \phi')$ and Cartesian coordinates $(x', y')$ from the receiver’s perspective are depicted, along with the transformed coordinates $(x, y)$ relative to the beam’s original center \cite{Dabiri2024TechRxiv}.}
    \label{fig:beam_misalignment}
\end{figure}

Under ideal alignment (and ignoring losses), one may have $u_{\text{MUX}}(r, \phi, z) = u_{\text{MUX}}(r', \phi', z)$. Substituting Equation~\eqref{eq:aperture} into Equation~\eqref{eq:modal_projection}, along with the definitions for the unmodulated basis function shows that the filter perfectly extracts the $\ell'$th component while yielding zero for all other modes $\ell \neq \ell'$ \cite{Dabiri2024TechRxiv}.

\begin{equation}
S_{p,\ell'}(r',z) =
\begin{cases}
    \dfrac{\sqrt{2\pi} \, p\ell \, u_{p,\ell}^{\text{filt}}(r',z)}{N_m}, & \ell' = \ell, \\
    0, & \text{otherwise}.
\end{cases}
\end{equation}

Hence, the filter recovers exactly the signal associated with mode $\ell'$ if there are no pointing errors or crosstalk. In real systems, however, offsets and partial truncation at the aperture degrade this orthogonality, increasing interference between different $\ell$ \cite{Dabiri2024TechRxiv}.

\begin{equation}
\begin{split}
    u_{p,\ell}^{\text{filt}}(r',z) &= \sqrt{\frac{2p!}{\pi (p + |\ell|)!}} 
    \frac{1}{w(z)}
    \left(\frac{\sqrt{2} r'}{w(z)}\right)^{|\ell|} \\
    &\quad \times L_p^{|\ell|} \left(\frac{2 r'^2}{w^2(z)}\right) \\
    &\quad \times \exp\left(-\frac{r'^2}{w^2(z)}\right) \\
    &\quad \times \exp\left(-ik \frac{r'^2}{2 R(z)} + i \psi(z) \right).
\end{split}
\end{equation}

We model the impact of pointing error by expressing the LG field in Cartesian coordinates and introducing a lateral offset $\mathbf{r}_{ch} = (x_{ch}, y_{ch})$. In the absence of misalignment, the mode is denoted by \cite{Dabiri2024TechRxiv}:

\begin{equation}
\begin{split}
    u_{p,\ell}(x,y,z) &= \sqrt{\frac{2p!}{\pi (p + |\ell|)!}} \frac{1}{w(z)} 
    \left(\frac{\sqrt{2} \sqrt{x^2 + y^2}}{w(z)}\right)^{|\ell|} \\
    &\quad \times L_p^{|\ell|} \left(\frac{2 (x^2 + y^2)}{w^2(z)}\right) \\
    &\quad \times \exp\left(-\frac{x^2 + y^2}{w^2(z)}\right) \\
    &\quad \times \exp\left(-i \ell \tan^{-1} \left(\frac{y}{x} \right) \right) \\
    &\quad \times \exp\left(-i k \frac{x^2 + y^2}{2 R(z)} + i \psi(z) \right).
\end{split}
\end{equation}

We incorporate misalignment by shifting the coordinates observed at the receiver aperture plane: $(x + x_{ch}, y + y_{ch})$. Thus, the aperture output for each mode $\ell$ becomes

\begin{equation}
    u_{\text{aperture}}(x', y', z) = \sum_{\ell} u_{p,\ell}(x' + x_{ch}, y' + y_{ch}, z).
\end{equation}

reflecting that each LG mode is now evaluated at shifted coordinates. Commonly, the aperture is described in polar form $(r', \phi')$ with radius $r_a$. Substituting $x' = r' \cos \phi'$, $y' = r' \sin \phi'$ yields an explicit formula where $(x' + x_{ch}, y' + y_{ch})$ appears in the argument of every radial, exponential, and phase term. An example is \cite{Dabiri2024TechRxiv}:

\noindent
\small

\begin{equation}
\begin{split}
    u_{\text{aperture}}(r', \phi', z) &= \sum_{\ell} 
    \sqrt{\frac{2p!}{\pi (p + |\ell|)!}} \frac{1}{w(z)} \\
    &\quad \times \left( \frac{\sqrt{2} \sqrt{(r' \cos \phi' + x_{\text{ch}})^2 + (r' \sin \phi' + y_{\text{ch}})^2}}{w(z)} \right)^{|\ell|} \\
    &\quad \times L_p^{|\ell|} \left( \frac{2((r' \cos \phi' + x_{\text{ch}})^2 + (r' \sin \phi' + y_{\text{ch}})^2)}{w(z)^2} \right) \\
    &\quad \times \exp \left( -\frac{(r' \cos \phi' + x_{\text{ch}})^2 + (r' \sin \phi' + y_{\text{ch}})^2}{w(z)^2} \right) \\
    &\quad \times \exp \left( -i \ell \tan^{-1} \left( \frac{r' \sin \phi' + y_{\text{ch}}}{r' \cos \phi' + x_{\text{ch}}} \right) \right) \\
    &\quad \times \exp \left( -i k \frac{(r' \cos \phi' + x_{\text{ch}})^2 + (r' \sin \phi' + y_{\text{ch}})^2}{2 R(z)} \right) \\
    &\quad \times \exp \left( i \psi(z) \right).
\end{split}
\end{equation}

\normalsize 
\noindent

valid for $r' \leq r_a$. Outside $r' > r_a$, the aperture blocks the beam, causing partial truncation. Hence any portion of the ring lying beyond $r_a$ is lost, introducing power penalties and potential mode crosstalk (since the ring shape is off-center and the overlap with each OAM mode changes). If $\mathbf{r}_{ch}$ is large compared to $r_a$, the misalignment can severely degrade OAM recognition and capacity. We begin by noting that, in an ideal scenario, each OAM mode $\ell$ remains orthogonal, and no offset is present between the beam center and the aperture center. Once we account for a lateral shift $\mathbf{r}_{ch} = (x_{ch}, y_{ch})$, however, a portion of the mode’s energy can “leak” into other demultiplexing paths. We capture this crosstalk using an integral expression that compares the shifted field from one mode $\ell$ against the reference mode $\ell'$. Concretely, we define \cite{Dabiri2024TechRxiv}:

\begin{equation}
    C_{\ell,\ell'}(x_{ch}, y_{ch}) = p_{\ell} \int_0^{r_a} \int_0^{2\pi} 
    \left| S_{p,\ell,\ell'}(r', \phi', z) \right|^2 r' \, d\phi' \, dr',
\end{equation}

The function $S_{p,\ell,\ell'}$ can be expanded in detail as follows \cite{Dabiri2024TechRxiv}:
\noindent
\scriptsize
\begin{equation}
\begin{split}
    S_{p,\ell,\ell'}(r', \phi', z) &= 
    \frac{1}{w(z')} \frac{1}{\sqrt{2\pi N_m}} \int_0^{2\pi} d\phi'' \\
    &\quad \times \sqrt{\frac{2p!}{\pi (p + |\ell|)!}} \\
    &\quad \times \left( \frac{\sqrt{2} \sqrt{(r' \cos \phi' + x_{\text{ch}})^2 
    + (r' \sin \phi' + y_{\text{ch}})^2}}{w(z')} \right)^{|\ell|} \\
    &\quad \times L_p^{|\ell|} \left( \frac{2((r' \cos \phi' + x_{\text{ch}})^2 
    + (r' \sin \phi' + y_{\text{ch}})^2)}{w^2(z')} \right) \\
    &\quad \times \exp \left( -\frac{(r' \cos \phi' + x_{\text{ch}})^2 
    + (r' \sin \phi' + y_{\text{ch}})^2}{w^2(z')} \right) \\
    &\quad \times \exp \left( -i \ell \tan^{-1} \left( \frac{r' \sin \phi' + y_{\text{ch}}}{r' \cos \phi' + x_{\text{ch}}} \right) 
    + i \ell' \phi' \right) \\
    &\quad \times \exp \left( - i k \frac{(r' \cos \phi' + x_{\text{ch}})^2 
    + (r' \sin \phi' + y_{\text{ch}})^2}{2 R(z')} \right) \\
    &\quad \times \exp \left( i \psi(z') \right).
\end{split}
\end{equation}

\normalsize 
\noindent

Numerous studies propose mathematical frameworks to describe random beam misalignment, whether through lateral displacement, angular tilt, or both. By employing Gaussian, Beta, or even more generalized statistical distributions, these works establish how imperfect beam alignment translates into mode coupling effects and power penalties. They also address how higher order OAM modes, though beneficial for increased channel capacity, are especially vulnerable to the slightest offsets, leading to significant crosstalk and mode degradation. Overall, this body of literature clarifies the key parameters such as beam waist, topological charge, receiver aperture, and propagation distance that govern a system’s sensitivity to pointing errors.

In \cite{Xie2015OAMFSO}, the authors analyze an OAM-based FSO system under misalignment. They show how larger transmit beam diameters and receiver apertures tolerate lateral shifts but become less resilient to angular tilt. The paper concludes that using larger mode spacing reduces crosstalk at the expense of higher power losses an acceptable trade-off when inter-mode interference dominates. \cite{Zhao2018SpiralSpectrum} examines the spiral-spectrum broadening of LG beams caused by lateral displacement and angular pointing errors. As the offset grows, the original OAM mode detection probability falls sharply. Beyond a certain threshold, the principal OAM state is barely recognizable, indicating strong crosstalk and high detection penalty under moderate-to-severe misalignment. \cite{Elamassie2023OAMPE} introduces a Beta-distributed approach for crosstalk coefficients in Gaussian-random pointing-error scenarios. It derives closed-form expressions for mode coupling and reveals that crosstalk intensifies sharply at higher OAM orders. Building on that, \cite{Elamassie2024Crosstalk} generalizes these findings by demonstrating that crosstalk coefficients follow Beta or Generalized Gamma distributions for varying mode indices and error strengths. These refined models yield more accurate system predictions, particularly for wide offsets or larger topological charges.

Another category of work focuses on developing models for analyzing system performance metrics such as BER and channel capacity under pointing errors. \cite{Djordjevic2023BER} evaluates the BER degradation in a coding-intensive scheme: an OAM-multiplexed FSO link employing quasi-cyclic Low-density parity-check (LDPC) and multidimensional modulation. Even under strong pointing misalignments, their coded system maintains sub-$10^{-6}$ BER for practical $E_b/N_0$ values, highlighting the importance of robust error correction in mitigating beam displacement. 

In \cite{Wang2020OAMIOP}, lateral displacement and angular tilt are modeled to see how they affect BER and channel capacity. The results show that lower-order OAM beams remain relatively robust, whereas higher-order beams suffer stronger crosstalk. Larger tilt angles can stall capacity gains unless the link operates at high SNR. Meanwhile, \cite{Zhang2022UnderwaterOAM} focuses on channel capacity in the presence of turbulence plus pointing errors, deriving closed-form expressions that help identify optimal beam parameters for LG links.

\cite{Sharma2021IsOWC} analyzes a mode-division-multiplexed inter-satellite optical wireless communication (OWC) system that uses Laguerre–Gaussian modes. It tracks capacity and BER for pointing errors from $1$ to $5$ µrad, finding that non-return-to-zero (NRZ) modulation outperforms return-to-zero (RZ) in crosstalk-limited situations, whereas pointing error worsens system fidelity exponentially as offset increases. Similarly, \cite{Sachdeva2023UltraHigh} demonstrates a $22,000$ km inter-satellite link using polarization-division multiplexed (PDM) 256-QAM plus OAM. They identify that high-order OAM states (e.g., LG$_{40,0}$) degrade more severely from pointing error, while fundamental modes (LG$_{0,0}$) are relatively stable. Expanding the receiver aperture mitigates these losses, hinting at an important design trade-off between aperture size, crosstalk, and mechanical complexity.

Another application domain appears in \cite{Li2024QuantumOAM}, dealing with high-dimensional quantum key distribution (HD-QKD). Here, the authors propose a channel model incorporating state-dependent pointing errors and turbulence, then quantify secure key rates (SKR) and quantum bit error rates (QBER). They conclude that even slight platform misalignment can considerably degrade quantum performance, especially under finite-key constraints. Furthermore, \cite{ElMeadawy2022NOAM} Proposes a hybrid NOAM-MPPM scheme for OAM-based FSO under Gamma–Gamma turbulence and pointing errors. It derives closed-form expressions for relevant channel parameters and shows that, under strong turbulence and misalignment, the new hybrid approach outperforms simpler modulation schemes in terms of BER. Most of these references build on the classical pointing-error model of \cite{farid2007outage}, which uses a simple Gaussian-based displacement assumption to characterize random misalignment. In real links, this baseline is often extended or modified to incorporate beam-waist effects, limited aperture, and advanced crosstalk formulations. \cite{Dabiri2024TechRxiv} introduces a refined approach by developing a coordinate-transformed OAM pointing error model. Instead of relying on traditional Gaussian-based pointing error assumptions, this study derives a mode-dependent cross-talk model that accounts for both beam waist variations and receiver aperture size, which were previously overlooked. The results show that conventional models significantly underestimate crosstalk effects in high-order OAM modes and that beam divergence plays a crucial role in determining system tolerance to pointing errors, particularly in inter-satellite links.

\begin{table*}[t]
\centering
\normalsize
\caption{Summary of Pointing-Error Modeling in OAM-Based FSO}

\begin{tabular}{|p{2.5cm}|p{4cm}|p{4cm}|p{5.5cm}|}
\hline
\textbf{Refs.} & \textbf{Main Focus} & \textbf{Model/Approach} & \textbf{Key Findings} \\ 
\hline \hline

\cite{Xie2015OAMFSO} 
& OAM-based FSO under misalignment 
& Beam size, receiver aperture, mode spacing vs. lateral/angle errors 
& Larger beam waist improves lateral offset tolerance but worsens tilt; larger mode spacing reduces crosstalk. \\ \hline

\cite{Zhao2018SpiralSpectrum} 
& Spiral-spectrum expansion of Laguerre--Gaussian modes 
& Offset/tilt effects on OAM detection probability and crosstalk 
& Major OAM modes rapidly lose recognizability beyond specific offset thresholds. \\ \hline

\cite{Elamassie2023OAMPE,Elamassie2024Crosstalk} 
& Crosstalk modeling under random pointing 
& Beta and Generalized Gamma distributions for crosstalk PDFs 
& Higher OAM orders are more vulnerable; Beta/Gen.-Gamma models better predict wide-offset behavior. \\ \hline

\cite{Djordjevic2023BER} 
& LDPC-coded OAM links with turbulence and pointing errors 
& Quasi-cyclic LDPC and multidimensional modulation 
& Maintains very low BER under moderate-to-strong misalignment, demonstrating strong coding gains. \\ \hline

\cite{Wang2020OAMIOP,Zhang2022UnderwaterOAM} 
& BER and capacity under lateral and angular misalignment 
& Analytical detection probability and capacity expressions 
& Lower-order modes exhibit higher robustness; large offsets require high SNR due to crosstalk. \\ \hline

\cite{Sharma2021IsOWC,Sachdeva2023UltraHigh} 
& Inter-satellite OWC under microradian offsets 
& Laguerre--Gaussian modes with BER/capacity analysis 
& Higher-order beams degrade faster; fundamental modes remain stable; aperture size improves tolerance. \\ \hline

\cite{Li2024QuantumOAM} 
& High-dimensional QKD with pointing error and turbulence 
& Atmospheric channel modeling for quantum protocols 
& Even small offsets significantly reduce secure key rates, limiting practical QKD links. \\ \hline

\cite{ElMeadawy2022NOAM} 
& Hybrid NOAM-MPPM under turbulence and pointing error 
& Closed-form BER and crosstalk expressions 
& Outperforms simpler modulations; beam waist and mode choice are critical to mitigate crosstalk. \\ \hline

\cite{Dabiri2024TechRxiv} 
& Coordinate-transformed OAM pointing-error model 
& Accounts for beam waist and receiver aperture effects 
& Conventional single-variance models may underestimate crosstalk at high OAM orders. \\ \hline

\cite{Hu2023Multimodal,Sharma2022HGIsOWC} 
& LG vs. HG beam comparison under misalignment 
& Lateral displacement, tilt sensitivity, and eye-diagram analysis 
& HG beams tolerate lateral shifts better but are more sensitive to tilt; advanced modulation improves performance. \\ \hline

\cite{Liu2022HGPropagation,Liu2022HGAverage} 
& Average irradiance of HG beams with offset 
& Statistical averaging with turbulence and misalignment 
& As offset increases, HG beams approach Gaussian profiles; higher orders resist broadening. \\ \hline

\cite{Kiasaleh2016StatisticalHG} 
& Residual jitter modeling for HG beams 
& Small-variance pointing-error assumptions 
& Residual jitter reduces peak intensity and degrades mode orthogonality. \\ \hline

\end{tabular}
\label{tab:pointing_error_models}
\end{table*}

\cite{Hu2023Multimodal} and \cite{Sharma2022HGIsOWC} compare HG versus LG beams under different error types. For instance, \cite{Hu2023Multimodal} notes that HG modes exhibit better tolerance against horizontal/vertical displacements but are more sensitive to angular tilt, whereas LG modes see more intense crosstalk under lateral offset but handle limited apertures better. \cite{Sharma2022HGIsOWC} addresses a multichannel HG-based inter-satellite link, investigating pointing-error increases up to $2$ µrad. Its eye-diagram results confirm that advanced modulation (e.g., alternate-mark-inversion) can outperform simpler pulse formats when moderate pointing error is introduced. Two further works, \cite{Liu2022HGPropagation} and \cite{Liu2022HGAverage}, propose closed-form expressions for average irradiance in HG beams under arbitrary pointing errors. They treat the beam-tracking offset as a statistically distributed displacement, then compute final intensity. Both show that as offset grows, HG beams effectively morph toward a near-Gaussian shape in their long-term average. Higher-order HG modes can be less vulnerable to broadening in some conditions, especially if a sufficiently large beam waist is used. Lastly, \cite{Kiasaleh2016StatisticalHG} refines a statistical model for residual jitter in HG beams, revealing that standard deviation of pointing error correlates directly with peak-intensity drop, which in turn harms orthogonality and system throughput. Table~\ref{tab:pointing_error_models} reviews several pointing-error modeling studies in the context of orbital angular momentum–based free-space optical systems. Each entry highlights the main focus, the modeling or analytical approach, and a concise summary of the primary conclusions or findings.

\subsection{Mitigation technique for OAM pointing error}
In tandem with modeling efforts, a parallel line of research focuses on mitigating pointing errors once they are characterized. Numerous solutions involve adaptive optics, beam shaping, larger receiver apertures, or channel-coding techniques to preserve transmission integrity. Some works target mechanical tracking systems that continuously re-center the beam, while others rely on advanced digital signal processing (ADSP) or machine learning (ML) methods that can dynamically compensate for misalignment-induced crosstalk. Collectively, these strategies underscore the trade-offs between complexity, power consumption, and achievable link robustness, guiding system designers to adopt a combination of optical, electronic, and algorithmic measures to maintain reliable OAM links despite real-world jitter and drift.

\begin{table*}[t]
\centering
\caption{Summary of Pointing-Error Mitigation Techniques in OAM-Based FSO}
\label{tab:techniques}

\small
\setlength{\tabcolsep}{4pt}
\renewcommand{\arraystretch}{1.15}

\begin{tabular}{|p{2.6cm}|p{2.6cm}|p{3.6cm}|p{3.6cm}|p{3.2cm}|p{1.8cm}|}
\hline
\textbf{Category} & \textbf{Technique} & \textbf{Mechanism} & \textbf{Advantages} & \textbf{Challenges} & \textbf{Refs.} \\
\hline \hline

Acquisition, Tracking, and Pointing (ATP)
& Gimbal Systems + FSMs
& Coarse pointing via gimbals; fine tip--tilt correction with mirrors
& Robust mechanical solution; covers large angular range
& Bulkier hardware; limited speed for high-frequency jitter
& \cite{Billaud2020Compensation,Kiasaleh2017Tracking} \\
\hline

Wavefront Engineering and Beam Shaping
& Adaptive Optics (AO)
& Deformable mirrors or SLMs correct wavefront distortions in real time
& Addresses tilt and turbulence; improves link capacity
& Expensive and complex; requires stable feedback loop
& \cite{Kiasaleh2017Tracking,Gangwar2023Mitigation} \\
\cline{2-6}

& Multi-Plane Light Conversion (MPLC)
& Mode-based optical transformation
& Passive crosstalk reduction; effective for multi-OAM channels
& Insertion loss for large mode sets; precision design required
& \cite{Billaud2020Compensation,Gangwar2023Mitigation} \\
\cline{2-6}

& Aperture Shaping and Larger Beam Waist
& Gaussian or tapered aperture to reduce offset sensitivity
& Lower sensitivity to misalignment; simple and low power
& Link-specific design; wider beams may suffer divergence
& \cite{Kiasaleh2016StatisticalHG,AlAbabneh2024Tolerance} \\
\hline

Digital Mitigation Techniques
& MIMO DSP
& Digital matrix inversion to mitigate OAM mode coupling
& Powerful software-based correction for multi-mode links
& High computational cost; real-time tracking required
& \cite{Gangwar2023Mitigation,Ramakrishnan2024CLEO} \\
\cline{2-6}

& FEC (LDPC or Similar)
& Forward error correction for overlap-induced bit errors
& Maintains low BER under moderate misalignment
& Adds latency and overhead; ineffective for severe crosstalk
& \cite{Djordjevic2023BER} \\
\hline

Machine Learning--Based Recognition
& CNN Demodulators
& Classification of distorted OAM intensity patterns
& High accuracy; tolerant to partial beam capture
& Training overhead; real-time GPU computation needed
& \cite{Gong2021CNN} \\
\hline

Hybrid Approaches and Self-Alignment
& Combined AO with CNN or MIMO with aperture control
& Integrates physical beam shaping with digital or ML techniques
& Robust across dynamic conditions
& Increased system complexity
& \cite{Gangwar2023Mitigation} \\
\hline

\end{tabular}
\end{table*}

\begin{enumerate}

\item \textit{Acquisition, Tracking, And Pointing Systems:}
Gimbal Systems can coarsely align the transmitter with the receiver, compensating for platform tilt and large angular displacements \cite{Billaud2020Compensation}. Once coarse pointing is established, additional fine-tracking mechanisms reduce residual jitter.  Several studies adopt quad-detector arrays at the receiver focal plane to generate error signals that drive FSMs in real time \cite{Billaud2020Compensation,Kiasaleh2017Tracking}. This feedback loop continuously corrects for small beam wander, reducing lateral offsets below a few microradians. For long-distance links, especially inter-satellite channels, mechanical or FSM-based ATP is essential to maintain the beam’s position within the receiver aperture.  One recent approach \cite{Gangwar2023Mitigation} uses a radial shearing interferometric technique to detect misalignment down to micrometers, enabling near-automatic realignment. Such advanced optical metrology can guide servo controls that self-correct lateral offsets, potentially paving the way to fully self-aligning OAM-based links.

\item \textit{Wavefront Engineering And Beam Shaping:}
Adaptive optics rely on deformable mirrors or spatial light modulators to correct wavefront errors. By measuring distorted beams (often via wavefront sensors) and applying real-time corrections, adaptive optics (AO) can address pointing error and atmospheric turbulence simultaneously \cite{Kiasaleh2017Tracking,Gangwar2023Mitigation}. However, AO systems can be complex and power-intensive multi-plane light conversion (MPLC)-based demultiplexers which use repeated phase plates passively convert an incoming misaligned OAM beam to a near-Gaussian or simpler modal basis \cite{Billaud2020Compensation,Gangwar2023Mitigation}. This technique tolerates moderate lateral shifts and tilts, reducing crosstalk at the receiver without needing dynamic feedback. Some authors propose tapered or Gaussian aperture profiles that concentrate power in the beam’s central region \cite{AlAbabneh2024Tolerance}. Even if a portion of the beam misses the receiver, a relatively high fraction of power remains captured. Similarly, increasing beam waist lowers intensity gradients across the receiver plane, reducing pointing sensitivity \cite{Liu2022HGPropagation,Liu2022HGAverage}.

\item \textit{Digital Mitigation Techniques:}
In multi-mode OAM systems, crosstalk from pointing errors can be treated as a mixing matrix that couples one spatial mode to another \cite{Gangwar2023Mitigation}. MIMO DSP algorithms invert or equalize this matrix, restoring orthogonality in software. This approach is powerful for multi-channel links but can be computationally expensive, especially at high data rates \cite{Ramakrishnan2024CLEO}. LDPC and similar forward error correction (FEC) methods \cite{Djordjevic2023BER} improve system tolerance to misalignments by correcting bit errors induced by partial beam overlap or crosstalk. While FEC does not remove the underlying alignment issue, it can salvage link performance until pointing error becomes severe.

\item \textit{Machine Learning–Based Recognition:}
Recent work uses convolutional neural networks (CNNs) to identify OAM states from distorted intensity patterns \cite{Gong2021CNN}. Even when only part of the ring-shaped beam is captured (due to pointing errors or limited apertures), CNNs can surpass 90–98\% classification accuracy. This significantly relaxes alignment requirements, trading hardware complexity for GPU/ML resources. Combining mechanical or AO-based coarse alignment with CNN-based fine recognition could enable robust, self-adjusting OAM systems. Ongoing research extends these networks to handle variable turbulence levels and multiple misalignment types (angular plus lateral). Table~\ref{tab:techniques} summarizes notable mitigation strategies proposed in the literature to alleviate pointing errors in OAM-based free-space optical communication. Each approach is highlighted with its key mechanism, practical benefits, and typical implementation challenges or trade-offs.
\end{enumerate}

\subsection{Quantum Communication}
Another important technology is quantum networks (QNs), which will play a crucial role in facilitating the distribution of quantum states, or qubits, across distant nodes. This ability is fundamental for unlocking novel applications in areas like quantum sensing, which allows for unprecedented precision in measurements; distributed quantum computing, where quantum processors collaborate over large distances; and quantum communication protocols, such as QKD, which offers unparalleled security for data transmission \cite{Chehimi2023Metaverse,Degen2017QuantumSensing,Chehimi2023QFL,Chehimi2022PhysicsInformed}. In most advanced quantum network designs, optical fiber channels are the main choice for transmitting quantum information. However, when faced with geographical challenges like natural obstacles or regions lacking infrastructure, deploying fiber networks can be impractical. In such situations, FSO communication provides a highly effective alternative. FSO transmits quantum signals wirelessly through open air between ground stations or via satellites in space, offering a more adaptable solution\cite{Pirandola2021LimitsFSQ}.

Quantum-based FSO systems, while offering numerous benefits, also inherit all the inherent challenges of classical FSO, with some issues becoming even more pronounced. For example, misalignment, which can already be a problem in classical FSO communication, becomes significantly more challenging in quantum-based FSO. In quantum systems, the transmission involves fragile quantum states like single photons or entangled particles, which are far more sensitive to errors than classical signals. Even slight misalignment between the transmitter and receiver can lead to the loss of these quantum states, resulting in higher error rates and reduced key generation rates in protocols like QKD. This increased sensitivity makes precise pointing and tracking mechanisms critical in maintaining the integrity of quantum communication links. Here are some direct effects of pointing errors on quantum communication \cite{Dabiri2025EntanglementLimits}:

\begin{enumerate}
\item \textit{Photon Loss and Decreased Transmission Efficiency:}
Quantum communication depends on the precise transmission of individual photons or entangled photon pairs. Even slight deviations in the pointing direction can cause photons to miss the detector, resulting in photon loss and since quantum signals are typically weak, a minor pointing errors can result in a substantial loss of information. Unlike classical FSO systems that tolerate beam divergence due to signal redundancy, quantum systems cannot afford such losses since the signals are transmitted at low intensity and rely on each individual photon.
	
\item \textit{Reduced Entanglement Fidelity:}
Entanglement fidelity is crucial in quantum communication, particularly for tasks such as quantum teleportation or quantum networking. Pointing errors can misalign the transmission of entangled photons, which degrades the fidelity of the entangled states between the sender and receiver. Lower fidelity leads to unreliable quantum communication, making it difficult to accurately perform quantum tasks.

\item \textit{Decoherence and Signal Distortion:}
Decoherence refers to the loss of quantum coherence, a property essential for superposition and entanglement. Pointing errors can introduce decoherence as the beam deviates from its intended path and interacts with environmental factors such as scattering or absorption in the atmosphere. This causes signal distortion and further degrades the quantum information being transmitted. 

\item \textit{Limited Transmission Distance:}
In FSO quantum communication, the range of transmission is highly dependent on the precision of beam alignment. Pointing errors exacerbate beam spreading and misalignment, which significantly reduces the effective range of communication. Over long distances, particularly in satellite-based quantum communication, small pointing errors can lead to complete communication failure, as the quantum signal may entirely miss the intended receiver. Precise alignment is crucial for ensuring long-distance quantum communication remains viable.

To evaluate the effect of pointing errors on quantum communication, we can draw from the established impact of pointing errors in classical FSO systems (as discussed \cite{Alshaer2021HybridQKD}). In these classical systems, pointing errors typically lead to increased symbol error rate (SER) and BER. This classical understanding provides a foundation that can be adapted to assess the effect of pointing errors on quantum metrics in quantum communication systems. Among the most important metrics in quantum communication are the QBER, SKR and fidelity.

QBER represents the fraction of qubits received incorrectly. A higher QBER compromises the security of quantum communication because it can indicate signal degradation or even potential eavesdropping. In quantum systems, the impact of errors is far more severe than in classical systems, where errors typically result in a decrease in signal quality but not necessarily in a breach of security. SKR represents the rate at which Alice and Bob can securely generate shared cryptographic keys after accounting for errors and potential eavesdropping. The SKR is derived from the raw transmission rate but reduced by factors such as the QBER and the overhead for error correction and privacy amplification. Pointing errors increase photon loss and QBER, directly impacting the SKR by reducing the number of usable qubits. A higher QBER or significant photon loss caused by pointing misalignment decreases the SKR. In extreme cases, when pointing errors are too severe, the SKR may fall to zero, meaning no secure key can be established.

Fidelity measures how closely the quantum state received at the destination matches the quantum state that was originally transmitted. High fidelity is critical in applications like quantum teleportation or entanglement distribution, where the accuracy of quantum state transfer is key to success as the pointing error increases, it becomes harder to maintain the integrity of the transmitted states, leading to lower fidelity. 
\end{enumerate}

A number of works have investigated pointing‐error models in tandem with atmospheric turbulence or other channel effects. Among these, \cite{Alshaer2021HybridQKD} considers a radial misalignment superimposed on a Gamma–Gamma turbulence model for an M‐ary pulse‐position modulation (MPPM)–BB84 scheme. Their framework treats beam wandering as a bivariate Gaussian displacement, demonstrating that modest angular offsets can substantially reduce both raw and secret key rates for links of up to a few kilometers. Overall, the authors show that MPPM’s inherent time binning helps maintain a stable QBER and SKR, with results indicating robust operation under moderate misalignment, especially if the average photon number is carefully tuned. Another approach to sustaining entanglement or QKD in difficult FSO conditions appears in \cite{Chehimi2024RISQKD}. Although it does not propose a specific misalignment distribution, this work leverages RIS to mitigate the combined effects of link blockages, atmospheric loss, and pointing errors for FSO quantum networks. Numerical evaluations reveal that even with pointing uncertainty, an RIS can maintain entanglement fidelity well above 0.8, preserving high success probabilities for photon transmission across moderate distances. Meanwhile, \cite{FanYuan2019AlignmentQKD} sharpens the standard “misalignment error” model used in decoy‐state QKD by introducing a “light‐leakage ratio,” aimed at capturing partial overlaps of multi‐photon pulses onto unintended detectors. Their extended formalism shows that under realistic pointing offsets, double‐click events can drive QBER upward to around 50\% when misalignment is pronounced. This underscores the significance of controlling or correcting beam displacement in multi‐photon QKD. By contrast, \cite{Wang2018TFQKD} tackles large misalignment within twin‐field QKD, presenting a “sending‐or‐not‐sending” protocol that tolerates up to 45\% error rate while preserving secure key distribution at distances as far as 500–700 km—a range difficult to achieve with baseline protocols under similar pointing conditions.

For ground‐based or mobile systems, \cite{Wang2024VehicularQKD} moves away from the typical zero‐mean radial offset assumption by devising a LOS attitude measurement technique intended for vehicular quantum communication terminals. By directly measuring the LOS orientation, they cut out mechanical structure errors, stabilizing misalignment on moving platforms. Experimental trials demonstrate star‐pointing accuracy improved by roughly 93\%, suggesting that LOS‐based feedback is a practical remedy for quantum communication on trucks or automobiles. Measurement‐device‐independent QKD (MDI‐QKD) typically addresses detection loopholes, but \cite{Lu2022MDIQKD} also tackles basis misalignment, modeling partial amplitude or phase offsets in the X and Z bases. Simulation results report stable phase error rates below 3\% at up to 45° of mismatch, validating the MDI approach’s robustness. Looking at terrestrial free‐space entanglement distribution over 144 km, \cite{Ursin2007Entanglement} largely measures misalignment in real time. Their final data attribute multi‐dB losses to seemingly sub‐microradian pointing drifts, emphasizing that meticulous real‐time beam tracking is crucial. Similar real‐world measurement occurs in \cite{Nauerth2013AirGround}, where air‐to‐ground QKD is tested over about 20 km. With advanced tracking and a carefully calibrated gimbal, they keep QBER around 4.8\% despite platform vibrations. In \cite{Vallone2015Satellite}, corner‐cube retroreflectors (CCRs) in orbit reflect single photons back to a ground station, again implying that even small misalignment from the CCR normal drastically reduces reflection efficiency. They achieve QBERs around 4.6\% over 85 s of feasible link time, showing that good alignment, in conjunction with an updated CCR coating, helps preserve a workable signal. The ground‐to‐satellite uplink regime is scrutinized in \cite{Zuo2021Teleportation}, which lumps pointing errors under an ``excess noise'' model for continuous-variable teleportation, revealing that wavefront mismatch can degrade nighttime fidelity unless the angular offset remains below about 50--60~$\mu$rad---particularly for orbits exceeding 300--500~km.

In the context of small satellites, \cite{Mazzarella2020QUARC} examines the Quantum Cubesat constellation design. While relying on classical radial offset assumptions, their simulation results show that if pointing error is kept under a few dozen microradians, a low‐cost 6U CubeSat can maintain QBER below 5\% for coverage distances up to 1,500 km. Consequently, an active APT system is identified as essential for stable QKD service. In a related refinement of downlink vs. uplink beamwidth issues, \cite{Bourgoin2013LEO} revisits an earlier satellite QKD design, concluding that narrower beams can yield higher flux if alignment is near‐perfect, but misalignment as small as 5–10~$\mu$$m$ can add several dB of extra loss.

Specialty encodings also bring their own misalignment models. In \cite{DAmbrosio2012AlignmentFree}, the authors propose a partial “alignment free” approach by combining spin and OAM. While purely rotational misalignments are mitigated, they note that significant transverse beam displacement—i.e., classical pointing error still lowers detection fidelity. Meanwhile, \cite{Dequal2021CVQKD} examines a continuous‐variable downlink scenario where pointing offset is integrated into an overall fading distribution. Their simulation indicates that if misalignment surpasses about 20–30~$\mu$ at high altitude, the secure key rate quickly approaches zero. Further focusing on OAM, \cite{Li2024HDQKD} links displacement to cross‐talk integrals among OAM modes. The authors find higher‐order OAM states are especially prone to partial overlap into neighboring modes when offset arises, limiting dimension‐based gains in QKD if the radial offset is large. Similarly, \cite{Zhao2019PointingQKD} generalizes pointing error to nonzero‐mean Gaussians, analyzing how separate azimuth and elevation angles each degrade the legitimate receiver’s and eavesdropper’s power. They thereby identify design sweet spots for telescopes, so as to minimize eavesdropper advantage in the presence of substantial misalignment.

Additional distinctions emerge in \cite{Lee2024SPAD}, where the broadening of the beam spot on the single‐photon avalanche diode (SPAD) is tied to partial offset, leading to bigger timing jitter at high detection rates. A parallel line of research in \cite{Cheng2025UplinkCVQKD} addresses ground‐to‐satellite uplink for continuous‐variable QKD, explicitly modeling how dynamic orbits cause time‐varying pointing offsets. The authors show that adaptive beam steering and a well‐chosen telescope diameter can retain a feasible key rate over a full satellite pass. Similarly, \cite{Wang2013GroundSatellite} and \cite{Bourgoin2015MovingReceiver} rely on measured offset data from field trials (a turntable, hot‐air balloons, or a truck at satellite‐equivalent speed), each concluding that the standard radial offset concept generally holds in practice, but local mechanical or environmental factors might produce short bursts of extra-large misalignment.

Collective1ly, these references underscore that while many researchers adopt a conventional bivariate Gaussian model for pointing errors, real‐world quantum links can deviate from this assumption—especially in dynamic or platform‐specific contexts (e.g., high‐order OAM states, multi‐photon pulses, or UAVs). Nonetheless, every study reaffirms the central role of accurate misalignment modeling and robust beam steering. Whether it is MPPM–BB84 with partial alignment in \cite{Alshaer2021HybridQKD}, entanglement distribution across 144 km in \cite{Ursin2007Entanglement}, or small‐sat constellation coverage in \cite{Mazzarella2020QUARC}, the consistent message is that precise pointing‐error modeling is a cornerstone for predicting, controlling, and, ultimately, enabling reliable free‐space quantum communication.

\subsection{Mitigation Technique}
\begin{enumerate}
\item \textit{Active Beam Steering and Tracking Loops:}
Active beam steering is one of the most effective and widely implemented solutions for correcting pointing errors in free-space quantum communication systems. This technique uses real-time feedback mechanisms such as beacon signals, quadrant detectors, or imaging sensors—to continuously monitor beam deviation, followed by rapid correction through mechanical actuators like FSMs or gimbals. In more advanced systems, cascaded control loops are used, where a coarse tracking loop stabilizes large-scale motion and a fine loop suppresses fast jitter or vibration. For example, the dual-loop system described in \cite{Ruddenklau2022PointingLoop} integrates FSM dynamics with outer-loop feedback to suppress residual angular jitter in dynamic platforms. Similarly, real-time optical tracking in \cite{Bourgoin2015MovingReceiver} and gimbal-based stabilization in \cite{Nauerth2013AirGround} maintain alignment during movement, enabling stable key rates even on mobile receivers. These solutions are particularly suited to long-distance links such as satellite-to-ground \cite{Ursin2007Entanglement}, UAV-based links \cite{Nauerth2013AirGround}, and ground-to-vehicle communication \cite{Bourgoin2015MovingReceiver}, where platform movement and turbulence introduce continuous misalignment.

\item \textit{LOS Sensor-Based Pointing:}
To eliminate structural and mechanical errors that affect pointing accuracy, LOS-based alignment systems directly sense the optical axis using onboard inertial sensors placed along the beam path. Unlike traditional models that estimate pointing based on mechanical structure, this method captures real-time beam orientation, enhancing alignment precision. In \cite{Wang2024VehicularQKD}, the LOS sensor approach applied to vehicular QKD reduced pointing error by more than 90\%, demonstrating its effectiveness in environments prone to mechanical vibration or motion. This method is especially beneficial for mobile terrestrial platforms, such as vehicles or drones, where mechanical flex or shifting payloads challenge standard pointing models.

\item \textit{Beam Shaping and Aperture Optimization:}
Beam shaping techniques modify the divergence or waist size of the transmitted beam to reduce sensitivity to small misalignments. A wider beam naturally covers a larger area at the receiver aperture, tolerating lateral deviations, albeit at the cost of peak intensity. Aperture optimization balances the trade-off between beam width, diffraction loss, and pointing tolerance. As shown in \cite{Bourgoin2013LEO} and \cite{Cheng2025UplinkCVQKD}, systems with moderate divergence achieve robust performance even under orbital motion or turbulence. In \cite{Naughton2019InterSat}, the authors design a high-precision nanosatellite system that achieves pointing tolerances within 10~$\mu$$m$ using optimized 80 mm apertures and stable beam parameters. These techniques are particularly suited for satellite-based QKD, both in uplink and inter-satellite configurations, where maintaining tight beam alignment over long distances is critical but difficult to guarantee.

\item \textit{Protocol-Level Robustness:}
Several QKD protocols have been designed or adapted to tolerate transient or partial misalignment. In twin-field QKD, the sending-or-not-sending scheme introduced in \cite{Wang2018TFQKD} selectively filters out rounds affected by poor alignment, maintaining security at the cost of reduced throughput. Similarly, \cite{FanYuan2019AlignmentQKD} presents a modified decoy-state security proof that incorporates light leakage due to misalignment in multi-photon pulses. For continuous variable schemes, \cite{Zuo2021Teleportation} employs quantum catalysis and postselection to recover high-fidelity teleportation in the presence of wavefront distortion and phase noise. These strategies are well suited for long range QKD systems, especially where frequent short-term pointing fluctuations occur such as satellite passes or atmospheric scintillation and when hardware tracking alone cannot fully correct the misalignment.

\item \textit{Specialized Optical Components:}
Passive optical components offer elegant solutions to reduce alignment sensitivity. CCRs, as used in \cite{Vallone2015Satellite}, reflect incident photons back along their original path regardless of the exact incident angle, thus simplifying alignment in satellite-to-ground links. While CCRs have limited angular tolerance, they significantly enhance photon return efficiency within their acceptance range. Additionally, q-plates, as demonstrated in \cite{DAmbrosio2012AlignmentFree}, generate spin–OAM hybrid states that exhibit partial immunity to rotational misalignment. These optical techniques are ideal for low-complexity or passive satellite systems where active tracking is infeasible and for high-dimensional QKD systems that benefit from rotation-resilient encoding.

\item \textit{Adaptive Beamwidth Control:}
Dynamic adaptation of beam divergence enables systems to respond in real time to changes in pointing accuracy or atmospheric stability. When misalignment is low, a narrow beam ensures high coupling efficiency; during high jitter periods, widening the beam reduces the risk of signal dropout. Theoretical models in \cite{Bourgoin2013LEO} and \cite{Dequal2021CVQKD} show that adjusting beamwidth based on feedback data can improve link reliability without significant sacrifice in key rate. This strategy is well suited to satellite QKD systems, where link conditions change rapidly due to orbital motion, atmospheric effects, or APT system delays.

\item \textit{Mechanical Platform Stabilization:}
Rather than relying solely on optical correction, some systems aim to suppress the origin of pointing error mechanical vibration and instability through platform design. In \cite{Mazzarella2020QUARC}, the QUARC CubeSat is equipped with a highly stable attitude control system that keeps QBER below 5\% over long distance links (~1,500 km). Similarly, full scale experiments in \cite{Wang2013GroundSatellite} and mobile receivers in \cite{Bourgoin2015MovingReceiver} show that vibration damped payloads or stabilized mounts enhance link availability and reduce jitter. This approach is especially beneficial for spaceborne platforms like nanosatellites, balloon payloads, or turntable-based experiments where optical feedback systems may be constrained by size, power, or latency.

\item \textit{High-Tolerance System Design:}
Some studies focus on designing quantum communication systems that inherently tolerate larger pointing errors without failure. In \cite{Naughton2019InterSat}, the authors present a nanosatellite QKD design with 10~$\mu$$m$ pointing tolerance by combining appropriately sized optics, real-time tracking control, and carefully defined beam parameters. Meanwhile, \cite{Ruddenklau2022PointingLoop} introduces a cascaded tracking system with bandwidth-aware control, improving robustness against vibration and jitter in dynamic environments. These approaches are essential for inter-satellite links, CubeSat constellations, or scalable architectures where maintaining sub-microradian alignment through all phases of operation is challenging.

\end{enumerate}

\begin{table*}[t]
\centering
\small
\caption{Summary of Recent Studies on Quantum Communication Highlighting Research Objectives, Key Metrics, and Mitigation Techniques}
\label{tab:quantum_comm_summary}

\begin{tabular}{|p{1.2cm}|p{4cm}|p{4cm}|p{4cm}|}
\hline
\textbf{Ref.} & \textbf{Research Objective} & \textbf{Key Metrics} & \textbf{Mitigation Techniques} \\
\hline \hline

\cite{Alshaer2021HybridQKD} & Improve BB84 performance with MPPM & SKR, QBER, raw key rate & MPPM time binning, optimized photon rate \\
\hline
\cite{Chehimi2024RISQKD} & Use RIS to support FSO entanglement & Entanglement fidelity, success probability & RIS placement for rerouting under misalignment \\
\hline
\cite{FanYuan2019AlignmentQKD} & Model misalignment in WCP QKD & QBER, SKR, double-click probability & Modified leakage model for misalignment \\
\hline
\cite{Wang2018TFQKD} & Misalignment-tolerant twin-field QKD & QBER, SKR, max distance & Sending-or-not-sending protocol \\
\hline
\cite{Wang2024VehicularQKD} & LOS-based pointing model for vehicles & Pointing RMSE, alignment accuracy & Direct LOS sensing removes mechanical error \\
\hline
\cite{Lu2022MDIQKD} & Robust MDI-QKD to basis mismatch & Phase error, SKR, distance & -- \\
\hline
\cite{Ursin2007Entanglement} & 144 km entanglement-based QKD demo & Coincidence rate, QBER, visibility & Active beam tracking \\
\hline
\cite{Nauerth2013AirGround} & Air-to-ground QKD experiment & QBER, link uptime & Optical gimbal, tracking loop \\
\hline
\cite{Vallone2015Satellite} & Satellite retroreflector feasibility & QBER, photon return rate & Corner-cube retroreflectors \\
\hline
\cite{Zuo2021Teleportation} & CV teleportation over uplink & Fidelity, channel noise & Postselection, quantum catalysis \\
\hline
\cite{Mazzarella2020QUARC} & Constellation design for QKD CubeSats & QBER, coverage, SKR & APT system for stable beam alignment \\
\hline
\cite{Bourgoin2013LEO} & Beamwidth vs. pointing trade-off in LEO & Photon loss, SKR & Beamwidth tuning \\
\hline
\cite{DAmbrosio2012AlignmentFree} & Alignment-free OAM state design & Fidelity under rotation & Q-plate for rotational invariance \\
\hline
\cite{Dequal2021CVQKD} & Satellite-to-ground CV-QKD model & SKR, fading, SNR & -- \\
\hline
\cite{Li2024HDQKD} & Misalignment effects on OAM-QKD & Crosstalk, QBER & -- \\
\hline
\cite{Zhao2019PointingQKD} & Generalized pointing-error model & Rx/Eve power ratio & -- \\
\hline
\cite{Lee2024SPAD} & Beam offset effect on SPAD timing & Timing jitter, detection efficiency & -- \\
\hline
\cite{Cheng2025UplinkCVQKD} & Uplink CV-QKD with orbit dynamics & SKR, SNR, orbital angle & Adaptive aperture and tracking \\
\hline
\cite{Wang2013GroundSatellite} & Full-scale mobile QKD test & QBER, stability & Real-time tracking \\
\hline
\cite{Bourgoin2015MovingReceiver} & QKD to moving ground receiver & QBER, key rate & Closed-loop optical tracking \\
\hline
\cite{Naughton2019InterSat} & Pointing-tolerance design for nanosatellites & Pointing error, link budget & High-precision tracking with 80 mm aperture \\
\hline
\cite{Ruddenklau2022PointingLoop} & Cascaded tracking control system & Jitter, loop bandwidth & Dual-loop FSM-based tracking \\
\hline

\end{tabular}
\end{table*}

Table~\ref{tab:quantum_comm_summary} provides a consolidated summary of recent studies on quantum communication, highlighting their research objectives, key performance metrics, and employed mitigation techniques. The table enables a comparative view of how different quantum communication scenarios address misalignment, pointing errors, and system robustness.

\subsection{Optical LEO Mega-Constellations as a Global Routing Infrastructure}
Large-scale LEO mega-constellations have emerged as a foundational element of next-generation global connectivity, fundamentally reshaping the design and operation of wide-area communication networks. Unlike traditional satellite systems with a small number of high-altitude platforms, mega-constellations deploy hundreds to thousands of satellites across coordinated orbital planes, forming a dense and continuously moving spaceborne network. This architectural transition is being pursued by major providers such as SpaceX Starlink, OneWeb, Amazon Kuiper, and Telesat Lightspeed, as well as emerging national initiatives, with the objective of delivering low-latency, high-capacity, and globally accessible services that complement and, in some cases, surpass terrestrial infrastructure. A defining characteristic of modern LEO constellations is the shift away from gateway-centric connectivity. Rather than relying primarily on isolated satellite–ground links, these systems increasingly leverage inter-satellite links (ISLs) to route traffic across the constellation before downlinking near the destination. This effectively transforms the constellation into a fully connected spaceborne mesh network, where routing becomes a first-order system function that directly governs latency, throughput, and service continuity on a global scale \cite{Jia2025TCOMM_SatComputing,Qin2025TAES_StarlinkTiming,Kozhaya2024TAES_OneWeb}.

Routing is especially critical because traffic patterns in mega-constellations are inherently long-range and intercontinental. Data may traverse thousands of kilometers through space, crossing multiple orbital planes and relay satellites before reaching its endpoint. In this context, routing decisions determine not only end-to-end delay, but also load distribution, relay utilization, and robustness against congestion or link disruptions. Consequently, routing must be performed with a global, geometry-aware perspective rather than relying solely on local connectivity. The importance of routing is further amplified by the highly dynamic nature of LEO constellations. Satellite positions, link distances, and neighborhood relationships evolve continuously due to orbital motion, causing network topology to change on the order of seconds. Unlike largely static terrestrial backbones, routing in LEO mega-constellations must adapt in real time to time-varying connectivity, requiring mechanisms that are scalable, responsive, and capable of maintaining performance guarantees under persistent topological evolution \cite{Luan2024LWC_MegaConstellationRouting}.

\begin{figure*}[h!]
	\begin{center} 
		\includegraphics[width=6 in]{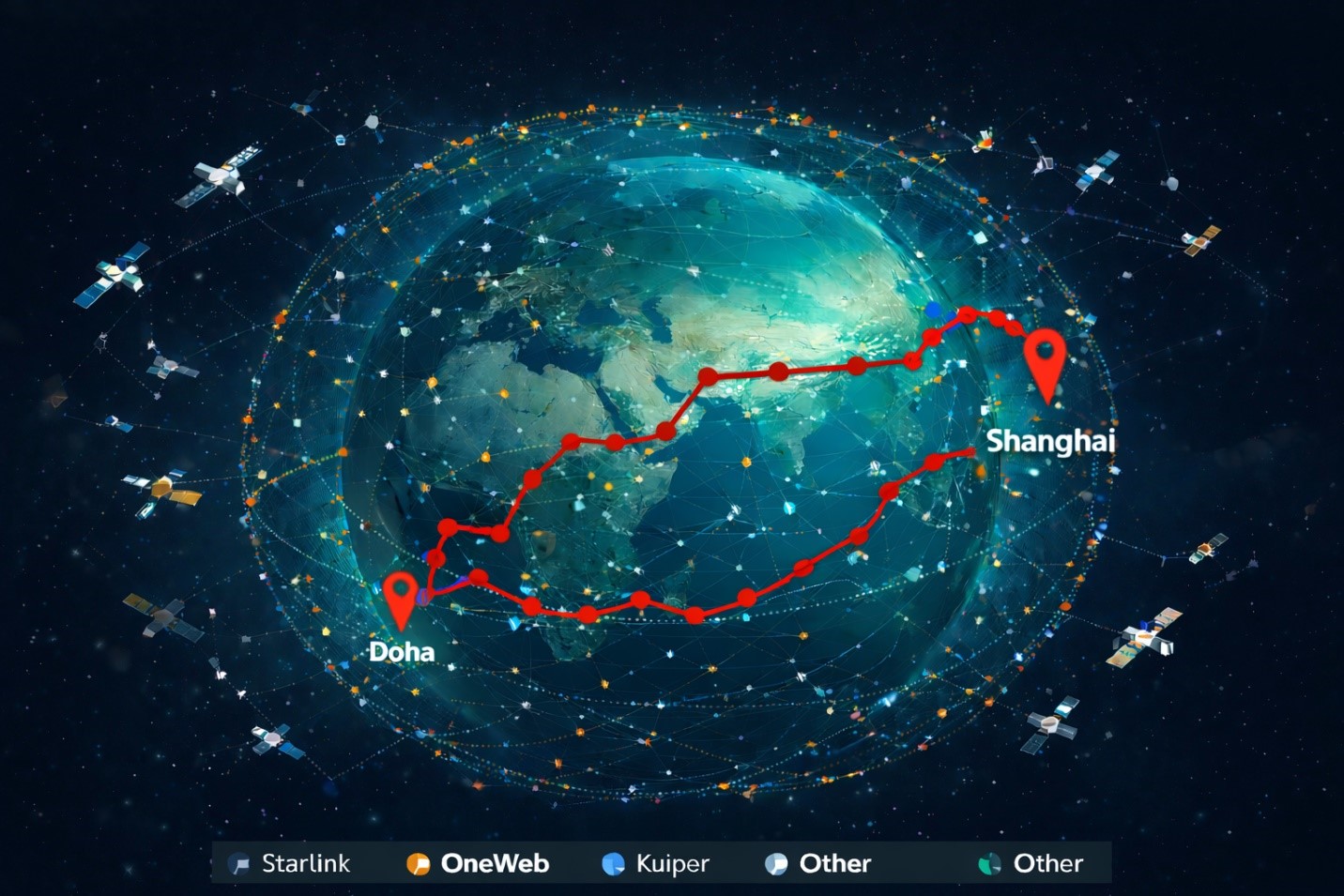}
		\caption{Conceptual illustration of multi-hop routing in optical LEO mega-constellations, showing multiple feasible end-to-end paths between distant ground terminals (e.g., Doha–Shanghai) enabled by dense satellite deployments across different providers.}

		\label{fig:mega}
	\end{center}
\end{figure*}

Effective routing also depends on cooperative behavior among satellites, which act as coordinated forwarding agents rather than independent endpoints. Such cooperation enables flexible path selection across orbital planes, spatial reuse of links, and efficient exploitation of the constellation’s global footprint. In this sense, a mega-constellation functions as a distributed space-based network infrastructure, where routing policies implicitly encode cooperation strategies to achieve system-wide objectives such as latency minimization and traffic balancing. Overall, optical LEO mega-constellations represent a paradigm shift toward network-centric space systems, in which routing plays a role analogous to backbone routing in the terrestrial Internet. Understanding how routing operates under the physical, geometric, and dynamic constraints of these constellations is therefore essential for evaluating the performance, scalability, and robustness of next-generation spaceborne communication networks.

As illustrated in Fig.~\ref{fig:mega}, the dense deployment of satellites across multiple LEO mega-constellations gives rise to a large number of feasible end-to-end routing paths between distant geographic locations. For a given source–destination pair, traffic may be forwarded through different sequences of satellites, potentially spanning multiple orbital planes and, in future scenarios, even involving satellites operated by different providers. These alternative routes differ in hop count, spatial trajectory, and relay utilization, leading to markedly different end-to-end performance characteristics. In such an environment, routing is not merely a connectivity mechanism but a critical control function that determines how effectively the constellation’s global resources are exploited. Selecting appropriate routing paths among the many available options is therefore essential to ensure low latency, balanced load distribution, and robust service delivery across the highly dynamic and heterogeneous mega-constellation landscape.

One of the most influential factors shaping routing decisions in optical LEO mega-constellations is beam pointing accuracy. Unlike RF-based links, optical inter-satellite links rely on extremely narrow beams, making the existence and quality of a link inherently dependent on precise alignment between transmitter and receiver. As satellites move along their orbits and experience platform vibrations and tracking imperfections, the feasibility of individual inter-satellite links becomes time-varying rather than binary. Consequently, routing in optical mega-constellations is constrained not only by network topology and geometry but also by whether sufficient pointing accuracy can be maintained along candidate paths. This elevates pointing from a purely physical-layer concern to a network-layer determinant, directly influencing which routes are available at any given time and how traffic can be propagated across the constellation.

From a system-level perspective, incorporating pointing considerations into routing is essential for efficient resource utilization in next-generation mega-constellations. Routing decisions that ignore pointing constraints may select long or cross-plane links that appear optimal geometrically but are costly or unreliable in practice, leading to frequent re-routing, increased relay overhead, or underutilization of available capacity. By contrast, routing strategies that account for pointing conditions can better balance hop count, link stability, and spatial reuse, enabling the constellation to operate closer to its physical limits without sacrificing reliability. In the era of dense, multi-provider mega-constellations supporting latency-critical global services, such physics-aware routing is key to optimizing spectral resources, minimizing unnecessary relays, and sustaining scalable and robust end-to-end connectivity.

\subsection{Direct-to-Device Non-Terrestrial Networks}

D2D-NTNs are rapidly emerging as a key paradigm in the convergence of satellite communications with terrestrial 5G and future 6G ecosystems. The core vision of D2D-NTNs is to enable seamless, ubiquitous connectivity directly between non-terrestrial platforms and unmodified user devices, eliminating the need for dedicated satellite terminals or intermediate ground relays. By leveraging LEO satellite constellations and HAPS, D2D-NTNs aim to extend cellular-grade services beyond the reach of terrestrial infrastructure, providing continuous coverage over remote, maritime, aerial, and underserved regions. This paradigm is increasingly aligned with 3GPP NTN roadmaps, positioning satellites and HAPS as integral components of future mobile networks rather than standalone access technologies. A key enabler of D2D-NTNs is the evolution of user-device antenna architectures, particularly the use of compact microstrip patch elements arranged in small phased arrays within handheld terminals. At Ku- and Ka-band frequencies, omnidirectional antennas are inadequate, making directional gain and electronic beam steering necessary. To this end, modern smartphones integrate low-profile microstrip patch arrays typically placed along the device edges which provide moderate directional gain and beam steering capability while remaining compatible with compact consumer form factors \cite{GarciaCabeza2025MCOM_DirectToCell}.

The importance of D2D-NTNs is further underscored by strong industrial momentum and ecosystem adoption. Several major stakeholders are actively transitioning their products and network architectures toward direct satellite-to-device connectivity. Notable examples include Starlink’s Direct-to-Cell initiative, which targets LTE-class services for standard smartphones, and Thales Alenia Space’s 5G D2D programs, which focus on tight satellite terrestrial integration. In parallel, chipset vendors and handset manufacturers are increasingly incorporating NTN-capable RF front-ends and patch-array antenna modules to support direct connectivity with non-terrestrial platforms. This industry-wide shift reflects the strategic role of D2D-NTNs in enabling always-on global access, emergency communications, and resilient mobile services, positioning direct satellite-to-device links as a foundational capability of the 6G era rather than a niche extension of legacy satellite systems \cite{Inserra2025LAWP_SidelobeReduction}.

\begin{figure}
    \centering
    \includegraphics[width=0.5\textwidth]{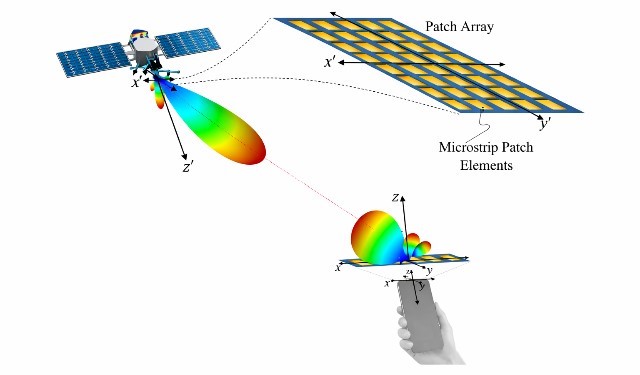}
    \caption{Geometry of the Proposed Non-Terrestrial High-Throughput Link Between the HAP/Satellite and the Mobile Terminal.}
    \label{fig:D2D}
\end{figure}

As shown in Fig.~\ref{fig:D2D}, a direct non-terrestrial link is formed between a LEO satellite or HAP and a handheld device using directional patch-array antennas at both ends. The platform steers a narrow beam toward the user via a planar microstrip patch array, while the smartphone employs a compact patch array integrated into its chassis to receive and align with the incident signal. The figure illustrates the geometric alignment between the two arrays and highlights the role of beam steering and antenna orientation in enabling high-throughput direct-to-device links. In D2D-NTNs, pointing accuracy and its modeling are critical determinants of link feasibility and performance due to the use of narrow, electronically steered beams at both the spaceborne platform and the handheld device. Unlike fixed satellite terminals, user devices exhibit continuous orientation variations caused by hand motion, body blockage, and user interaction, making the effective antenna gain highly sensitive to angular misalignment. As a result, pointing errors directly translate into rapid fluctuations in received power, link margin, and interference levels, particularly at Ku- and Ka-band frequencies where beamwidths are small. Accurate modeling of pointing behavior capturing device orientation dynamics, array radiation patterns, and relative geometry between transmitter and receiver is therefore essential for realistic performance evaluation and system design. Without such models, assessments of coverage, throughput, and reliability in D2D-NTNs risk being overly optimistic, underscoring pointing-aware modeling as a foundational requirement for robust 6G direct satellite-to-device connectivity.

\section{Future Directions and Open Challenges}
Despite significant progress in understanding and mitigating pointing errors in directional communication systems, several fundamental challenges remain open and motivate future research across modeling, analysis, and system design.

\begin{enumerate}
\item \textit{Unified dynamic pointing-error models across technologies:}
Most existing pointing-error models are either static or technology-specific, often assuming simplified distributions or decoupled error sources. Future work should develop unified dynamic models that jointly capture platform motion, mechanical vibrations, tracking latency, and control-loop dynamics in a time-varying manner. Such models should be applicable across FSO, mmWave, and THz systems, enabling consistent comparison and facilitating cross-technology design methodologies.

\item \textit{Coupled modeling of pointing errors and beam management procedures:}
In mmWave and THz systems, pointing errors are closely intertwined with beam training, tracking, and recovery mechanisms, yet these processes are often treated independently. Future studies should explicitly model the interaction between pointing-error dynamics and beam management protocols, including training overhead, misalignment recovery delay, and feedback imperfections, to better understand end-to-end performance and reliability.

\item \textit{Data-driven and hybrid physics--learning approaches:}
While learning-based methods have shown promise for alignment estimation and mitigation, most existing approaches rely on data-driven models without explicit physical interpretability. A promising direction is the development of hybrid physics-informed learning frameworks that combine analytical pointing-error models with data-driven adaptation. Such approaches could improve robustness, generalization across deployment scenarios, and explainability of learning-based mitigation strategies.

\item \textit{Pointing-error-aware system optimization and cross-layer design:}
Future research should move beyond performance evaluation toward pointing-error-aware optimization, where misalignment statistics explicitly inform system design choices such as beamwidth selection, coding and modulation, diversity schemes, and control parameters. Cross-layer designs that jointly optimize physical-layer transmission, beam management, and higher-layer protocols under pointing uncertainty remain largely unexplored.

\item \textit{Non-terrestrial and high-mobility scenarios:}
Pointing errors are particularly critical in non-terrestrial networks, including UAV-based links, satellite communications, and integrated aerial--terrestrial systems. Future work should focus on scenario-specific modeling and mitigation for high-mobility and long-distance links, accounting for platform dynamics, Doppler effects, and limited feedback, especially in THz and optical regimes.

\item \textit{Emerging directional paradigms and non-classical effects:}
In emerging technologies such as OAM based communications and free-space quantum optical systems, misalignment can induce effects beyond simple power loss, including mode coupling, crosstalk, and degradation of quantum state fidelity. Further research is needed to extend pointing-error models and mitigation techniques to capture these non-classical effects and to design alignment-robust spatial and quantum modulation schemes.

\item \textit{Experimental validation and benchmarking:}
There is a clear need for experimental datasets and benchmarking frameworks that enable validation of pointing-error models and mitigation strategies under realistic conditions. Standardized experimental setups and open datasets would facilitate reproducibility, accelerate progress, and support fair comparison of competing techniques across technologies.

\item \textit{Integration of pointing and routing in optical LEO networks:}
Despite rapid progress in optical LEO mega-constellation deployment, the joint design of routing and beam pointing remains an open and underexplored research problem. Current system architectures often treat pointing as a local physical-layer mechanism and routing as a higher-layer network function, leading to suboptimal resource utilization when these processes are decoupled. As constellations scale in size, density, and heterogeneity potentially involving multi-provider coexistence the impact of pointing constraints on route availability, path stability, and network-wide performance becomes increasingly pronounced. Developing unified, cross-layer frameworks that explicitly integrate pointing behavior into routing decisions is therefore a critical future direction. Such frameworks must capture time-varying alignment feasibility, inter-plane asymmetries, and resource trade-offs while remaining scalable to thousands of satellites. Addressing this gap is essential for enabling robust, efficient, and globally optimized routing in the next era of optical space networks.

\item \textit{Pointing-aware design for D2D-NTN:}
Despite recent advances, pointing-aware design for direct-to-device non-terrestrial networks remains a largely open research area. Existing studies often model pointing effects in isolation or under simplified assumptions, while practical D2D-NTN deployments involve tightly coupled interactions between device orientation dynamics, beam steering limits, user mobility, and network-level resource management. Developing unified frameworks that jointly capture user-side pointing behavior, platform-side beam control, and their impact on coverage and reliability is a key future direction. Moreover, scalable modeling approaches are needed to translate individual device-level pointing effects into system-level insights for dense user populations and multi-platform scenarios. Addressing these challenges is essential to enable robust, energy-efficient, and globally scalable direct satellite-to-device services in future 6G non-terrestrial networks.

\end{enumerate}

Addressing these challenges will be essential for translating highly directional communication concepts into robust and scalable 6G systems, and positions pointing-error modeling and mitigation as a foundational research theme for future wireless and optical networks.

\section{Conclusions}
Highly directional transmission is central to 6G performance gains, yet this survey finds that pointing errors and misalignment constitute a shared, first-order impairment that fundamentally limits the reliability of beam-centric links across FSO, mmWave, THz systems, while existing results remain difficult to compare because pointing error is defined, modeled, and mitigated differently across technology silos. Our synthesis shows that consistent interpretation and transferable insight require a unified chain that begins with standardized definitions and taxonomy, proceeds through compatible modeling choices including geometric and antenna-pattern/sector abstractions as well as statistical and dynamic distribution modeling that captures tracking and stabilizer imperfections and platform motion, and then evaluates impact using performance measures that reflect both link quality degradation and alignment-management overhead. We further find that mitigation effectiveness is inseparable from the assumed pointing-error model and time scale: hardware stabilization and acquisition–tracking–pointing dominate when drift and jitter are primary, beam management and tracking procedures govern robustness in mmWave/THz, and diversity, signal processing, and learning-assisted control provide complementary gains when uncertainty and nonstationarity are significant. Finally, the reviewed evidence indicates that emerging technologies intensify alignment sensitivity and can introduce effects beyond simple gain loss, motivating explicit treatment of pointing errors and mitigation in OAM-based and quantum optical systems within the same cross-technology framework.

\bibliographystyle{IEEEtran}
\bibliography{References}

@article{Jia2025TCOMM_SatComputing,
  author  = {X. Jia and D. Zhou and M. Sheng and Y. Shi and S. Ji and J. Li},
  title   = {Satellite Computing Network Construction: Optimal Computing Node Deployment in Multi-Layer {LEO} Mega-Constellations},
  journal = {IEEE Transactions on Communications},
  year    = {2025},
  doi     = {10.1109/TCOMM.2025.3638621},
  note    = {Early Access}
}

@article{Qin2025TAES_StarlinkTiming,
  author  = {W. Qin and A. M. Graff and Z. L. Clements and Z. M. Komodromos and T. E. Humphreys},
  title   = {Timing Properties of the {Starlink} {Ku}-Band Downlink},
  journal = {IEEE Transactions on Aerospace and Electronic Systems},
  year    = {2025},
  doi     = {10.1109/TAES.2025.3627180},
  note    = {Early Access}
}

@article{Kozhaya2024TAES_OneWeb,
  author  = {S. Kozhaya and Z. M. Kassas},
  title   = {A First Look at the {OneWeb} {LEO} Constellation: Beacons, Beams, and Positioning},
  journal = {IEEE Transactions on Aerospace and Electronic Systems},
  volume  = {60},
  number  = {5},
  pages   = {7528--7534},
  year    = {2024},
  month   = oct,
  doi     = {10.1109/TAES.2024.3410252}
}

@article{Luan2024LWC_MegaConstellationRouting,
  author  = {S. Luan and L. Wang and Y. Liu and N. Sun and R. Zhang},
  title   = {A Fast Percolation--Dijkstra {SPF} Method for Mega-Constellation Satellite Network Routing},
  journal = {IEEE Wireless Communications Letters},
  year    = {2024},
  doi     = {10.1109/LWC.2024.3432569},
  note    = {Early Access}
}

@article{GarciaCabeza2025MCOM_DirectToCell,
  author  = {J. Garcia-Cabeza and J. Albert-Smet and Z. Frias and L. Mendo and S. A. Azcoitia and E. Yraola},
  title   = {Direct-to-Cell: A First Look into {Starlink}'s Direct Satellite-to-Device Radio Access Network Through Crowdsourced Measurements},
  journal = {IEEE Communications Magazine},
  year    = {2025},
  doi     = {10.1109/MCOM.001.2500350},
  note    = {Early Access}
}

@article{Inserra2025LAWP_SidelobeReduction,
  author  = {D. Inserra and Q. Ding and G. Li and G. Wen},
  title   = {Antenna Array Sidelobe Level Reduction With Largely Spaced Co-Located Multiple Radiating Mode Microstrip Patch Antennas},
  journal = {IEEE Antennas and Wireless Propagation Letters},
  year    = {2025},
  doi     = {10.1109/LAWP.2025.3629822},
  note    = {Early Access}
}

@article{Li2024OJCOMS_LaserStabilizers,
  author  = {J. Li and S. Huang and M. Dehghani Soltani and H. Haas and M. Safari},
  title   = {Laser-Based Indoor Mobile Wireless Communication Aided by Stabilizers},
  journal = {IEEE Open Journal of the Communications Society},
  volume  = {5},
  pages   = {7525--7541},
  year    = {2024},
  doi     = {10.1109/OJCOMS.2024.3420102}
}

@article{Peng2021GimbalPointing,
  author    = {C. Peng and others},
  title     = {Modeling and Correction of Pointing Errors in Gimbals-Type Optical Communication Terminals on Motion Platforms},
  journal   = {IEEE Photonics Journal},
  volume    = {13},
  number    = {3},
  pages     = {1--15},
  year      = {2021},
  month     = jun,
  articleno = {6600915},
  doi       = {10.1109/JPHOT.2021.3078227}
}

@article{Guo2025FSMCompensation,
  author  = {X. Guo and Y. Lin and D. Pang and X. Li and Y. Song and K. Dong},
  title   = {Beam-Deviation Compensation of {FSM} via Beam-Tracing Modeling in Dynamic Optomechanical Systems},
  journal = {IEEE Photonics Technology Letters},
  year    = {2025},
  doi     = {10.1109/LPT.2025.3642279},
  note    = {Early Access}
}

@article{Dabiri2025LWC_AAVFSO,
  author  = {M. T. Dabiri and M. Hasna and S. Althunibat and K. Qaraqe},
  title   = {How Secure Are {AAV}-Based {FSO} Links With Modulating Retroreflectors?},
  journal = {IEEE Wireless Communications Letters},
  volume  = {14},
  number  = {3},
  pages   = {606--610},
  year    = {2025},
  month   = mar,
  doi     = {10.1109/LWC.2024.3511338}
}

@article{Dabiri2025TCOMM_MRTracking,
  author  = {M. T. Dabiri and M. Hasna and S. Althunibat and K. Qaraqe},
  title   = {Modulating Retroreflector-Based Satellite-to-Ground Optical Links: Joint Communications and Tracking},
  journal = {IEEE Transactions on Communications},
  volume  = {73},
  number  = {3},
  pages   = {1950--1962},
  year    = {2025},
  month   = mar,
  doi     = {10.1109/TCOMM.2024.3462665}
}

@article{Dabiri2025LWC_IRS,
  author  = {M. T. Dabiri and M. Hasna and K. Qaraqe},
  title   = {From Idealized Optical {IRS} Models to Realistic Lens-Based Architectures},
  journal = {IEEE Wireless Communications Letters},
  pages   = {1--1},
  year    = {2025},
  doi     = {10.1109/LWC.2025.3620255},
  note    = {Early Access}
}

@article{Yao2011OAM,
  author  = {A. M. Yao and M. J. Padgett},
  title   = {Orbital Angular Momentum: Origins, Behavior and Applications},
  journal = {Advances in Optics and Photonics},
  volume  = {3},
  number  = {2},
  pages   = {161--204},
  year    = {2011},
  doi     = {10.1364/AOP.3.000161}
}

@article{Affan2021OAM6G,
  author  = {A. Affan and S. Mumtaz and H. M. Asif and L. Musavian},
  title   = {Performance Analysis of Orbital Angular Momentum (OAM): A 6G Waveform Design},
  journal = {IEEE Communications Letters},
  volume  = {25},
  number  = {12},
  pages   = {3985--3989},
  year    = {2021},
  doi     = {10.1109/LCOMM.2021.3115041}
}

@article{Noor2022OAMReview,
  author  = {S. K. Noor and others},
  title   = {A Review of Orbital Angular Momentum Vortex Waves for the Next Generation Wireless Communications},
  journal = {IEEE Access},
  volume  = {10},
  pages   = {89465--89484},
  year    = {2022},
  doi     = {10.1109/ACCESS.2022.3197653}
}

@article{Cheng2019OAMWireless,
  author  = {W. Cheng and W. Zhang and H. Jing and S. Gao and H. Zhang},
  title   = {Orbital Angular Momentum for Wireless Communications},
  journal = {IEEE Wireless Communications},
  volume  = {26},
  number  = {1},
  pages   = {100--107},
  year    = {2019}
}

@article{ref6,
  title={{Capacity limits of spatially multiplexed free-space communication}},
  author={Zhao, N. and Li, X. and Li, G. and Kahn, J. M.},
  journal={Nature Photonics},
  volume={9},
  number={12},
  pages={822--826},
  year={2015},
  publisher={Nature Publishing Group}
}

@article{ref10,
  title={{Orbital angular momentum of light and the transformation of Laguerre--Gaussian laser modes}},
  author={Allen, L. and Beijersbergen, M. W. and Spreeuw, R. J. C. and Woerdman, J. P.},
  journal={Physical Review A},
  volume={45},
  number={11},
  pages={8185--8189},
  year={1992},
  publisher={American Physical Society}
}

@article{McGloin2005Bessel,
  author  = {D. McGloin and K. Dholakia},
  title   = {Bessel Beams: Diffraction in a New Light},
  journal = {Contemporary Physics},
  volume  = {46},
  number  = {1},
  pages   = {15--28},
  year    = {2005}
}

@article{GutierrezVega2000Mathieu,
  author  = {J. C. Gutierrez-Vega and M. Iturbe-Castillo and S. Chavez-Cerda},
  title   = {Alternative Formulation for Invariant Optical Fields: Mathieu Beams},
  journal = {Optics Letters},
  volume  = {25},
  number  = {20},
  pages   = {1493--1495},
  year    = {2000}
}

@article{Xie2015OAMFSO,
  author  = {G. Xie and others},
  title   = {Performance Metrics and Design Considerations for a Free-Space Optical Orbital-Angular-Momentum-Multiplexed Communication Link},
  journal = {Optica},
  volume  = {2},
  number  = {4},
  pages   = {357--365},
  year    = {2015}
}

@inproceedings{Zhao2018SpiralSpectrum,
  author    = {Q. Zhao and S. Hao and Y. Wang and X. Wan and C. Xu},
  title     = {Spiral Spectrum of Laguerre--Gaussian Beams Under Atmospheric Turbulence and Pointing Errors},
  booktitle = {International Conference on Electronics Technology (ICET)},
  year      = {2018},
  pages     = {1--6},
  publisher = {IEEE}
}

@inproceedings{Elamassie2023OAMPE,
  author    = {M. Elamassie and H. Tadayyoni and M. Uysal},
  title     = {Characterization of Orbital Angular Momentum-Multiplexed FSO Channel in the Presence of Pointing Errors},
  booktitle = {31st Signal Processing and Communications Applications Conference (SIU)},
  year      = {2023},
  pages     = {1--4},
  publisher = {IEEE}
}

@inproceedings{Elamassie2024Crosstalk,
  author    = {M. Elamassie and M. Yaseen and S. S. Ikki and M. Uysal},
  title     = {Modeling Crosstalk Coefficients in Orbital Angular Momentum-Multiplexed FSO Channels Under Gaussian Pointing Errors},
  booktitle = {32nd Signal Processing and Communications Applications Conference (SIU)},
  year      = {2024},
  pages     = {1--4},
  publisher = {IEEE}
}

@inproceedings{Djordjevic2023BER,
  author    = {G. T. Djordjevic and I. B. Djordjevic},
  title     = {Effect of Pointing Errors on BER Performance of Multidimensional LDPC-Coded OAM Modulation With Direct Detection Over Turbulent FSO Channels},
  booktitle = {23rd International Conference on Transparent Optical Networks (ICTON)},
  year      = {2023},
  pages     = {1--4},
  publisher = {IEEE}
}

@article{Wang2020OAMIOP,
  author  = {H. Wang},
  title   = {Performance of Free-Space Optical Communication Based on Orbital Angular Momentum With Pointing Errors},
  journal = {IOP Conference Series: Materials Science and Engineering},
  volume  = {711},
  number  = {1},
  pages   = {012080},
  year    = {2020},
  doi     = {10.1088/1757-899X/711/1/012080}
}

@article{Zhang2022UnderwaterOAM,
  author  = {Y. Zhang and Q. Yan and L. Yu and Y. Zhu},
  title   = {Information Capacity of Turbulent and Absorptive Underwater Wireless Link With Perfect Laguerre--Gaussian Beam and Pointing Errors},
  journal = {Journal of Marine Science and Engineering},
  volume  = {10},
  number  = {12},
  pages   = {1957},
  year    = {2022}
}

@article{Sharma2021IsOWC,
  author  = {A. Sharma and J. Malhotra and S. Chaudhary and V. Thappa},
  title   = {Analysis of $2 \times 10$ {Gbps} {MDM}-Enabled Inter-Satellite Optical Wireless Communication Under the Impact of Pointing Errors},
  journal = {Optik},
  volume  = {227},
  pages   = {165250},
  year    = {2021},
  doi     = {10.1016/j.ijleo.2020.165250}
}

@article{Sachdeva2023UltraHigh,
  author  = {S. Sachdeva and others},
  title   = {Ultra-High Capacity Optical Satellite Communication System Using {PDM}-256-{QAM} and Optical Angular Momentum Beams},
  journal = {Sensors},
  volume  = {23},
  number  = {2},
  pages   = {786},
  year    = {2023},
  doi     = {10.3390/s23020786}
}

@article{Li2024QuantumOAM,
  author  = {J. Li and X. Wang and H. Yu and J. Tang and Y. Liu and Y. Cao and Z. Deng and D. Wu and H. Hu and Y. Wang and others},
  title   = {State-Dependent Misalignment and Turbulence Effects on High-Dimensional Quantum Key Distribution With Orbital Angular Momentum},
  journal = {New Journal of Physics},
  volume  = {26},
  number  = {5},
  pages   = {053034},
  year    = {2024}
}

@article{ElMeadawy2022NOAM,
  author  = {S. A. El-Meadawy and others},
  title   = {Proposal of Hybrid {NOAM}-{MPPM} Technique for Gamma--Gamma Turbulence Channel With Pointing Error and Different Deep Learning Techniques},
  journal = {IEEE Access},
  volume  = {10},
  pages   = {10295--10309},
  year    = {2022}
}

@article{Dabiri2024TechRxiv,
  author    = {M. T. Dabiri and M. Ghanbari and M. Hasna},
  title     = {Advancing {OAM}-Based {FSO} Systems: Tackling Pointing Errors for Next-Generation Space and Terrestrial Links},
  journal   = {IEEE Photonics Journal},
  volume    = {17},
  number    = {4},
  pages     = {1--14},
  year      = {2025},
  month     = aug,
  articleno = {7301714},
  doi       = {10.1109/JPHOT.2025.3575365}
}

@article{Hu2023Multimodal,
  author  = {Z. Hu and others},
  title   = {Aiming for High-Capacity Multi-Modal Free-Space Optical Transmission Leveraging Complete Modal Basis Sets},
  journal = {Optics Communications},
  volume  = {541},
  pages   = {129531},
  year    = {2023}
}

@article{Sharma2022HGIsOWC,
  author  = {S. Sharma and G. K. Walia and H. Kaur},
  title   = {A Multichannel Hermite--Gaussian Intensity Profiles Based Inter-Satellite Optical Wireless Communication Using Transmitter Diversity},
  journal = {Optical and Quantum Electronics},
  volume  = {55},
  number  = {2},
  year    = {2023},
  doi     = {10.1007/s11082-022-04383-3}
}

@article{Liu2022HGPropagation,
  author  = {X. Liu and D. Jiang and Y. Zhang and L. Kong and Q. Zeng and K. Qin},
  title   = {Propagation Characteristics of Hermite--Gaussian Beam Under Pointing Error in Free Space},
  journal = {Photonics},
  volume  = {9},
  number  = {7},
  pages   = {478},
  year    = {2022},
  doi     = {10.3390/photonics9070478}
}

@article{Liu2022HGAverage,
  author  = {X. Liu and D. Jiang and L. Kong and Y. Zhang and Q. Zeng and K. Qin},
  title   = {Average Irradiance of Hermite--Gaussian Beam Under Pointing Error in Atmospheric Turbulence},
  journal = {Optical Engineering},
  volume  = {61},
  number  = {10},
  pages   = {108102},
  year    = {2022},
  doi     = {10.1117/1.OE.61.10.108102}
}

@article{Kiasaleh2016StatisticalHG,
  author  = {K. Kiasaleh},
  title   = {Statistical Profile of Hermite--Gaussian Beam in the Presence of Residual Spatial Jitter in {FSO} Communications},
  journal = {IEEE Communications Letters},
  volume  = {20},
  number  = {4},
  pages   = {656--659},
  year    = {2016},
  doi     = {10.1109/LCOMM.2016.2517624}
}

@misc{Billaud2020Compensation,
  author       = {A. Billaud and D. Allioux and N. Laurenchet and P. Jian and O. Pinel and G. Labroille},
  title        = {Pointing Error Compensation for Inter-Satellite Communication Using Multi-Plane Light Conversion Spatial Demultiplexer},
  howpublished = {arXiv preprint},
  year         = {2020},
  doi          = {10.48550/arXiv.2012.10247}
}

@article{Kiasaleh2017Tracking,
  author  = {K. Kiasaleh},
  title   = {Spatial Beam Tracking for Hermite--Gaussian-Based Free-Space Optical Communications},
  journal = {Optical Engineering},
  volume  = {56},
  number  = {7},
  pages   = {076106},
  year    = {2017},
  doi     = {10.1117/1.OE.56.7.076106}
}

@article{Gangwar2023Mitigation,
  author  = {S. Gangwar and R. Dwivedi and V. K. Jaiswal and R. Mehrotra and S. Saha and P. Sharma},
  title   = {Mitigation of Lateral Misalignment in the Optical Setup for the Generation of Perfect Vortex Beam by the Radial Shearing Self-Interferometric Technique},
  journal = {Journal of Modern Optics},
  volume  = {70},
  number  = {2},
  pages   = {100--113},
  year    = {2023},
  doi     = {10.1080/09500340.2023.2192812}
}

@article{AlAbabneh2024Tolerance,
  author  = {N. Al-Ababneh and H. Aldiabat},
  title   = {Improvement of Misalignment Tolerance in Free-Space Optical Interconnects},
  journal = {International Journal of Electrical and Computer Engineering},
  volume  = {14},
  number  = {1},
  pages   = {426--434},
  year    = {2024},
  doi     = {10.11591/ijece.v14i1.pp426-434}
}

@inproceedings{Ramakrishnan2024CLEO,
  author    = {M. Ramakrishnan and others},
  title     = {Demonstration of Mitigation of Both Turbulence and Misalignment in a 50-{Gbit/s}/Channel Free-Space Mode-Division-Multiplexed Optical Link Using Complex Beamforming},
  booktitle = {Conference on Lasers and Electro-Optics (CLEO)},
  year      = {2024},
  pages     = {SF1L.5},
  publisher = {Optica Publishing Group}
}

@inproceedings{Gong2021CNN,
  author    = {B. Gong and S. Cai and Z. Xiao and X. Wang and L. Li and Z. Zhang},
  title     = {Recognition of {OAM} State Using {CNN}-Based Deep Learning for {OAM} Shift Keying {FSO} System With Pointing Error and Limited Receiving Aperture},
  booktitle = {Conference on Lasers and Electro-Optics (CLEO)},
  address   = {San Jose, CA, USA},
  year      = {2021},
  pages     = {1--2},
  publisher = {IEEE}
}

@inproceedings{Chehimi2023Metaverse,
  author    = {M. Chehimi and O. Hashash and W. Saad},
  title     = {The Roadmap to a Quantum-Enabled Wireless Metaverse: Beyond the Classical Limits},
  booktitle = {Fifth International Conference on Advances in Computational Tools for Engineering Applications (ACTEA)},
  year      = {2023},
  pages     = {7--12},
  publisher = {IEEE}
}

@article{Degen2017QuantumSensing,
  author  = {C. L. Degen and F. Reinhard and P. Cappellaro},
  title   = {Quantum Sensing},
  journal = {Reviews of Modern Physics},
  volume  = {89},
  number  = {3},
  pages   = {035002},
  year    = {2017}
}

@article{Chehimi2023QFL,
  author  = {M. Chehimi and S. Y.-C. Chen and W. Saad and D. Towsley and M. Debbah},
  title   = {Foundations of Quantum Federated Learning Over Classical and Quantum Networks},
  journal = {IEEE Network},
  year    = {2023},
  month   = nov
}

@article{Chehimi2022PhysicsInformed,
  author  = {M. Chehimi and W. Saad},
  title   = {Physics-Informed Quantum Communication Networks: A Vision Towards the Quantum Internet},
  journal = {IEEE Network},
  volume  = {36},
  number  = {5},
  pages   = {134--142},
  year    = {2022},
  month   = sep
}

@article{Pirandola2021LimitsFSQ,
  author  = {S. Pirandola},
  title   = {Limits and Security of Free-Space Quantum Communications},
  journal = {Physical Review Research},
  volume  = {3},
  number  = {1},
  pages   = {013279},
  year    = {2021}
}

@article{Alshaer2021HybridQKD,
  author  = {N. Alshaer and M. E. Nasr and T. Ismail},
  title   = {Hybrid {MPPM}-{BB84} Quantum Key Distribution Over {FSO} Channel Considering Atmospheric Turbulence and Pointing Errors},
  journal = {IEEE Photonics Journal},
  volume  = {13},
  number  = {6},
  pages   = {1--9},
  year    = {2021},
  articleno = {7600109},
  doi     = {10.1109/JPHOT.2021.3119767}
}

@misc{Chehimi2024RISQKD,
  author       = {M. Chehimi and others},
  title        = {Reconfigurable Intelligent Surface (RIS)-Assisted Entanglement Distribution in {FSO} Quantum Networks},
  howpublished = {arXiv preprint},
  year         = {2024},
  doi          = {10.48550/arXiv.2401.10823}
}

@article{FanYuan2019AlignmentQKD,
  author  = {G.-J. Fan-Yuan and others},
  title   = {Modeling Alignment Error in Quantum Key Distribution Based on a Weak Coherent Source},
  journal = {Physical Review Applied},
  volume  = {12},
  number  = {6},
  pages   = {064044},
  year    = {2019},
  doi     = {10.1103/PhysRevApplied.12.064044}
}

@article{Wang2018TFQKD,
  author  = {X.-B. Wang and Z.-W. Yu and X.-L. Hu},
  title   = {Twin-Field Quantum Key Distribution With Large Misalignment Error},
  journal = {Physical Review A},
  volume  = {98},
  number  = {6},
  pages   = {062323},
  year    = {2018},
  doi     = {10.1103/PhysRevA.98.062323}
}

@article{Wang2024VehicularQKD,
  author  = {Z. Wang and others},
  title   = {Pointing Model for Vehicular Quantum Communication Terminals Based on Line-of-Sight Attitude Measurement},
  journal = {IEEE Photonics Journal},
  volume  = {16},
  number  = {4},
  pages   = {1--8},
  year    = {2024},
  articleno = {7600108},
  doi     = {10.1109/JPHOT.2024.3428932}
}

@article{Lu2022MDIQKD,
  author  = {F.-Y. Lu and others},
  title   = {Unbalanced-Basis-Misalignment-Tolerant Measurement-Device-Independent Quantum Key Distribution},
  journal = {Optica},
  volume  = {9},
  number  = {8},
  pages   = {886--894},
  year    = {2022},
  doi     = {10.1364/OPTICA.454228}
}

@article{Ursin2007Entanglement,
  author  = {R. Ursin and others},
  title   = {Entanglement-Based Quantum Communication Over 144 km},
  journal = {Nature Physics},
  volume  = {3},
  number  = {7},
  pages   = {481--486},
  year    = {2007},
  doi     = {10.1038/nphys629}
}

@article{Nauerth2013AirGround,
  author  = {S. Nauerth and others},
  title   = {Air-to-Ground Quantum Communication},
  journal = {Nature Photonics},
  volume  = {7},
  number  = {5},
  pages   = {382--386},
  year    = {2013},
  doi     = {10.1038/nphoton.2013.46}
}

@article{Vallone2015Satellite,
  author  = {G. Vallone and others},
  title   = {Experimental Satellite Quantum Communications},
  journal = {Physical Review Letters},
  volume  = {115},
  number  = {4},
  pages   = {040502},
  year    = {2015},
  doi     = {10.1103/PhysRevLett.115.040502}
}

@article{Zuo2021Teleportation,
  author  = {Z. Zuo and Y. Wang and Q. Liao and Y. Guo},
  title   = {Overcoming the Uplink Limit of Satellite-Based Quantum Communication With Deterministic Quantum Teleportation},
  journal = {Physical Review A},
  volume  = {104},
  number  = {2},
  pages   = {022615},
  year    = {2021},
  doi     = {10.1103/PhysRevA.104.022615}
}

@article{Mazzarella2020QUARC,
  author  = {L. Mazzarella and others},
  title   = {{QUARC}: Quantum Research CubeSat---A Constellation for Quantum Communication},
  journal = {Cryptography},
  volume  = {4},
  number  = {1},
  pages   = {7},
  year    = {2020},
  doi     = {10.3390/cryptography4010007}
}

@article{Bourgoin2013LEO,
  author  = {J.-P. Bourgoin and others},
  title   = {A Comprehensive Design and Performance Analysis of Low Earth Orbit Satellite Quantum Communication},
  journal = {New Journal of Physics},
  volume  = {15},
  number  = {2},
  pages   = {023006},
  year    = {2013},
  doi     = {10.1088/1367-2630/15/2/023006}
}

@article{DAmbrosio2012AlignmentFree,
  author  = {V. D'Ambrosio and others},
  title   = {Complete Experimental Toolbox for Alignment-Free Quantum Communication},
  journal = {Nature Communications},
  volume  = {3},
  number  = {1},
  pages   = {961},
  year    = {2012},
  doi     = {10.1038/ncomms1951}
}

@article{Dequal2021CVQKD,
  author  = {D. Dequal and others},
  title   = {Feasibility of Satellite-to-Ground Continuous-Variable Quantum Key Distribution},
  journal = {npj Quantum Information},
  volume  = {7},
  number  = {1},
  pages   = {3},
  year    = {2021},
  doi     = {10.1038/s41534-020-00336-4}
}

@article{Li2024HDQKD,
  author  = {J. Li and others},
  title   = {State-Dependent Misalignment and Turbulence Effects on High-Dimensional Quantum Key Distribution With Orbital Angular Momentum},
  journal = {New Journal of Physics},
  volume  = {26},
  number  = {5},
  pages   = {053034},
  year    = {2024},
  doi     = {10.1088/1367-2630/ad49c3}
}

@article{Zhao2019PointingQKD,
  author  = {H. Zhao and M.-S. Alouini},
  title   = {On the Performance of Quantum Key Distribution {FSO} Systems Under a Generalized Pointing Error Model},
  journal = {IEEE Communications Letters},
  volume  = {23},
  number  = {10},
  pages   = {1801--1805},
  year    = {2019},
  doi     = {10.1109/LCOMM.2019.2929037}
}

@article{Lee2024SPAD,
  author  = {A. Lee and A. T. Castillo and C. Whitehill and R. Donaldson},
  title   = {The Impact of Spot-Size on Single-Photon Avalanche Diode Timing-Jitter and Quantum Key Distribution},
  journal = {IET Quantum Communication},
  volume  = {5},
  number  = {4},
  pages   = {443--449},
  year    = {2024},
  doi     = {10.1049/qtc2.12091}
}

@article{Cheng2025UplinkCVQKD,
  author  = {J. Cheng and others},
  title   = {Feasibility and Parameter Optimization of Ground-to-Satellite Uplink Continuous-Variable Quantum Key Distribution},
  journal = {New Journal of Physics},
  volume  = {27},
  number  = {2},
  pages   = {023011},
  year    = {2025},
  doi     = {10.1088/1367-2630/adac83}
}

@article{Wang2013GroundSatellite,
  author  = {J.-Y. Wang and others},
  title   = {Direct and Full-Scale Experimental Verifications Towards Ground--Satellite Quantum Key Distribution},
  journal = {Nature Photonics},
  volume  = {7},
  number  = {5},
  pages   = {387--393},
  year    = {2013},
  doi     = {10.1038/nphoton.2013.89}
}

@article{Bourgoin2015MovingReceiver,
  author  = {J.-P. Bourgoin and others},
  title   = {Free-Space Quantum Key Distribution to a Moving Receiver},
  journal = {Optics Express},
  volume  = {23},
  number  = {26},
  pages   = {33437--33447},
  year    = {2015},
  doi     = {10.1364/OE.23.033437}
}

@article{Naughton2019InterSat,
  author  = {D. Naughton and R. Bedington and S. Barraclough and T. Islam and D. Griffin and B. Smith},
  title   = {Design Considerations for an Optical Link Supporting Intersatellite Quantum Key Distribution},
  journal = {Optical Engineering},
  volume  = {58},
  number  = {1},
  pages   = {016106},
  year    = {2019},
  doi     = {10.1117/1.OE.58.1.016106}
}

@inproceedings{Ruddenklau2022PointingLoop,
  author    = {R. Ruddenklau and E. Peev and F. Moll},
  title     = {Development of a Cascaded Fine Pointing and Tracking Control Loop for Quantum Key Distribution via Free-Space Optical Communications},
  booktitle = {Proc. SPIE 12203, Environmental Effects on Light Propagation and Adaptive Systems V},
  pages     = {122030L},
  year      = {2022},
  doi       = {10.1117/12.2636111}
}

@inproceedings{Dabiri2024BalloonUAV,
  author    = {Dabiri, Mohammad-Taghi and Hasna, Mazen and Althunibat, Sadeq and Qaraqe, Khalid A. and Alouini, Mohamed-Slim},
  title     = {A Balloon-Based UAV-Aided Non-Terrestrial Sectorized Network for Post Disaster Cellular Coverage: A Dynamic Environment Perspective},
  booktitle = {2024 7th International Conference on Advanced Communication Technologies and Networking (CommNet)},
  address   = {Rabat, Morocco},
  year      = {2024},
  pages     = {1--7},
  doi       = {10.1109/CommNet63022.2024.10793319}
}

@inproceedings{Dabiri2023THzFSO,
  author    = {Dabiri, Mohammad-Taghi and Hasna, Mazen and Khattab, Tamer},
  title     = {THz vs. FSO: An Outage Probability and Channel Capacity Performance Comparison Study},
  booktitle = {2023 International Symposium on Networks, Computers and Communications (ISNCC)},
  address   = {Doha, Qatar},
  year      = {2023},
  pages     = {1--6},
  doi       = {10.1109/ISNCC58260.2023.10323755}
}

@article{Fayad2025COMST,
  author  = {A. Fayad and T. Cinkler and J. Rak},
  title   = {Toward 6G Optical Fronthaul: A Survey on Enabling Technologies and Research Perspectives},
  journal = {IEEE Communications Surveys \& Tutorials},
  volume  = {27},
  number  = {1},
  pages   = {629--666},
  year    = {2025},
  month   = feb,
  doi     = {10.1109/COMST.2024.3408090}
}

@ARTICLE{Ozturk2025IEEE_Network,
  author  = {Ozturk, M. and Salamatmoghadasi, M. and Yanikomeroglu, H.},
  title   = {Integrating Terrestrial and Non-Terrestrial Networks for Sustainable {6G} Operations: A Latency-Aware Multi-Tier Cell-Switching Approach},
  journal = {arXiv preprint},
  year    = {2025},
  doi     = {10.48550/ARXIV.2508.10849}
}

@ARTICLE{Guirado2025TGCN,
  author  = {Guirado, R. and Verdecia-Pe{\~n}a, R. and Perez-Palomino, G. and Carrasco, E. and Alonso, J. I.},
  title   = {Prototype and Performance Assessment of mmWave Active {RIS}-Enhanced {6G} Wireless Communications},
  journal = {IEEE Transactions on Green Communications and Networking},
  year    = {2025},
  pages   = {1--1},
  doi     = {10.1109/TGCN.2025.3587012}
}

@inproceedings{Dabiri2025HardLimiter,
  author    = {M. T. Dabiri and M. Hasna and S. Althunibat and K. Qaraqe},
  title     = {Revolutionizing Low-Latency Satellite Networks: Harnessing Optical Hard Limiter Technology for Next-Gen Inter-Satellite Links},
  booktitle = {2025 Joint European Conference on Networks and Communications \& 6G Summit (EuCNC/6G Summit)},
  address   = {Poznan, Poland},
  year      = {2025},
  pages     = {775--780},
  doi       = {10.1109/EuCNC/6GSummit63408.2025.11037091},
  publisher = {IEEE}
}

@article{Dabiri2025JSAC_MRRUAV,
  author  = {M. T. Dabiri and M. Hasna},
  title   = {A Novel {MRR}-{UAV}-Based Relay With Optical Network Coding: A Comparative Study With Optical {IRS} and Conventional {UAV} Relaying},
  journal = {IEEE Journal on Selected Areas in Communications},
  volume  = {43},
  number  = {5},
  pages   = {1607--1620},
  year    = {2025},
  month   = may,
  doi     = {10.1109/JSAC.2025.3543516}
}

@ARTICLE{Liu2025NPJ,
  author  = {Liu, M. and Kazemi, H. and Safari, M. and Tavakkolnia, I. and Haas, H.},
  title   = {A Comprehensive Comparison Between Terahertz and Optical Wireless Communications},
  journal = {npj Wireless Technology},
  volume  = {1},
  number  = {1},
  month   = nov,
  year    = {2025},
  doi     = {10.1038/s44459-025-00002-1}
}

@article{Zaman2021JMASS,
  author  = {I. U. Zaman and A. Eltawil and O. Boyraz},
  title   = {Wireless Communication Technologies in Omnidirectional CubeSat Crosslink: Feasibility Study and Performance Analysis},
  journal = {IEEE Journal on Miniaturization for Air and Space Systems},
  volume  = {2},
  number  = {3},
  pages   = {157--166},
  year    = {2021},
  month   = sep,
  doi     = {10.1109/JMASS.2021.3079102}
}

@article{Dabiri2025TAES_THzUAV,
  author  = {M. T. Dabiri and M. Hasna and S. Althunibat and K. Qaraqe},
  title   = {Joint {THz} Communication and {3-D} Map Reconstruction Using Swarm {UAV}s for Maximum {LoS} Coverage},
  journal = {IEEE Transactions on Aerospace and Electronic Systems},
  volume  = {61},
  number  = {4},
  pages   = {9511--9526},
  year    = {2025},
  month   = aug,
  doi     = {10.1109/TAES.2025.3549013}
}

@inproceedings{Jahed2023OWCFSO,
  author    = {M. S. Jahed and M. Ghanbari and S. M. S. Sadough},
  title     = {Bit Error Rate Analysis for a Mixed Underwater {OWC}-{FSO} Relaying System in the Presence of Pointing Error},
  booktitle = {31st International Conference on Electrical Engineering (ICEE)},
  address   = {Tehran, Iran},
  year      = {2023},
  pages     = {512--516},
  doi       = {10.1109/ICEE59167.2023.10334840},
  publisher = {IEEE}
}

@article{Wang2016LPT,
  author  = {P. Wang and others},
  title   = {Multihop FSO Over Exponentiated Weibull Fading Channels With Nonzero Boresight Pointing Errors},
  journal = {IEEE Photonics Technology Letters},
  volume  = {28},
  number  = {16},
  pages   = {1747--1750},
  year    = {2016},
  month   = aug,
  doi     = {10.1109/LPT.2016.2567439}
}

@article{Kaushal2017COMST,
  author  = {H. Kaushal and G. Kaddoum},
  title   = {Optical Communication in Space: Challenges and Mitigation Techniques},
  journal = {IEEE Communications Surveys \& Tutorials},
  volume  = {19},
  number  = {1},
  pages   = {57--96},
  year    = {2017},
  doi     = {10.1109/COMST.2016.2603518}
}

@article{Priyadarshani2024TAES,
  author  = {R. Priyadarshani and M.-S. Alouini},
  title   = {Earth-to-HAP FSO Communication With Spatial Diversity and Channel Correlation},
  journal = {IEEE Transactions on Aerospace and Electronic Systems},
  volume  = {60},
  number  = {1},
  pages   = {304--319},
  year    = {2024},
  month   = feb,
  doi     = {10.1109/TAES.2023.3326123}
}

@article{Pradhan2019TVT,
  author  = {C. Pradhan and A. Li and L. Zhuo and Y. Li and B. Vucetic},
  title   = {Beam Misalignment Aware Hybrid Transceiver Design in mmWave MIMO Systems},
  journal = {IEEE Transactions on Vehicular Technology},
  volume  = {68},
  number  = {10},
  pages   = {10306--10310},
  year    = {2019},
  month   = oct,
  doi     = {10.1109/TVT.2019.2936545}
}

@article{Singh2024LAWP,
  author  = {N. Singh and others},
  title   = {Channel Modeling for 60 GHz Fixed mmWave O2I and O2O Uplink with Angular Misalignment},
  journal = {IEEE Antennas and Wireless Propagation Letters},
  volume  = {23},
  number  = {5},
  pages   = {1653--1657},
  year    = {2024},
  month   = may,
  doi     = {10.1109/LAWP.2024.3365568}
}

@ARTICLE{Liu2025TAP,
  author  = {Liu, Z. and Liao, S. and Xue, Q.},
  title   = {Distance and Misalignment Estimations of {OAM} Waves Based on Eigenfield Mode Analysis},
  journal = {IEEE Transactions on Antennas and Propagation},
  year    = {2025},
  pages   = {1--1},
  doi     = {10.1109/TAP.2025.3622372}
}

@article{Dabiri2025LCOMM_OAMSecurity,
  author  = {M. Taghi Dabiri and M. Hasna and S. Althunibat and K. A. Qaraqe},
  title   = {{OAM}-Based Optical Wireless Security: Leveraging Beam Misalignment and Crosstalk for Eavesdropping Resistance},
  journal = {IEEE Communications Letters},
  volume  = {29},
  number  = {9},
  pages   = {2003--2007},
  year    = {2025},
  month   = sep,
  doi     = {10.1109/LCOMM.2025.3581505}
}

@misc{Dabiri2025OAMISL,
      title={Compact Analytical Model for Real-Time Evaluation of OAM-Based Inter-Satellite Links}, 
      author={Mohammad Taghi Dabiri and Mazen Hasna},
      year={2025},
      eprint={2506.20823},
      archivePrefix={arXiv},
      primaryClass={eess.SP},
      url={https://arxiv.org/abs/2506.20823}, 
}

@article{ArteagaDiaz2022ACCESS,
  author  = {P. Arteaga-D{\'\i}az and D. Cano and V. Fernandez},
  title   = {Practical Side-Channel Attack on Free-Space QKD Systems With Misaligned Sources and Countermeasures},
  journal = {IEEE Access},
  volume  = {10},
  pages   = {82697--82705},
  year    = {2022},
  doi     = {10.1109/ACCESS.2022.3196677}
}

@misc{Dabiri2025UAVQuantumFSO,
  author       = {M. T. Dabiri and M. Hasna and S. Al-Kuwari and K. Qaraqe},
  title        = {A Unified Framework for {UAV}-Based Free-Space Quantum Links: Beam Shaping and Adaptive Field-of-View Control},
  howpublished = {arXiv preprint},
  year         = {2025},
  doi          = {10.48550/arXiv.2506.20336}
}

@misc{Dabiri2025EntanglementLimits,
  author       = {M. T. Dabiri and M. Hasna and S. Al-Kuwari and K. Qaraqe},
  title        = {Physical Limits of Entanglement-Based Quantum Key Distribution over Long-Distance Satellite Links},
  howpublished = {arXiv preprint},
  year         = {2025},
  doi          = {10.48550/arXiv.2506.20798}
}

@article{Kaymak2018COMST,
  author  = {Y. Kaymak and R. Rojas-Cessa and J. Feng and N. Ansari and M. Zhou and T. Zhang},
  title   = {A Survey on Acquisition, Tracking, and Pointing Mechanisms for Mobile Free-Space Optical Communications},
  journal = {IEEE Communications Surveys \& Tutorials},
  volume  = {20},
  number  = {2},
  pages   = {1104--1123},
  year    = {2018},
  doi     = {10.1109/COMST.2018.2804323}
}

@article{Hamza2019COMST_Class,
  author  = {A. S. Hamza and J. S. Deogun and D. R. Alexander},
  title   = {Classification Framework for Free Space Optical Communication Links and Systems},
  journal = {IEEE Communications Surveys \& Tutorials},
  volume  = {21},
  number  = {2},
  pages   = {1346--1382},
  year    = {2019},
  doi     = {10.1109/COMST.2018.2876805}
}

@article{AlGailani2021ACCESS,
  author  = {S. A. Al-Gailani and others},
  title   = {A Survey of Free Space Optics (FSO) Communication Systems, Links, and Networks},
  journal = {IEEE Access},
  volume  = {9},
  pages   = {7353--7373},
  year    = {2021},
  doi     = {10.1109/ACCESS.2020.3048049}
}

@article{LePham2022COMST,
  author  = {H. D. Le and A. T. Pham},
  title   = {Link-Layer Retransmission-Based Error-Control Protocols in FSO Communications: A Survey},
  journal = {IEEE Communications Surveys \& Tutorials},
  volume  = {24},
  number  = {3},
  pages   = {1602--1633},
  year    = {2022},
  doi     = {10.1109/COMST.2022.3175509}
}

@article{Tarhouni2025OJCOMS,
  author  = {F. Tarhouni and R. Wang and M.-S. Alouini},
  title   = {Free Space Optical Mesh Networks: A Survey},
  journal = {IEEE Open Journal of the Communications Society},
  volume  = {6},
  pages   = {642--655},
  year    = {2025},
  doi     = {10.1109/OJCOMS.2025.3525468}
}

@article{Altakhaineh2025ACCESS,
  author  = {A. T. Altakhaineh and others},
  title   = {Outdoor Free Space Optical Systems: Motivations, Challenges, Contributions in Environmental Conditions, and Future Directions---A Systematic Survey},
  journal = {IEEE Access},
  volume  = {13},
  pages   = {49121--49161},
  year    = {2025},
  doi     = {10.1109/ACCESS.2025.3549776}
}

@article{Karmous2025COMST,
  author  = {S. Karmous and N. Adem and M. Atiquzzaman and S. Samarakoon},
  title   = {How Can Optical Communications Shape the Future of Deep Space Communications? A Survey},
  journal = {IEEE Communications Surveys \& Tutorials},
  volume  = {27},
  number  = {1},
  pages   = {725--747},
  year    = {2025},
  doi     = {10.1109/COMST.2024.3403873}
}

@article{Hemadeh2018COMST,
  author  = {I. A. Hemadeh and K. Satyanarayana and M. El-Hajjar and L. Hanzo},
  title   = {Millimeter-Wave Communications: Physical Channel Models, Design Considerations, Antenna Constructions, and Link-Budget},
  journal = {IEEE Communications Surveys \& Tutorials},
  volume  = {20},
  number  = {2},
  pages   = {870--913},
  year    = {2018},
  doi     = {10.1109/COMST.2017.2783541}
}

@article{He2021JPROC,
  author  = {S. He and others},
  title   = {A Survey of Millimeter-Wave Communication: Physical-Layer Technology Specifications and Enabling Transmission Technologies},
  journal = {Proceedings of the IEEE},
  volume  = {109},
  number  = {10},
  pages   = {1666--1705},
  year    = {2021},
  month   = oct,
  doi     = {10.1109/JPROC.2021.3107494}
}

@article{Li2022COMST,
  author  = {J. Li and others},
  title   = {Mobility Support for Millimeter Wave Communications: Opportunities and Challenges},
  journal = {IEEE Communications Surveys \& Tutorials},
  volume  = {24},
  number  = {3},
  pages   = {1816--1842},
  year    = {2022},
  doi     = {10.1109/COMST.2022.3176802}
}

@article{Xue2024COMST,
  author  = {Q. Xue and others},
  title   = {A Survey of Beam Management for mmWave and THz Communications Towards 6G},
  journal = {IEEE Communications Surveys \& Tutorials},
  volume  = {26},
  number  = {3},
  pages   = {1520--1559},
  year    = {2024},
  doi     = {10.1109/COMST.2024.3361991}
}

@article{Tan2024COMST,
  author  = {J. Tan and others},
  title   = {Beam Alignment in mmWave V2X Communications: A Survey},
  journal = {IEEE Communications Surveys \& Tutorials},
  volume  = {26},
  number  = {3},
  pages   = {1676--1709},
  year    = {2024},
  doi     = {10.1109/COMST.2024.3383093}
}

@article{Adnan2025ACCESS,
  author  = {M. Adnan and A. Silva and L. Krzymien and R. Dinis},
  title   = {A Survey on Positioning for mmWave Distributed MIMO Systems: Challenges, Solution Techniques, and Applications},
  journal = {IEEE Access},
  volume  = {13},
  pages   = {205882--205914},
  year    = {2025},
  doi     = {10.1109/ACCESS.2025.3638143}
}

@article{Han2022COMST,
  author  = {C. Han and others},
  title   = {Terahertz Wireless Channels: A Holistic Survey on Measurement, Modeling, and Analysis},
  journal = {IEEE Communications Surveys \& Tutorials},
  volume  = {24},
  number  = {3},
  pages   = {1670--1707},
  year    = {2022},
  doi     = {10.1109/COMST.2022.3182539}
}

@article{Sharma2025OJCOMS,
  author  = {S. Sharma and P. K. Singya and K. Deka and C. Adjih and M. Sharma},
  title   = {Terahertz Communication: State-of-the-Art and Future Directions},
  journal = {IEEE Open Journal of the Communications Society},
  volume  = {6},
  pages   = {6281--6322},
  year    = {2025},
  doi     = {10.1109/OJCOMS.2025.3592365}
}

@article{Thomas2025OJCOMS,
  author  = {S. Thomas and J. Singh Virdi and A. Babakhani and I. P. Roberts},
  title   = {A Survey on Advancements in THz Technology for 6G: Systems, Circuits, Antennas, and Experiments},
  journal = {IEEE Open Journal of the Communications Society},
  volume  = {6},
  pages   = {1998--2016},
  year    = {2025},
  doi     = {10.1109/OJCOMS.2025.3549710}
}

@article{Ahmed2024ACCESS,
  author  = {M. Ahmed and others},
  title   = {A Survey on RIS Advances in Terahertz Communications: Emerging Paradigms and Research Frontiers},
  journal = {IEEE Access},
  volume  = {12},
  pages   = {173867--173901},
  year    = {2024},
  doi     = {10.1109/ACCESS.2024.3482564}
}

@article{Chukhno2024COMST,
  author  = {N. Chukhno and others},
  title   = {Models, Methods, and Solutions for Multicasting in 5G/6G mmWave and Sub-THz Systems},
  journal = {IEEE Communications Surveys \& Tutorials},
  volume  = {26},
  number  = {1},
  pages   = {119--159},
  year    = {2024},
  doi     = {10.1109/COMST.2023.3319354}
}

@article{Zhang2026TNSE,
  author  = {H. Zhang and others},
  title   = {Terahertz Sensing, Communication, and Networking: A Survey},
  journal = {IEEE Transactions on Network Science and Engineering},
  volume  = {13},
  pages   = {501--521},
  year    = {2026},
  doi     = {10.1109/TNSE.2025.3584922}
}

@book{saleh2019fundamentals,
  title={Fundamentals of Photonics},
  author={Saleh, B. E. and Teich, M. C.},
  edition={3rd},
  year={2019},
  publisher={Wiley},
  address={Hoboken, NJ, USA}
}

@article{farid2007outage,
  title={Outage capacity optimization for free-space optical links with pointing errors},
  author={Farid, Ahmed A and Hranilovic, Steve},
  journal={Journal of Lightwave technology},
  volume={25},
  number={7},
  pages={1702--1710},
  year={2007},
  publisher={IEEE}
}

@article{boulogeorgos2019analytical,
	title={{Analytical performance assessment of THz wireless systems}},
	author={Boulogeorgos, Alexandros-Apostolos A and Papasotiriou, Evangelos N and Alexiou, Angeliki},
	journal={IEEE Access},
	volume={7},
	pages={11436--11453},
	year={2019},
	publisher={IEEE}
}

@article{singya2022hybrid,
	title={{Hybrid FSO/THz-based backhaul network for mmWave terrestrial communication}},
	author={Singya, Praveen Kumar and Makki, Behrooz and D’Errico, Antonio and Alouini, Mohamed-Slim},
	journal={IEEE Transactions on Wireless Communications},
	year={2022},
	publisher={IEEE}
}

@book{balanis2016antenna,
	title={Antenna theory: analysis and design},
	author={Balanis, Constantine A},
	year={2016},
	publisher={John wiley \& sons}
}

@article{niu2015survey,
	title={Technical Specification Group Radio Access Network; Study of Radio Frequency {(RF)} and Electromagnetic Compatibility {(EMC)} requirements for Active Antenna Array System {(AAS)} base station},
	author={{3GPP TR 37.840 v12.1.0}},
	journal={Tech. Rep.},
	year={2013},
	publisher={Springer}
}

@article{dabiri2024non,
  title={Non-Orthogonal Multiple Access Scheme Using Directional {THz} Antennas Under Positioning Errors},
  author={Dabiri, Mohammad Taghi and Althunibat, Saud and Hasna, Mazen and Qaraqe, Khalid},
  journal={IEEE Transactions on Vehicular Technology},
  year={2024},
  publisher={IEEE}
}

@article{dabiri20203d,
  title={{3D channel characterization and performance analysis of UAV-assisted millimeter wave links}},
  author={Dabiri, Mohammad Taghi and Rezaee, Mohsen and Yazdanian, Vahid and Maham, Behrouz and Saad, Walid and Hong, Choong Seon},
  journal={IEEE Transactions on Wireless Communications},
  volume={20},
  number={1},
  pages={110--125},
  year={2020},
  publisher={IEEE}
}

@article{dabiri2020analytical,
  title={{Analytical channel models for millimeter wave UAV networks under hovering fluctuations}},
  author={Dabiri, Mohammad Taghi and Safi, Hossein and Parsaeefard, Saeedeh and Saad, Walid},
  journal={IEEE Transactions on Wireless Communications},
  volume={19},
  number={4},
  pages={2868--2883},
  year={2020},
  publisher={IEEE}
}

@article{safi2024cubesat,
  title={{CubeSat}-Enabled Free-Space Optics: Joint Data Communication and Fine Beam Tracking},
  author={Safi, Hossein and Dabiri, Mohammad Taghi and Cheng, Julian and Tavakkolnia, Iman and Haas, Harald},
  journal={arXiv preprint arXiv:2406.18598},
  year={2024}
}

@article{dabiri2024modulating2,
  title={Modulating Retroreflector-based Satellite-to-Ground Optical Links: Joint Communications and Tracking},
  author={Dabiri, Mohammad Taghi and Hasna, Mazen and Althunibat, Saud and Qaraqe, Khalid},
  journal={IEEE Transactions on Communications},
  year={2024},
  publisher={IEEE}
}

@article{azari2022thz,
  title={{THz-empowered UAVs in 6G: Opportunities, challenges, and trade-offs}},
  author={Azari, M Mahdi and Solanki, Sourabh and Chatzinotas, Symeon and Bennis, Mehdi},
  journal={IEEE Communications Magazine},
  volume={60},
  number={5},
  pages={24--30},
  year={2022},
  publisher={IEEE}
}

@article{yang2022impact,
  title={{Impact of rotary-wing UAV wobbling on millimeter-wave air-to-ground wireless channel}},
  author={Yang, Songjiang and Zhang, Zitian and Zhang, Jiliang and Zhang, Jie},
  journal={IEEE Transactions on Vehicular Technology},
  volume={71},
  number={9},
  pages={9174--9185},
  year={2022},
  publisher={IEEE}
}

@article{berman2007beam,
  title={Beam wandering in the atmosphere: The effect of partial coherence},
  author={Berman, GP and Chumak, AA and Gorshkov, VN},
  journal={Physical Review E—Statistical, Nonlinear, and Soft Matter Physics},
  volume={76},
  number={5},
  pages={056606},
  year={2007},
  publisher={APS}
}

@article{briantcev2023beam,
  title={Beam wander prediction with recurrent neural networks},
  author={Briantcev, Dmitrii and Cox, Mitchell A and Trichili, Abderrahmen and Ooi, Boon S and Alouini, Mohamed-Slim},
  journal={Optics Express},
  volume={31},
  number={18},
  pages={28859--28873},
  year={2023},
  publisher={Optica Publishing Group}
}

@article{sridhar2022performance,
  title={{Performance evaluation of FSO system under atmospheric turbulence and noise}},
  author={Sridhar, B and Sridhar, S and Nanchariah, V},
  journal={Journal of The Institution of Engineers (India): Series B},
  volume={103},
  number={6},
  pages={2085--2095},
  year={2022},
  publisher={Springer}
}

@article{trung2021performance,
  title={{Performance of UAV-to-ground FSO communications with APD and pointing errors}},
  author={Trung, Ha Duyen},
  journal={Applied System Innovation},
  volume={4},
  number={3},
  pages={65},
  year={2021},
  publisher={MDPI}
}

@article{du2010measurements,
  title={Measurements of angle-of-arrival fluctuations over an 11.8 km urban path},
  author={Du, Wenhe and Tan, Liying and Ma, Jing and Yu, Siyuan and Jiang, Yijun},
  journal={Laser and Particle Beams},
  volume={28},
  number={1},
  pages={91--99},
  year={2010},
  publisher={Cambridge University Press}
}

@article{gao2023integrated,
  title={{Integrated sensing and communications with joint beam-squint and beam-split for mmWave/THz massive MIMO}},
  author={Gao, Feifei and Xu, Liangyuan and Ma, Shaodan},
  journal={IEEE Transactions on Communications},
  volume={71},
  number={5},
  pages={2963--2976},
  year={2023},
  publisher={IEEE}
}

@article{wan2021hybrid,
  title={{Hybrid precoding and combining for millimeter wave/sub-THz MIMO-OFDM systems with beam squint effects}},
  author={Wan, Qian and Fang, Jun and Chen, Zhi and Li, Hongbin},
  journal={IEEE Transactions on Vehicular Technology},
  volume={70},
  number={8},
  pages={8314--8319},
  year={2021},
  publisher={IEEE}
}

@article{zhang2020hybrid,
  title={{Hybrid precoding design for wideband THz massive MIMO-OFDM systems with beam squint}},
  author={Zhang, Ruizhe and Hao, Wanming and Sun, Gangcan and Yang, Shouyi},
  journal={IEEE Systems Journal},
  volume={15},
  number={3},
  pages={3925--3928},
  year={2020},
  publisher={IEEE}
}

@article{zhang2009effects,
  title={Effects of angular misalignment in interferometric detection of distributed polarization coupling},
  author={Zhang, Hongxia and Xu, Tianhua and Jia, Dagong and Jing, Wencai and Liu, Kun and Zhang, Yimo},
  journal={Measurement science and technology},
  volume={20},
  number={9},
  pages={095112},
  year={2009},
  publisher={IOP Publishing}
}

@article{core2006cross,
  title={Cross polarization interference cancellation for fiber optic systems},
  author={Core, Mark T},
  journal={Journal of lightwave technology},
  volume={24},
  number={1},
  pages={305},
  year={2006},
  publisher={IEEE}
}

@inproceedings{lankl1988cross,
  title={Cross-polarization interference cancellation in the presence of delay effects},
  author={Lankl, B and Nossek, JA and Sebald, G},
  booktitle={IEEE International Conference on Communications,-Spanning the Universe.},
  pages={1355--1361},
  year={1988},
  organization={IEEE}
}

@article{liu2023broadband,
  title={Broadband, Low-Crosstalk, and Massive-Channels {OAM} Modes De/Multiplexing Based on Optical Diffraction Neural Network},
  author={Liu, Zhibing and Gao, Shecheng and Lai, Zhaoyu and Li, Yuru and Ao, Zhaohuan and Li, Jinpei and Tu, Jiajing and Wu, Yanxiong and Liu, Weiping and Li, Zhaohui},
  journal={Laser \& Photonics Reviews},
  volume={17},
  number={4},
  pages={2200536},
  year={2023},
  publisher={Wiley Online Library}
}

@article{liu2018multiplexed,
  title={{Multiplexed OAM wave communication with two-OAM-mode antenna systems}},
  author={Liu, Dandan and Gui, Liangqi and Zhang, Zixiao and Chen, Han and Song, Guochao and Jiang, Tao},
  journal={IEEE Access},
  volume={7},
  pages={4160--4166},
  year={2018},
  publisher={IEEE}
}

@article{wang2023synthesizing,
  title={{Synthesizing the crosstalk between OAM modes of vortex beam by simultaneously propagating a probe vortex beam in free space}},
  author={Wang, Xiaohui and Wang, Yang and Mao, Shuai and Yu, Yongze and Gu, Haoyu and Deng, Dongdong and Song, Yingxiong and Pang, Fufei and Zhuang, Liyun and Yang, Song and others},
  journal={Optics \& Laser Technology},
  volume={165},
  pages={109622},
  year={2023},
  publisher={Elsevier}
}

@article{jaeken2011peripheral,
  title={Peripheral aberrations in the human eye for different wavelengths: off-axis chromatic aberration},
  author={Jaeken, Bart and Lundstr{\"o}m, Linda and Artal, Pablo},
  journal={JOSA A},
  volume={28},
  number={9},
  pages={1871--1879},
  year={2011},
  publisher={Optica Publishing Group}
}

@article{huang2017free,
  title={Free-space optical communication impaired by angular fluctuations},
  author={Huang, Shenjie and Safari, Majid},
  journal={IEEE Transactions on Wireless Communications},
  volume={16},
  number={11},
  pages={7475--7487},
  year={2017},
  publisher={IEEE}
}

@inproceedings{lopresti2007mitigating,
  title={{Mitigating angular misalignment from atmospheric effects in FSO links}},
  author={LoPresti, Peter G and Refai, Hazem and Sluss, James J},
  booktitle={Atmospheric Propagation IV},
  volume={6551},
  pages={209--216},
  year={2007},
  organization={SPIE}
}

@article{bhatnagar2016performance,
  title={{Performance analysis of gamma--gamma fading FSO MIMO links with pointing errors}},
  author={Bhatnagar, Manav R and Ghassemlooy, Zabih},
  journal={Journal of Lightwave technology},
  volume={34},
  number={9},
  pages={2158--2169},
  year={2016},
  publisher={IEEE}
}

@article{sandalidis2008ber,
  title={{BER performance of FSO links over strong atmospheric turbulence channels with pointing errors}},
  author={Sandalidis, Harilaos G and Tsiftsis, Theodoros A and Karagiannidis, George K and Uysal, Murat},
  journal={IEEE Communications Letters},
  volume={12},
  number={1},
  pages={44--46},
  year={2008},
  publisher={IEEE}
}

@article{agheli2022high,
  title={{High-speed trains access connectivity through RIS-assisted FSO communications}},
  author={Agheli, Pouya and Beyranvand, Hamzeh and Emadi, Mohammad Javad},
  journal={Journal of Lightwave Technology},
  volume={40},
  number={21},
  pages={7084--7094},
  year={2022},
  publisher={IEEE}
}

@inproceedings{khallaf2019uav,
  title={{UAV-based FSO communications for high speed train backhauling}},
  author={Khallaf, Haitham S and Uysal, Murat},
  booktitle={2019 IEEE Wireless Communications and Networking Conference (WCNC)},
  pages={1--6},
  year={2019},
  organization={IEEE}
}

@article{taheri2017provisioning,
  title={{Provisioning internet access using FSO in high-speed rail networks}},
  author={Taheri, Mina and Ansari, Nirwan and Feng, Jianghua and Rojas-Cessa, Roberto and Zhou, Mengchu},
  journal={IEEE Network},
  volume={31},
  number={4},
  pages={96--101},
  year={2017},
  publisher={IEEE}
}

@article{khallaf2021comprehensive,
  title={{Comprehensive study on UAV-based FSO links for high-speed train backhauling}},
  author={Khallaf, Haitham S and Uysal, Murat},
  journal={Applied Optics},
  volume={60},
  number={27},
  pages={8239--8247},
  year={2021},
  publisher={Optica Publishing Group}
}

@article{dabiri2019blind,
  title={{Blind signal detection under synchronization errors for FSO links with high mobility}},
  author={Dabiri, Mohammad Taghi and Sadough, Seyed Mohammad Sajad and Khalighi, Mohammad Ali},
  journal={IEEE Transactions on Communications},
  volume={67},
  number={10},
  pages={7006--7015},
  year={2019},
  publisher={IEEE}
}

@article{dabiri2018channel,
  title={{Channel modeling and parameter optimization for hovering UAV-based free-space optical links}},
  author={Dabiri, Mohammad Taghi and Sadough, Seyed Mohammad Sajad and Khalighi, Mohammad Ali},
  journal={IEEE Journal on Selected Areas in Communications},
  volume={36},
  number={9},
  pages={2104--2113},
  year={2018},
  publisher={IEEE}
}

@article{dabiri2019tractable,
  title={{Tractable optical channel modeling between UAVs}},
  author={Dabiri, Mohammad Taghi and Sadough, Seyed Mohammad Sajad and Ansari, Imran Shafique},
  journal={IEEE Transactions on Vehicular Technology},
  volume={68},
  number={12},
  pages={11543--11550},
  year={2019},
  publisher={IEEE}
}

@article{dabiri2024coverage,
  title={{LoS Coverage Analysis for UAV-based THz Communication Networks: Towards 3D Visualization of Wireless Networks}},
  author={Dabiri, Mohammad T and Hasna, Mazen and Althunibat, Saud and Qaraqe, Khalid},
  journal={IEEE Transactions on Aerospace and Electronic Systems},
  year={2024},
  publisher={IEEE}
}

@article{lyu2019network,
  title={{Network-connected UAV: 3-D system modeling and coverage performance analysis}},
  author={Lyu, Jiangbin and Zhang, Rui},
  journal={IEEE Internet of Things Journal},
  volume={6},
  number={4},
  pages={7048--7060},
  year={2019},
  publisher={IEEE}
}

@article{sabzehali20213d,
  title={{3D placement and orientation of mmWave-based UAVs for guaranteed LoS coverage}},
  author={Sabzehali, Javad and Shah, Vijay K and Dhillon, Harpreet S and Reed, Jeffrey H},
  journal={IEEE Wireless Communications Letters},
  volume={10},
  number={8},
  pages={1662--1666},
  year={2021},
  publisher={IEEE}
}

@inproceedings{verbeke2016vibration,
  title={{Vibration analysis of a UAV multirotor frame}},
  author={Verbeke, Jon and Debruyne, Stijn},
  booktitle={Proceedings of isma 2016 international conference on noise and vibration engineering},
  pages={2329--2337},
  year={2016},
  organization={KATHOLIEKE UNIV LEUVEN, DEPT WERKTUIGKUNDE}
}

@article{plasencia2012modelling,
  title={Modelling and Analysis of Vibrations in a {UAV} Helicopter with a Vision System},
  author={Plasencia, G Nicol{\'a}s Marichal and Rodr{\'\i}guez, Mar{\'\i}a Tom{\'a}s and Rivera, Salvador Castillo and L{\'o}pez, {\'A}ngela Hern{\'a}ndez},
  journal={International Journal of Advanced Robotic Systems},
  volume={9},
  number={5},
  pages={220},
  year={2012},
  publisher={SAGE Publications Sage UK: London, England}
}

@inproceedings{banerjee2023vibration,
  title={{Vibration Anomaly Indicator in UAVs in presence of Wind}},
  author={Banerjee, Portia and Ghimire, Rajeev and Hale, Elizabeth},
  booktitle={AIAA AVIATION 2023 Forum},
  pages={3860},
  year={2023}
}

@article{ansari2015performance,
  title={Performance analysis of free-space optical links over m{\'a}laga (M) turbulence channels with pointing errors},
  author={Ansari, Imran Shafique and Yilmaz, Ferkan and Alouini, Mohamed-Slim},
  journal={IEEE Transactions on Wireless Communications},
  volume={15},
  number={1},
  pages={91--102},
  year={2015},
  publisher={IEEE}
}

@article{borah2009pointing,
  title={Pointing error effects on free-space optical communication links in the presence of atmospheric turbulence},
  author={Borah, Deva K and Voelz, David G},
  journal={Journal of Lightwave Technology},
  volume={27},
  number={18},
  pages={3965--3973},
  year={2009},
  publisher={IEEE}
}

@article{trung2014pointing,
  title={{Pointing error effects on performance of free-space optical communication systems using SC-QAM signals over atmospheric turbulence channels}},
  author={Trung, Ha Duyen and Pham, Anh T and others},
  journal={AEU-International Journal of Electronics and Communications},
  volume={68},
  number={9},
  pages={869--876},
  year={2014},
  publisher={Elsevier}
}

@article{balaji2018performance,
  title={{Performance evaluation of FSO system using wavelength and time diversity over malaga turbulence channel with pointing errors}},
  author={Balaji, KA and Prabu, K},
  journal={Optics Communications},
  volume={410},
  pages={643--651},
  year={2018},
  publisher={Elsevier}
}

@article{varotsos2018df,
  title={{DF relayed subcarrier FSO links over Malaga turbulence channels with phase noise and non-zero boresight pointing errors}},
  author={Varotsos, George K and Nistazakis, Hector E and Gappmair, Wilfried and Sandalidis, Harilaos G and Tombras, George S},
  journal={Applied Sciences},
  volume={8},
  number={5},
  pages={664},
  year={2018},
  publisher={MDPI}
}

@inproceedings{Ghanbari2023MISOFSO,
  author    = {M. Ghanbari and M. S. Jahed and S. M. Sajad Sadough},
  title     = {On the Interaction Between Meteorological Conditions and Performance Optimization in {MISO} Free-Space Optical Communication},
  booktitle = {31st International Conference on Electrical Engineering (ICEE)},
  address   = {Tehran, Iran},
  year      = {2023},
  pages     = {479--483},
  doi       = {10.1109/ICEE59167.2023.10334716},
  publisher = {IEEE}
}

@article{yang2014free,
  title={Free-space optical communication with nonzero boresight pointing errors},
  author={Yang, Fan and Cheng, Julian and Tsiftsis, Theodoros A},
  journal={IEEE Transactions on Communications},
  volume={62},
  number={2},
  pages={713--725},
  year={2014},
  publisher={IEEE}
}

@article{boluda2016novel,
  title={Novel approximation of misalignment fading modeled by Beckmann distribution on free-space optical links},
  author={Boluda-Ruiz, Rub{\'e}n and Garc{\'\i}a-Zambrana, Antonio and Castillo-V{\'a}zquez, Carmen and Castillo-V{\'a}zquez, Beatriz},
  journal={Optics express},
  volume={24},
  number={20},
  pages={22635--22649},
  year={2016},
  publisher={Optica Publishing Group}
}

@inproceedings{bhatnagar2015performance,
  title={{Performance evaluation of FSO MIMO links in Gamma-Gamma fading with pointing errors}},
  author={Bhatnagar, Manav R and Ghassemlooy, Zabih},
  booktitle={2015 IEEE International Conference on Communications (ICC)},
  pages={5084--5090},
  year={2015},
  organization={IEEE}
}

@article{najafi2020statistical,
  title={{Statistical modeling of the FSO fronthaul channel for UAV-based communications}},
  author={Najafi, Marzieh and Ajam, Hedieh and Jamali, Vahid and Diamantoulakis, Panagiotis D and Karagiannidis, George K and Schober, Robert},
  journal={IEEE Transactions on Communications},
  volume={68},
  number={6},
  pages={3720--3736},
  year={2020},
  publisher={IEEE}
}

@article{dabiri2020channel,
  title={{Channel modeling for UAV-based optical wireless links with nonzero boresight pointing errors}},
  author={Dabiri, Mohammad Taghi and Rezaee, Mohsen and Ansari, Imran Shafique and Yazdanian, Vahid},
  journal={IEEE Transactions on Vehicular Technology},
  volume={69},
  number={12},
  pages={14238--14246},
  year={2020},
  publisher={IEEE}
}

@article{elamassie2023multi,
  title={Multi-Hop Airborne {FSO} Systems With Relay Selection Over Outdated Log-Normal Turbulence Channels},
  author={Elamassie, Mohammed and Al-Shaikhi, Ali A and Sait, Sadiq M and Uysal, Murat},
  journal={IEEE Transactions on Vehicular Technology},
  year={2023},
  publisher={IEEE}
}

@article{khallaf2021composite,
  title={{Composite fading model for aerial MIMO FSO links in the presence of atmospheric turbulence and pointing errors}},
  author={Khallaf, Haitham S and Kato, Kazutoshi and Mohamed, Ehab Mahmoud and Sait, Sadiq M and Yanikomeroglu, Halim and Uysal, Murat},
  journal={IEEE Wireless Communications Letters},
  volume={10},
  number={6},
  pages={1295--1299},
  year={2021},
  publisher={IEEE}
}

@article{bashir2022energy,
  title={{Energy optimization of a laser-powered hovering-UAV relay in optical wireless backhaul}},
  author={Bashir, Muhammad Salman and Alouini, Mohamed-Slim},
  journal={IEEE Transactions on Wireless Communications},
  volume={22},
  number={5},
  pages={3216--3230},
  year={2022},
  publisher={IEEE}
}

@article{safi2020analytical,
  title={{Analytical channel model and link design optimization for ground-to-HAP free-space optical communications}},
  author={Safi, Hossein and Dargahi, Akbar and Cheng, Julian and Safari, Majid},
  journal={Journal of Lightwave Technology},
  volume={38},
  number={18},
  pages={5036--5047},
  year={2020},
  publisher={IEEE}
}

@article{le2024fso,
  title={{FSO-Based HAP-Assisted Multi-UAV Backhauling Over F Channels With Imperfect CSI}},
  author={Le, Hoang D and Nguyen, Tinh V and Mai, Vuong and Pham, Anh T},
  journal={IEEE Transactions on Vehicular Technology},
  year={2024},
  publisher={IEEE}
}

@article{ghanbari2024outage,
  title={{Outage probability minimization by optimal beamwidth selection for UAV-to-HAP FSO links under pointing inaccuracies}},
  author={Ghanbari, Meysam and Ataee, Mahdi and Sadough, Seyed Mohammad Sajad},
  journal={Optik},
  volume={312},
  pages={171953},
  year={2024},
  publisher={Elsevier}
}

@article{xu2024outage,
  title={{Outage Probability and Average BER of UAV-Assisted RF/FSO System for Space-Air-Ground Integrated Networks under Angle-of-Arrival Fluctuations}},
  author={Xu, Guanjun and Lu, Shuyuan and Qu, Lin and Zhang, Qinyu and Song, Zhaohui and Ai, Bo},
  journal={IEEE Internet of Things Journal},
  year={2024},
  publisher={IEEE}
}

@inproceedings{ghanbari2022outage,
  title={{Outage performance analysis for UAV-based mixed underwater-FSO communication under pointing errors}},
  author={Ghanbari, Meysam and Ataee, Mahdi and Sadough, Seyed Mohammad Sajad},
  booktitle={2022 4th West Asian Symposium on Optical and Millimeter-wave Wireless Communications (WASOWC)},
  pages={1--5},
  year={2022},
  organization={IEEE}
}

@article{saber2024security,
  title={{Security Analysis of Integrated HAP-Based FSO and UAV-Enabled RF Downlink Communications}},
  author={Saber, Mohammad Javad and Hasna, Mazen},
  journal={IEEE Open Journal of the Communications Society},
  year={2024},
  publisher={IEEE}
}

@article{alshaer2021reliability,
  title={{Reliability and security analysis of an entanglement-based QKD protocol in a dynamic ground-to-UAV FSO communications system}},
  author={Alshaer, Nancy and Moawad, Ahmed and Ismail, Tawfik},
  journal={IEEE Access},
  volume={9},
  pages={168052--168067},
  year={2021},
  publisher={IEEE}
}

@article{paul2019jamming,
  title={Jamming in free space optical systems: Mitigation and performance evaluation},
  author={Paul, Pratiti and Bhatnagar, Manav R and Jaiswal, Anshul},
  journal={IEEE transactions on communications},
  volume={68},
  number={3},
  pages={1631--1647},
  year={2019},
  publisher={IEEE}
}

@article{saxena2023analysis,
  title={{Analysis of jamming effects in IRS assisted UAV dual-hop FSO communication systems}},
  author={Saxena, Prakriti and Chung, Yeon Ho},
  journal={IEEE Transactions on Vehicular Technology},
  volume={72},
  number={7},
  pages={8956--8971},
  year={2023},
  publisher={IEEE}
}

@article{dabiri2022pointing,
  title={{Pointing error modeling of mmWave to THz high-directional antenna arrays}},
  author={Dabiri, Mohammad Taghi and Hasna, Mazen},
  journal={IEEE Wireless Communications Letters},
  volume={11},
  number={11},
  pages={2435--2439},
  year={2022},
  publisher={IEEE}
}

@article{badarneh2023channel,
  title={{Channel modeling and performance analysis of directional THz links under pointing errors and $\alpha$-$\mu$ distribution}},
  author={Badarneh, Osamah S and Dabiri, Mohammad T and Hasna, Mazen},
  journal={IEEE Communications Letters},
  volume={27},
  number={3},
  pages={812--816},
  year={2023},
  publisher={IEEE}
}

@article{dabiri2022general,
  title={A general model for pointing error of high frequency directional antennas},
  author={Dabiri, Mohammad Taghi and Hasna, Mazen and Zorba, Nizar and Khattab, Tamer and Qaraqe, Khalid A},
  journal={IEEE Open Journal of the Communications Society},
  volume={3},
  pages={1978--1990},
  year={2022},
  publisher={IEEE}
}

@article{badarneh2023general,
  title={{A General Framework for UAV-aided THz communications subject to generalized geometric loss}},
  author={Badarneh, Osamah S and El Bouanani, Faissal and Almehmadi, Fares},
  journal={IEEE Transactions on Vehicular Technology},
  volume={72},
  number={11},
  pages={14589--14600},
  year={2023},
  publisher={IEEE}
}

@inproceedings{bhardwaj2024generalized,
  title={{A generalized statistical model for THz wireless channel with random atmospheric absorption}},
  author={Bhardwaj, Pranay and Khanna, Raghav and Zafaruddin, SM},
  booktitle={2024 IEEE Wireless Communications and Networking Conference (WCNC)},
  pages={1--6},
  year={2024},
  organization={IEEE}
}

@article{bhardwaj2023exact,
  title={{Exact performance analysis of THz link under transceiver hardware impairments}},
  author={Bhardwaj, Pranay and Zafaruddin, Syed Mohammad},
  journal={IEEE Communications Letters},
  volume={27},
  number={8},
  pages={2197--2201},
  year={2023},
  publisher={IEEE}
}

@article{yadav2023terahertz,
  title={{Terahertz Communications Between Unstable Transceivers With Pointing Errors over FTR Fading}},
  author={Yadav, Anil and Mallik, Ranjan K},
  journal={IEEE Open Journal of the Communications Society},
  year={2023},
  publisher={IEEE}
}

@article{almeida2024cascaded,
  title={Cascaded and Sum of Cascaded Over $\boldsymbol{\alpha }$–$\boldsymbol{\mathcal {F}}$ Fading Channels With Pointing Error Impairment},
  author={Almeida, Pedro HD and Silva, Hugerles S and Dias, Ugo S and de Souza, Rausley AA and Fonseca, Iguatemi E and Li, Yonghui},
  journal={IEEE Open Journal of Vehicular Technology},
  year={2024},
  publisher={IEEE}
}

@article{jemaa2024performance,
  title={{Performance Analysis of Outdoor THz Links under Mixture Gamma Fading with Misalignment}},
  author={Jemaa, Hakim and Tarboush, Simon and Sarieddeen, Hadi and Alouini, Mohamed-Slim and Al-Naffouri, Tareq Y},
  journal={IEEE Communications Letters},
  year={2024},
  publisher={IEEE}
}

@article{dabiri2024enabling,
  title={{Enabling Flexible Arial Backhaul Links for Post Disasters: A Design using UAV Swarms and Distributed Charging Stations}},
  author={Dabiri, Mohammad T and Hasna, Mazen and Zorba, Nizar and Khattab, Tamer},
  journal={IEEE Open Journal of Vehicular Technology},
  year={2024},
  publisher={IEEE}
}

@article{yadav2024performance,
  title={{A Performance Study of Relay-Assisted Double-Hop Hybrid RF-THz Wireless Systems}},
  author={Yadav, Anil and Mallik, Ranjan K},
  journal={IEEE Transactions on Communications},
  year={2024},
  publisher={IEEE}
}

@inproceedings{chapala2024multiple,
  title={{Multiple RIS Transmissions With Multihop Relaying for THz Wireless System}},
  author={Chapala, Vinay Kumar and Durgada, Omkar R and Sood, Amay and Gupta, Vishal and Kapur, Rohan and Zafaruddin, SM},
  booktitle={2024 IEEE International Black Sea Conference on Communications and Networking (BlackSeaCom)},
  pages={7--12},
  year={2024},
  organization={IEEE}
}

@ARTICLE{8796365,
  author={Basar, Ertugrul and Di Renzo, Marco and De Rosny, Julien and Debbah, Merouane and Alouini, Mohamed-Slim and Zhang, Rui},
  journal={IEEE Access}, 
  title={Wireless Communications Through Reconfigurable Intelligent Surfaces}, 
  year={2019},
  volume={7},
  number={},
  pages={116753-116773},
  doi={10.1109/ACCESS.2019.2935192}}

@ARTICLE{9140329,
  author={Di Renzo, Marco and Zappone, Alessio and Debbah, Merouane and Alouini, Mohamed-Slim and Yuen, Chau and de Rosny, Julien and Tretyakov, Sergei},
  journal={IEEE Journal on Selected Areas in Communications}, 
  title={Smart Radio Environments Empowered by Reconfigurable Intelligent Surfaces: How It Works, State of Research, and The Road Ahead}, 
  year={2020},
  volume={38},
  number={11},
  pages={2450-2525},
  doi={10.1109/JSAC.2020.3007211}}

@ARTICLE{8811733,
  author={Wu, Qingqing and Zhang, Rui},
  journal={IEEE Transactions on Wireless Communications}, 
  title={Intelligent Reflecting Surface Enhanced Wireless Network via Joint Active and Passive Beamforming}, 
  year={2019},
  volume={18},
  number={11},
  pages={5394-5409},
  doi={10.1109/TWC.2019.2936025}}

@article{cui2014coding,
  title={Coding metamaterials, digital metamaterials and programmable metamaterials},
  author={Cui, Tie Jun and Qi, Mei Qing and Wan, Xiang and Zhao, Jie and Cheng, Qiang},
  journal={Light: science \& applications},
  volume={3},
  number={10},
  pages={e218--e218},
  year={2014},
  publisher={Nature Publishing Group}
}

@ARTICLE{8741198,
  author={Huang, Chongwen and Zappone, Alessio and Alexandropoulos, George C. and Debbah, Mérouane and Yuen, Chau},
  journal={IEEE Transactions on Wireless Communications}, 
  title={Reconfigurable Intelligent Surfaces for Energy Efficiency in Wireless Communication}, 
  year={2019},
  volume={18},
  number={8},
  pages={4157-4170},
  doi={10.1109/TWC.2019.2922609}}

@ARTICLE{10243495,
  author={Chepuri, Sundeep Prabhakar and Shlezinger, Nir and Liu, Fan and Alexandropoulos, George C. and Buzzi, Stefano and Eldar, Yonina C.},
  journal={IEEE Signal Processing Magazine}, 
  title={Integrated Sensing and Communications With Reconfigurable Intelligent Surfaces: From signal modeling to processing}, 
  year={2023},
  volume={40},
  number={6},
  pages={41-62},
  doi={10.1109/MSP.2023.3279986}}

@INPROCEEDINGS{9473762,
  author={Fu, Min and Zhou, Yong and Shi, Yuanming},
  booktitle={2021 IEEE International Conference on Communications Workshops (ICC Workshops)}, 
  title={Reconfigurable Intelligent Surface for Interference Alignment in MIMO Device-to-Device Networks}, 
  year={2021},
  volume={},
  number={},
  pages={1-6},
  doi={10.1109/ICCWorkshops50388.2021.9473762}}

@ARTICLE{9690474,
  author={Deng, Ruoqi and Di, Boya and Zhang, Hongliang and Niyato, Dusit and Han, Zhu and Poor, H. Vincent and Song, Lingyang},
  journal={IEEE Wireless Communications}, 
  title={Reconfigurable Holographic Surfaces for Future Wireless Communications}, 
  year={2021},
  volume={28},
  number={6},
  pages={126-131},
  doi={10.1109/MWC.001.2100204}}

@ARTICLE{10557617,
  author={Al-Badarneh, Yazan H. and Alshawaqfeh, Mustafa K. and Badarneh, Osamah S. and Khattabi, Yazid M.},
  journal={IEEE Transactions on Cognitive Communications and Networking}, 
  title={Performance Analysis of {RIS}-Assisted Spectrum Sharing Systems}, 
  year={2025},
  volume={11},
  number={1},
  pages={465-474},
  doi={10.1109/TCCN.2024.3414395}}

@article{alshawaqfeh2025performance,
  title={On the performance of RIS-assisted energy harvesting over mixed dual-hop RF-FSO communications},
  author={Alshawaqfeh, Mustafa K and Badarneh, Osamah S and Al-Badarneh, Yazan H},
  journal={Physical Communication},
  pages={102615},
  year={2025},
  publisher={Elsevier}
}

@ARTICLE{9882367,
  author={Zhang, Bingxin and Yang, Kun and Wang, Kezhi and Zhang, Guopeng},
  journal={IEEE Transactions on Vehicular Technology}, 
  title={Performance Analysis of RIS-Assisted Wireless Communications With Energy Harvesting}, 
  year={2023},
  volume={72},
  number={1},
  pages={1325-1330},
  doi={10.1109/TVT.2022.3205448}}

@article{bartoli2023spatial,
  title={Spatial multiplexing in near field MIMO channels with reconfigurable intelligent surfaces},
  author={Bartoli, Giulio and Abrardo, Andrea and Decarli, Nicolo and Dardari, Davide and Di Renzo, Marco},
  journal={IET Signal Processing},
  volume={17},
  number={3},
  pages={e12195},
  year={2023},
  publisher={Wiley Online Library}
}

@ARTICLE{10552229,
  author={Ioannou, Iacovos I. and Raspopoulos, Marios and Nagaradjane, Prabagarane and Christophorou, Christophoros and Ali Aziz, Waqar and Vassiliou, Vasos and Pitsillides, Andreas},
  journal={IEEE Access}, 
  title={DeepRISBeam: Deep Learning-Based RIS Beam Management for Radio Channel Optimization}, 
  year={2024},
  volume={12},
  number={},
  pages={81646-81681},
  doi={10.1109/ACCESS.2024.3411929}}

@ARTICLE{10559954,
  author={Toka, Mesut and Lee, Byungju and Seong, Jaehyup and Kaushik, Aryan and Lee, Juhwan and Lee, Jungwoo and Lee, Namyoon and Shin, Wonjae and Poor, H. Vincent},
  journal={IEEE Communications Magazine}, 
  title={RIS-Empowered LEO Satellite Networks for 6G: Promising Usage Scenarios and Future Directions}, 
  year={2024},
  volume={62},
  number={11},
  pages={128-135},
  doi={10.1109/MCOM.002.2300554}}

@ARTICLE{9703337,
  author={Pogaku, Arjun Chakravarthi and Do, Dinh-Thuan and Lee, Byung Moo and Nguyen, Nhan Duc},
  journal={IEEE Access}, 
  title={UAV-Assisted RIS for Future Wireless Communications: A Survey on Optimization and Performance Analysis}, 
  year={2022},
  volume={10},
  number={},
  pages={16320-16336},
  doi={10.1109/ACCESS.2022.3149054}}

@article{chen2022reconfigurable,
  title={Reconfigurable intelligent surface (RIS)-aided vehicular networks: Their protocols, resource allocation, and performance},
  author={Chen, Yuanbin and Wang, Ying and Zhang, Jiayi and Zhang, Ping and Hanzo, Lajos},
  journal={IEEE Vehicular Technology Magazine},
  volume={17},
  number={2},
  pages={26--36},
  year={2022},
  publisher={IEEE}
}

@ARTICLE{9530530,
  author={Chapala, Vinay Kumar and Zafaruddin, S. M.},
  journal={IEEE Communications Letters}, 
  title={Exact Analysis of RIS-Aided THz Wireless Systems Over $\alpha$-$\mu$ Fading With Pointing Errors}, 
  year={2021},
  volume={25},
  number={11},
  pages={3508-3512},
  doi={10.1109/LCOMM.2021.3110865}}

@ARTICLE{10361836,
  author={Zhou, Hao and Erol-Kantarci, Melike and Liu, Yuanwei and Poor, H. Vincent},
  journal={IEEE Communications Surveys and Tutorials}, 
  title={A Survey on Model-Based, Heuristic, and Machine Learning Optimization Approaches in RIS-Aided Wireless Networks}, 
  year={2024},
  volume={26},
  number={2},
  pages={781-823},
  doi={10.1109/COMST.2023.3340099}}

@article{yuan2022reconfigurable,
  title={Reconfigurable intelligent surface relay: Lessons of the past and strategies for its success},
  author={Yuan, Yifei and Wu, Dan and Huang, Yuhong and others},
  journal={IEEE Communications Magazine},
  volume={60},
  number={12},
  pages={117--123},
  year={2022},
  publisher={IEEE}
}

@article{yang2021energy,
  title={Energy-efficient wireless communications with distributed reconfigurable intelligent surfaces},
  author={Yang, Zhaohui and Chen, Mingzhe and Saad, Walid and Xu, Wei and Shikh-Bahaei, Mohammad and Poor, H Vincent and Cui, Shuguang},
  journal={IEEE Transactions on Wireless Communications},
  volume={21},
  number={1},
  pages={665--679},
  year={2021},
  publisher={IEEE}
}

@ARTICLE{7263349,
  author={Dai, Linglong and Wang, Bichai and Yuan, Yifei and Han, Shuangfeng and Chih-lin, I. and Wang, Zhaocheng},
  journal={IEEE Communications Magazine}, 
  title={Non-orthogonal multiple access for 5G: solutions, challenges, opportunities, and future research trends}, 
  year={2015},
  volume={53},
  number={9},
  pages={74-81},
  doi={10.1109/MCOM.2015.7263349}}

@ARTICLE{7676258,
  author={Islam, S. M. Riazul and Avazov, Nurilla and Dobre, Octavia A. and Kwak, Kyung-sup},
  journal={IEEE Communications Surveys and Tutorials}, 
  title={Power-Domain Non-Orthogonal Multiple Access (NOMA) in 5G Systems: Potentials and Challenges}, 
  year={2017},
  volume={19},
  number={2},
  pages={721-742},
  doi={10.1109/COMST.2016.2621116}}

@article{akbar2021noma,
  title={NOMA and 5G emerging technologies: A survey on issues and solution techniques},
  author={Akbar, Aamina and Jangsher, Sobia and Bhatti, Farrukh A},
  journal={Computer Networks},
  volume={190},
  pages={107950},
  year={2021},
  publisher={Elsevier}
}

@article{islam2021impact,
  title={Impact of Correlation and Pointing Error on Secure Outage Performance Over Arbitrary Correlated Nakagami-$m$ and $\mathcal{M}$-Turbulent Fading Mixed RF-FSO Channel},
  author={Islam, Sheikh Habibul and Badrudduza, ASM and Islam, SM Riazul and Shahid, Fardin Ibne and Ansari, Imran Shafique and Kundu, Milton Kumar and Yu, Heejung},
  journal={IEEE Photonics Journal},
  volume={13},
  number={2},
  pages={1--17},
  year={2021},
  publisher={IEEE}
}

@article{xiao2017millimeter,
  title={Millimeter wave communications for future mobile networks},
  author={Xiao, Ming and Mumtaz, Shahid and Huang, Yongming and Dai, Linglong and Li, Yonghui and Matthaiou, Michail and Karagiannidis, George K and Bj{\"o}rnson, Emil and Yang, Kai and Ghosh, Amitabha and others},
  journal={IEEE Journal on Selected Areas in communications},
  volume={35},
  number={9},
  pages={1909--1935},
  year={2017},
  publisher={IEEE}
}

@article{dong2019adaptive,
  title={Adaptive power allocation scheme for mobile NOMA visible light communication system},
  author={Dong, Zanyang and Shang, Tao and Li, Qian and Tang, Tang},
  journal={Electronics},
  volume={8},
  number={4},
  pages={381},
  year={2019},
  publisher={MDPI}
}

@article{liu2020robust,
  title={Robust adaptive beam tracking for mobile millimetre wave communications},
  author={Liu, Chunshan and Li, Min and Zhao, Lou and Whiting, Philip and Hanly, Stephen Vaughan and Collings, Iain B and Zhao, Minjian},
  journal={IEEE Transactions on Wireless Communications},
  volume={20},
  number={3},
  pages={1918--1934},
  year={2020},
  publisher={IEEE}
}

@article{iswarya2021survey,
  title={A survey on successive interference cancellation schemes in non-orthogonal multiple access for future radio access},
  author={Iswarya, N and Jayashree, LS},
  journal={Wireless Personal Communications},
  volume={120},
  number={2},
  pages={1057--1078},
  year={2021},
  publisher={Springer}
}

@article{giang2020hybrid,
  title={Hybrid NOMA/OMA-based dynamic power allocation scheme using deep reinforcement learning in 5G networks},
  author={Giang, Hoang Thi Huong and Hoan, Tran Nhut Khai and Thanh, Pham Duy and Koo, Insoo},
  journal={Applied Sciences},
  volume={10},
  number={12},
  pages={4236},
  year={2020},
  publisher={MDPI}
}

@inproceedings{suganuma2019hybrid,
  title={Hybrid multiple access using simultaneously NOMA and OMA},
  author={Suganuma, Hirofumi and Suenaga, Hiroaki and Maehara, Fumiaki},
  booktitle={2019 International Symposium on Intelligent Signal Processing and Communication Systems (ISPACS)},
  pages={1--2},
  year={2019},
  organization={IEEE}
}

@article{kara2021lightweight,
  title={A lightweight machine learning assisted power optimization for minimum error in NOMA-CRS over Nakagami-$ m $ channels},
  author={Kara, Ferdi and Kaya, Hakan and Yanikomeroglu, Halim},
  journal={IEEE transactions on Vehicular Technology},
  volume={70},
  number={10},
  pages={11067--11072},
  year={2021},
  publisher={IEEE}
}

@article{yang2021machine,
  title={Machine learning for user partitioning and phase shifters design in RIS-aided NOMA networks},
  author={Yang, Zhong and Liu, Yuanwei and Chen, Yue and Al-Dhahir, Naofal},
  journal={IEEE Transactions on Communications},
  volume={69},
  number={11},
  pages={7414--7428},
  year={2021},
  publisher={IEEE}
}

\end{document}